\documentclass[12pt,american,round,hidelinks]{article}

\usepackage[a4paper]{geometry}
\usepackage{babel}
\usepackage[T1]{fontenc}
\usepackage[latin9]{inputenc}

\usepackage{times}
\usepackage{microtype}

\usepackage{datetime}
\ddmmyyyydate

\usepackage{setspace}

\usepackage{enumitem}
\setlist[itemize]{leftmargin=\parindent, topsep=0.5em} 
\setlist{nosep} 

\usepackage{titlesec}
\titlespacing*{\paragraph}{0pt}{\medskipamount}{1em}

\usepackage[authoryear]{natbib}
\usepackage[pdfencoding=auto, psdextra]{hyperref}
\hypersetup{
  pdfborder={0 0 0},
  pdfborderstyle={},
  colorlinks=false}
\usepackage{bookmark}

\makeatletter

\usepackage{amsmath}
\usepackage{amsthm}
\usepackage{amssymb}
\usepackage{mathtools}
\usepackage{bbm}

\theoremstyle{definition}
  
\theoremstyle{plain}
  \newtheorem{lemma}{\protect\lemmaname}
\theoremstyle{plain}
  \newtheorem{proposition}{\protect\propositionname}
\providecommand{\definitionname}{Definition}
\providecommand{\lemmaname}{Lemma}
\providecommand{\propositionname}{Proposition}
\theoremstyle{definition}

\usepackage{float}
\usepackage{graphicx}

\usepackage{epstopdf}

\def\FIGURE#1#2#3{%
    \centering
  \begin{minipage}{ 0.9 \textwidth}
    {\centering \baselineskip=12pt
    #1 \\[-.5em]
    \caption[]{#2}}
  {\footnotesize \emph{Notes:} #3}
  \end{minipage}
}

\usepackage{caption}
\@ifundefined{showcaptionsetup}{}{%
 \PassOptionsToPackage{caption=false}{subfig}}
\usepackage[position=top]{subfig}

\makeatother

\usepackage{chngcntr}

\usepackage{subfiles}

\AtEndDocument{%
    \ifSubfilesClassLoaded{
        \begin{singlespace}
        \bibliographystyle{chicago}
        \bibliography{library}
        \end{singlespace}
    }{}

}

\usepackage[flushmargin]{footmisc}

\usepackage[flushleft]{threeparttable}
\usepackage{booktabs,multirow,multicol}

\usepackage{fancyhdr}
\graphicspath{{./tables_figures/}}

\begin{document}
\title{Trade-Offs Between Ranking Objectives: Descriptive Evidence and Structural Estimation}
\author{Rafael P. Greminger\thanks{UCL School of Management, University College London, \protect\href{mailto:r.greminger@ucl.ac.uk}{r.greminger@ucl.ac.uk}. This paper was previously circulated as ``Heterogeneous Position Effects and the Power of Rankings'' and ``Trade-Offs Between Ranking Objectives: Reduced-Form Evidence and Structural Estimation.'' I am deeply grateful to my advisors, Tobias Klein and Jaap Abbring, for their thoughtful guidance and support. I also thank Bart Bronnenberg, Jean-Pierre Dub\'e, Dorothee Hillrichs, Yufeng Huang, Maarten Janssen, Aaron Kaye, Jos\'{e} L. Moraga, Ilya Morozov, Stephan Seiler, Brad Shapiro, Raluca Ursu; seminar participants at Northwestern University, Rochester University, University of California Berkeley, University of Chicago, University of Frankfurt, University of Passau; and members of the Structural Econometrics Group in Tilburg for excellent comments. Finally, I am grateful to the Nederlandse Organisatie voor Wetenschappelijk Onderzoek (NWO) for financial support through Research Talent grant 406.18.568. } }
\date{This version: \today \\ \vspace{0.1em}
First version: June 30, 2021}
\maketitle
\begin{abstract} \small

  \noindent When designing product rankings, online retailers and platforms choose which outcome to maximize: revenues from commissions or markups, the number of transactions, or consumer welfare. These objectives need not align, creating potential trade-offs. This paper studies how rankings differ between objectives and quantifies the resulting trade-offs. I provide descriptive evidence showing that lower-priced and high-utility alternatives gain more demand when ranked higher, suggesting that ranking them higher increases transactions and consumer welfare but may decrease revenues. To quantify these trade-offs, I develop and estimate a structural demand model based on the search and discovery framework of \cite{Greminger2021} and construct rankings for each objective. The results show that these counterfactual rankings all increase consumer welfare, transactions, and platform revenues relative to a neutral benchmark and the status quo, and that trade-offs between these rankings are limited.

\end{abstract}

\newpage

\onehalfspacing

\section{Introduction}

Online retailers and search platforms typically present products on ranked product lists. The ranking of these lists influences which products consumers search and choose \citep[e.g.,][]{Ursu2018}, and thus affects three outcomes relevant for platforms: revenues from commissions or markups, the number of transactions, and consumer welfare. Platforms can design rankings to maximize any of these outcomes. These objectives need not align, creating potential trade-offs between them.

Understanding and quantifying these trade-offs is important for platforms designing rankings. When trade-offs are substantial, platforms have to carefully consider which objective to target. For example, suppose the revenue-maximizing ranking substantially decreases transactions and consumer welfare, whereas the transaction-maximizing ranking only slightly decreases revenues and consumer welfare. In this case, concerns about customer churn could lead a platform to implement the transaction-maximizing ranking, even when maximizing revenues is its main objective.\footnote{\cite{Donnelly2023} find that a large online platform considers consumer welfare when personalizing rankings.} In contrast, when trade-offs are limited, platforms can implement rankings targeting one objective with only limited effects on others.

In this paper, I study how rankings differ between the three objectives and quantify the resulting trade-offs. First, I show how heterogeneity in \emph{position effects}---how the position on the list affects the demand for an alternative---determines how rankings differ between objectives. I then develop an empirical framework that microfounds this heterogeneity and construct counterfactual rankings for the different objectives. Using the model, I quantify the effects of these rankings and find that the trade-offs between them are limited.

Rankings determine the three outcomes through position effects: products ranked higher are purchased more often \citep[e.g.,][]{Ghose2014, Ursu2018}. Platforms can thus steer demand toward alternatives by ranking them higher. This demand steering affects consumers by determining the products they buy. It also affects platform revenues through a \emph{price effect}: steering demand toward higher-priced products increases the average transaction price, and thus revenues from commissions or markups that are a share of transaction prices.\footnote{I only consider revenues through sales commissions or markups that are a percentage of the transaction price. Ecommerce platforms may also have other revenue sources (e.g., sponsored slots) that are outside the scope of this paper.} However, not all products gain demand equally when ranked higher. Such heterogeneity in position effects creates a \emph{demand effect}: ranking products with larger position effects higher increases transactions because these products gain disproportionately more demand.

How position effects vary across products determines how rankings differ between objectives. For example, if higher-priced products have larger position effects, the demand and price effects align: ranking higher-priced products higher increases both transactions and revenues. In contrast, if lower-priced products have larger position effects, ranking them higher increases transactions but may decrease revenues depending on the relative strength of the two effects. How such rankings affect consumer welfare also depends on which products have larger position effects: ranking products with larger position effects higher steers consumers toward these products, which benefits consumers only if these products also offer higher utility.

It is \emph{a priori} unclear which products have larger position effects. Depending on how consumers search and choose products on ranked product lists, lower-priced and high-utility products could have smaller or larger position effects. Yet, despite this importance for designing rankings, empirical evidence on how position effects vary across products on ranked lists remains limited.\footnote{Only \cite{Ghose2014} and \cite{Derakhshan2022} document some heterogeneity in position effects but do not show how position effects vary with price or other attributes. The literature on position effects in search advertising---a different context from ranked product lists---has more closely studied their heterogeneity and finds that smaller or less prominent advertisers have larger position effects \citep[e.g.,][]{Narayanan2015,Jeziorski2018}.}

I estimate this heterogeneity using click-stream data with exogenous ranking variation and obtain two main results. First, controlling for other attributes, lower-priced alternatives have larger position effects: they gain more searches and purchases when moved up on the list. Second, high-utility alternatives---those that are more often searched and purchased (reflecting appeal across price and other attributes)---also tend to have larger position effects.

These results suggest that the trade-offs between maximizing transactions and consumer welfare are limited: maximizing transactions requires ranking high-utility alternatives higher, which also benefits consumers by steering them toward these alternatives. The results also reveal that two opposing effects determine the trade-offs with the third objective, maximizing revenues. Ranking higher-priced products higher increases revenues per transaction but reduces transactions because lower-priced alternatives have larger position effects. The revenue-maximizing ranking therefore has to balance these demand and price effects, and their relative strength determines how it differs from the other two rankings.

Quantifying how these opposing effects balance and how different rankings affect the three objectives requires a structural demand model. To this end, I develop and estimate a structural search model that builds on the search and discovery framework of \cite{Greminger2021}. I show how this framework microfounds the estimated heterogeneity in position effects and introduce rankings targeting the different outcomes. Estimating the model and simulating the effects of these counterfactual rankings then allows me to quantify the trade-offs between them.

The results reveal that the proposed rankings substantially increase consumer welfare, transactions, and platform revenues relative to a neutral randomized ranking and to the platform's own ranking. The model captures that high-utility alternatives have larger position effects, so ranking them higher increases transactions and benefits consumers. This demand effect is strong enough that maximizing revenues also requires ranking high-utility alternatives higher rather than steering demand toward expensive low-utility alternatives. As a result, the proposed rankings improve all objectives independent of the one they were designed for, and trade-offs between these rankings are limited.

The descriptive evidence and counterfactual analysis together offer important insights for platforms and regulators. The finding that the proposed rankings improve all objectives suggests that platforms can implement rankings targeting one objective without substantial detrimental effects on others, making the choice of objective less consequential than prior results indicate \citep[][]{Ursu2018, Compiani2023}. It also suggests that platforms aiming to increase consumer welfare can implement rankings designed to maximize transactions, an outcome observed in the data and thus easier to target. I also find that trade-offs remain limited regardless of whether consumers update their beliefs in response to the ranking change, a result relevant for regulators: consumers not realizing that the ranking maximizes platform revenues does not imply that these rankings harm consumers.\footnote{In a recent report, the UK's Competition and Markets Authority raises the concern that firms can exploit ranking effects in ways that are not transparent to (or understood by) consumers and could come at their expense \citep[][pp. 22--23]{CMA2021}.}

\paragraph{Methodological Contributions.} This paper offers several methodological contributions. First, I develop an empirical version of \citeauthor{Greminger2021}'s \citeyearpar{Greminger2021} search and discovery model. In this model, consumers arrive on the platform and initially see only the first few alternatives on the ranked product list. They then sequentially decide between scrolling down the list to discover more products and viewing detail pages of previously discovered products to learn more about them. Notably, the model also captures that detail pages reveal vertical product attributes, which \cite{Compiani2023} show to be important for optimal rankings.

Rankings in the search and discovery model influence consumers' choices by determining which alternatives they discover---a mechanism I show predicts the observed heterogeneity in position effects. This contrasts with \cite{Weitzman1979} models, where rankings affect choices only by determining the cost of searching different alternatives. In a comparison, I show that the different mechanism lets the search and discovery model better capture position effects in purchases and their heterogeneity, and thus the demand effect of rankings. This difference translates to different counterfactual results, leading to more limited trade-offs between ranking objectives than suggested by prior work estimating Weitzman models on the same data \citep[][]{Ursu2018, Compiani2023}.

Second, I develop a computationally efficient estimation approach for the search and discovery model. Similar to the Weitzman model, estimating the search and discovery model via maximum likelihood is challenging because it requires computing the likelihood of a sequence of search decisions. Moreover, I do not observe this sequence: the data only reveal which products consumers searched and purchased, not how far they scrolled or the order of searches.

To address these challenges, I extend the generalized eventual purchase theorem of \cite{Greminger2021} and obtain a characterization of a consumer's searches and eventual purchase that is independent of the choice sequence and substantially simplifies the likelihood computation. Notably, my approach also applies to the Weitzman model, where it offers a computationally efficient alternative to existing simulated maximum likelihood estimation approaches \citep[e.g.,][]{Jiang2021}.

Finally, I develop ranking algorithms for three objectives: maximizing revenues, transactions, and consumer welfare. Comparing all possible rankings would yield the optimal ranking for each objective, but the number of possible rankings is typically too large for an exhaustive comparison. By introducing heuristic rankings and showing that they perform well, this paper also offers platforms effective ways to rank alternatives for these objectives.

For maximizing consumer welfare and transactions, I propose the \emph{Discovery-Value Ranking}, which ranks products by an index similar to the discovery value governing consumers' decisions on whether to discover more products. For maximizing revenues, I propose the \emph{Bottom-Up Ranking}, which iteratively ranks products from the bottom of the list. I show that both proposed rankings perform well in the empirical application, outperforming several other heuristics and substantially improving on the platform's own ranking.

\paragraph{Related Literature.} This paper contributes to the empirical literature on position effects and rankings. Prior work highlights the importance of position effects on ranked product lists \citep[e.g.,][]{Ghose2012, Ghose2014, DeLosSantos2017, Ursu2018} and quantifies how rankings affect different outcomes \citep[e.g.,][]{Zhang2021a, Donnelly2023, Korganbekova2023, Kaye2024, JameeiOsgouei2025}. Closest to this paper are \cite{Ursu2018} and \cite{Compiani2023}, who use the same data as this paper. \cite{Ursu2018} estimates a Weitzman model and shows that platforms can use utility-based rankings to increase consumer welfare, with mixed effects for revenues.\footnote{\cite{Ursu2018} analyzes ranking effects for four destinations separately. Platform revenues substantially decrease in one, slightly decrease in two more destinations, and substantially increase in another destination.} \cite{Compiani2023} generalize the Weitzman model to allow consumers to reveal vertical attributes on product detail pages and find substantial trade-offs between maximizing revenues and consumer welfare. While these papers document trade-offs between rankings, I complement them by showing \emph{how} heterogeneity in position effects determines these trade-offs through the opposing demand and price effects. Moreover, I develop and estimate a model that microfounds this heterogeneity and find more limited trade-offs.

My empirical framework also offers advantages over other search models used to quantify ranking effects.\footnote{By studying rankings, this paper differs from \cite{Zhang2023}, who use the search and discovery model to study how search routes affect search behavior, while abstracting from rankings.} It relaxes a limitation of ``top-down'' search models \citep[][]{Chan2015, Choi2019} by allowing consumers to search products discovered earlier. By integrating click and scroll decisions in a sequential framework, my model further accommodates consumers who continue scrolling after clicking, unlike two-stage models that separate these decisions \citep[][]{Derakhshan2022, Chung2024}. My model also differs from that of \cite{Gibbard2022, Gibbard2023} in that consumers discover products by scrolling down a product list, rather than deciding on an order in which to ``browse'' products.

My estimation approach adds to existing approaches for Weitzman models. Unlike existing approaches, it does not require specifying smoothing parameters \citep[][]{Honka2017, Ursu2018} or observing the search sequence \citep[][]{Jiang2021,Morozov2023,  Chung2023, Onzo2025}. It also avoids integrating over all possible search sequences \citep[][]{Compiani2023}, which can be used when the search order is not observed but is computationally costly unless consumers search very few alternatives. Other recent approaches use indirect inference \citep[][]{Sullivan2024} or neural networks \citep[][]{Wei2025}, which avoid computing a likelihood or posterior but require choosing auxiliary models.

Finally, the ranking algorithms I introduce contribute to the operations literature \citep[e.g.,][]{Ryzin1999}. Unlike prior work, I derive these algorithms for an empirical framework that microfounds a key factor determining differences between rankings and use it to quantify their effects.\footnote{Early work did not consider search \citep[e.g.,][]{Ryzin1999, Talluri2004}. \citet{UrsuDzyabura2018} focus on placing independent categories within a retailer. \citet{Chu2020} and \citet{Derakhshan2022} develop algorithms for search models that contradict the data and thus would be difficult to estimate. The former implies no position effects for consumers choosing the outside option, contradicting my data (see Online Appendix \ref{subsec:proof-nonincreasing-mu-h}). The latter implies consumers inspect all alternatives up to some rank, contradicting data where consumers skip the first alternative. \cite{Kosilova2025} derive optimal rankings for a Weitzman model under specific distributional assumptions, without empirical application.}

\section{Data and Descriptive Evidence\label{sec:Data-and-Descriptive}}

I use click-stream data from Expedia. The data can be obtained from Kaggle.com and contain information on clicks and purchases for 166,039 search sessions between November 2012 and June 2013.\footnote{The data are available at \url{https://www.kaggle.com/c/expedia-personalized-sort/data}.} A search session starts when a consumer submits a query for a hotel stay on Expedia. Following this query, Expedia presents a list of available options. The list displays several hotel characteristics (e.g., price per night, review score). Consumers interact with the list by scrolling down to reveal more hotels or by clicking on a hotel to view its detail page, which reveals more information and allows them to book the hotel. \citet{Ursu2018} provides a comprehensive discussion of the dataset, and I apply similar criteria to prepare the final sample (see Online Appendix \ref{subsec:Data-preparation}).

\subsection{Data Summary\label{subsec:Summary}}

The main feature of the data is that for about 30\% of search sessions, Expedia ranked hotels that fit the query randomly. As \cite{Ursu2018} highlights, this exogenous variation is necessary to identify position effects without confounding them with the effect of more desirable hotels being ranked higher. For the remaining 70\% of sessions in the sample, Expedia used its ranking algorithm to assign hotels to positions.

Table \ref{tab:summary_stats_rr} summarizes the dataset at the hotel and session level for consumers who observed the randomized ranking.\footnote{Online Appendix \ref{subsec:Data-preparation} provides a detailed description of each variable. The ``no reviews'' variable is a dummy indicating whether a hotel has no reviews. This is coded as a ``review score'' of zero in the raw data. However, given that it differs from both a ``review score'' of zero and a missing ``review score,'' I treat this dummy separately.} In total, there are 51,510 sessions in this sample. On average, there are 1.14 clicks per session, and about 8\% of sessions end with a hotel booking. The number of alternatives on the product list varies between 5 and 38 across sessions. This variation does not result from consumers not browsing further, but from queries for hotels in destinations or on dates where only a few hotels had available rooms. Some destinations may also offer more alternatives, but the data contain only the results displayed on the first page. As \citet{Ursu2018} notes, this imposes little restriction because position effects are identified from differences across positions.

\begin{table}[tb] \centering \footnotesize
\caption{Summary Statistics (Randomized Ranking)} \label{tab:summary_stats_rr}

\begin{threeparttable}
\begin{tabular}{lllccccc} 
\midrule

 && \multicolumn{1}{c}{N} & \multicolumn{1}{c}{Mean} & \multicolumn{1}{c}{Median} & \multicolumn{1}{c}{Std. Dev} & \multicolumn{1}{c}{Min.} & \multicolumn{1}{c}{Max.} \\
\midrule
\bfseries{Hotel-level} \\
\hspace{1em}Price (in \$)&&1,357,106&171.70&141.04&114.03&10.00&1000.00\\
\hspace{1em}Star rating&&1,333,734&\hphantom{00}3.34&\hphantom{00}3&\hphantom{00}0.89&\hphantom{0}1&\hphantom{0}5\\
\hspace{1em}Review score&&1,354,996&\hphantom{00}3.81&\hphantom{00}4.00&\hphantom{00}0.97&\hphantom{0}0.00&\hphantom{0}5.00\\
\hspace{1em}No reviews&&1,354,996&\hphantom{00}0.04&\hphantom{00}0&\hphantom{00}0.19&\hphantom{0}0&\hphantom{0}1\\
\hspace{1em}Chain&&1,357,106&\hphantom{00}0.62&\hphantom{00}1.00&\hphantom{00}0.48&\hphantom{0}0.00&\hphantom{0}1.00\\
\hspace{1em}Location score&&1,357,106&\hphantom{00}3.26&\hphantom{00}3&\hphantom{00}1.53&\hphantom{0}0&\hphantom{0}7\\
\hspace{1em}On promotion&&1,357,106&\hphantom{00}0.24&\hphantom{00}0&\hphantom{00}0.43&\hphantom{0}0&\hphantom{0}1\\
\midrule
\bfseries{Session-level} \\
\hspace{1em}Number of items&&51,510&\hphantom{0}26.35&\hphantom{0}31&\hphantom{00}8.46&\hphantom{0}5&\hphantom{00}38\\
\hspace{1em}Number of clicks&&51,510&\hphantom{00}1.14&\hphantom{00}1&\hphantom{00}0.66&\hphantom{0}1&\hphantom{00}25\\
\hspace{1em}Made booking&&51,510&\hphantom{00}0.08&\hphantom{00}0&\hphantom{00}0.27&\hphantom{0}0&\hphantom{0}1\\
\hspace{1em}Trip length (in days)&&51,510&\hphantom{00}3.07&\hphantom{00}2&\hphantom{00}2.42&\hphantom{0}1&\hphantom{00}40\\
\hspace{1em}Booking window (in days)&&51,510&\hphantom{0}53.67&\hphantom{0}31&\hphantom{0}62.49&\hphantom{0}0&\hphantom{0}498\\
\hspace{1em}Number of adults&&51,510&\hphantom{00}2.08&\hphantom{00}2&\hphantom{00}0.94&\hphantom{0}1&\hphantom{0}9\\
\hspace{1em}Number of children&&51,510&\hphantom{00}0.43&\hphantom{00}0&\hphantom{00}0.82&\hphantom{0}0&\hphantom{0}9\\
\hspace{1em}Number of rooms&&51,510&\hphantom{00}1.14&\hphantom{00}1&\hphantom{00}0.46&\hphantom{0}1&\hphantom{0}8\\
\midrule 
\end{tabular}
\begin{tablenotes}
\item \footnotesize{\emph{Notes:}  Summary statistics for sessions under the randomized ranking. }
\end{tablenotes} 
\end{threeparttable}
\end{table}

\subsection{Approach \label{subsec:Descriptive-evidence-of}}

I now use these data to show how position effects vary across alternatives. This heterogeneity determines how rankings affect transactions and thus how rankings targeting different objectives differ. For example, if alternative $A$ has a larger position effect than $B$, ranking $A$ above $B$ increases transactions because $A$'s demand gain exceeds $B$'s loss. How ranking $A$ above $B$ affects platform revenues and consumers depends on whether $A$ has a higher price and whether $A$ offers higher utility than $B$. Hence, which products have larger position effects determines the trade-offs between ranking objectives.

However, which products have larger position effects is \emph{a priori} unclear. Depending on how consumers search and choose products, lower-priced and high-utility alternatives could have either smaller or larger position effects. For example, if consumers can easily find high-utility alternatives even when ranked at the bottom, these alternatives will have smaller position effects. In contrast, if consumers struggle to find high-utility alternatives unless they appear at the top, these alternatives will have larger position effects.

To quantify this heterogeneity, I estimate the following linear probability model (LPM) separately for clicks and bookings:
\begin{equation}
\mathbb{P}(Y_{ij}=1|z_{ij}, pos_{ij}) =x_j'\beta- pos_{ij}\gamma - pos_{ij}x_j'\theta + \tau_i\ .\label{eq:lpm}
\end{equation}

Each observation is a hotel $j$ displayed in position $pos_{ij}=1,2,\dots$ in search session $i$. The binary outcome $Y_{ij}$ either indicates whether hotel $j$ was clicked or whether it was booked. The column vector $x_j$ gathers all seven hotel attributes listed in Table \ref{tab:summary_stats_rr} (e.g., price or star rating), and $\beta$ captures their direct effects on the outcome variable. Although a hotel's location score is only observable from the detail page, I include it in the click regression because consumers may imperfectly anticipate it from the information displayed on the list.\footnote{Excluding location score from the regression with clicks as the outcome does not change the results.} The session fixed effects $\tau_i$ control for unobservable factors determining both the query (e.g., destination, length of stay) and consumers' subsequent choices.\footnote{For example, consumers searching for a longer stay or a holiday destination may see hotels with different unobservable attributes on the list and may differ in how likely they are to book any hotel.} I include $pos_{ij}$ with a negative sign to simplify interpreting the estimated position effects. Finally, $z_{ij}$ gathers $x_j$ and $\tau_i$.

I define a hotel's \emph{position effect} as the effect of moving it up by one position, conditional on its attributes. In the linear probability model \eqref{eq:lpm}, this position effect is given by
\begin{align}
\text{Position effect}_{ij} &= \mathbb{P}(Y_{ij}=1|z_{ij}, pos_{ij}= h ) - \mathbb{P}(Y_{ij}=1|z_{ij}, pos_{ij}= h + 1) \nonumber \\
 & = \gamma + x_j'\theta \ , \label{eq:def-position-effect}
\end{align}
where $h$ is the current position of hotel $j$. With this definition, a larger position effect means the hotel gains more clicks or bookings when moved higher on the list. Moreover, position effects measure absolute changes in click-through rates and booking probabilities, the relevant metrics because they determine changes in the overall number of transactions and platform revenues when comparing rankings. To simplify exposition, I scale the position effects to percentage points throughout.

\subsection{Results on Heterogeneous Position Effects \label{subsec:Results-on-Heterogeneous}}

\begin{figure}[t]
  \FIGURE
  {\subfloat{\hspace{-2.em}\includegraphics[width=0.5\textwidth,trim=0 1em 0 1em,clip]{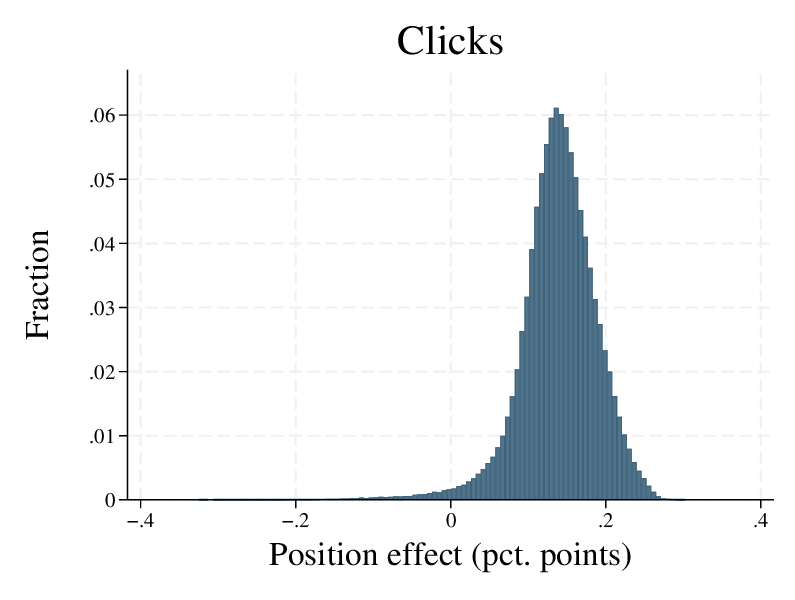}}
  \subfloat{\includegraphics[width=0.5\textwidth,trim=0 1em 0 1em,clip]{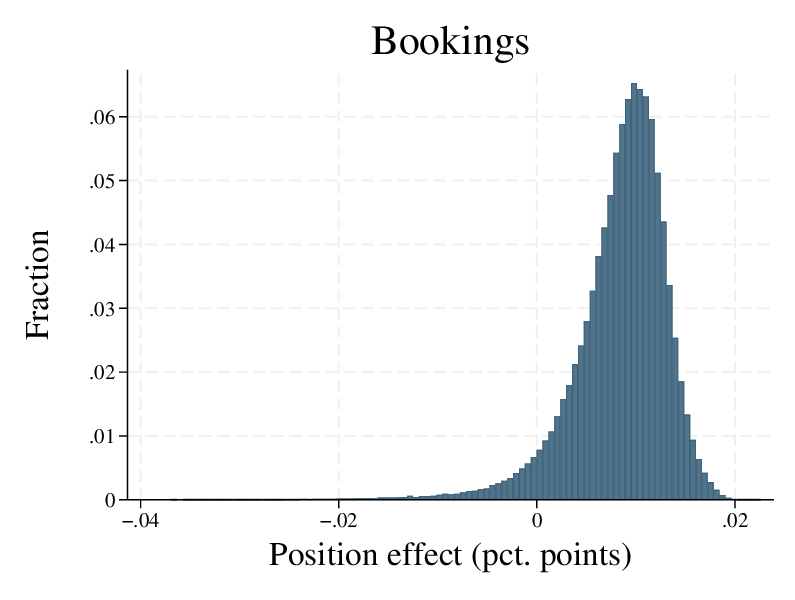}}}
  {Estimated Distribution of Position Effects \label{fig:Heterogeneous-position-effects-distribution}}
  {Estimated position effects for hotels in the data. }
\end{figure}

Figure \ref{fig:Heterogeneous-position-effects-distribution} shows the distribution of estimated position effects across hotels in the data. The left panel reveals substantial heterogeneity for clicks: position effects range from near zero to 0.25 percentage points, close to twice the average of 0.14 percentage points.\footnote{The estimated position effect is negative for some hotels. These cases likely result from imprecisely estimated interaction terms, as indicated by the standard errors in Table \ref{tab:reduced-form} in Online Appendix \ref{subsec:Robustness:-Position-effects}.} Though 0.14 percentage points may seem small, it is substantial relative to the average click-through rate. In an average session with 26 alternatives, moving a hotel from the middle to the top position increases its click-through rate by $13\times0.14=1.82$ percentage points, about 40\% of the average click-through rate of 4.39 percentage points.\footnote{The average within-hotel standard deviation of position is ten, implying that the randomized ranking commonly generates large moves.} This effect doubles for hotels at the upper end of the distribution or disappears for those at the lower end. Hence, position effects are economically meaningful for clicks, and their heterogeneity means some hotels gain substantially more clicks from higher rankings than others.

The right panel shows similar patterns for bookings. Position effects range from near zero to over 0.015 percentage points, almost twice the average of 0.008 percentage points. Such a position effect implies that moving a hotel from the middle to the top position increases its booking probability by $13\times0.008=0.104$ percentage points, about 38\% of the average booking probability of 0.27 percentage points. As for clicks, these results indicate that position effects are economically significant for bookings, with substantial heterogeneity across hotels.

\begin{figure}[t]
  \FIGURE
  {\subfloat{\hspace{-2.2em}\includegraphics[width=0.5\textwidth,trim=0 1em 0 0.5em,clip]{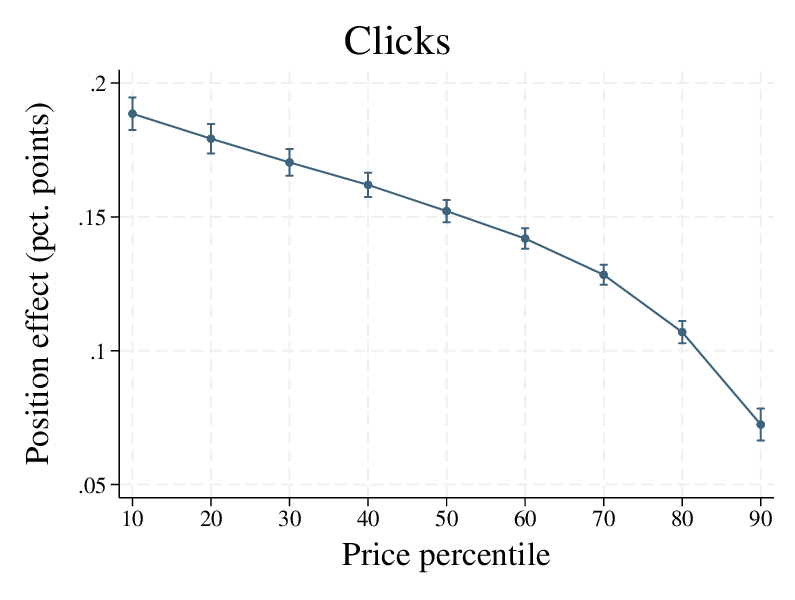}}\subfloat{\centering{}\includegraphics[width=0.5\textwidth,trim=0 1em 0 0.5em,clip]{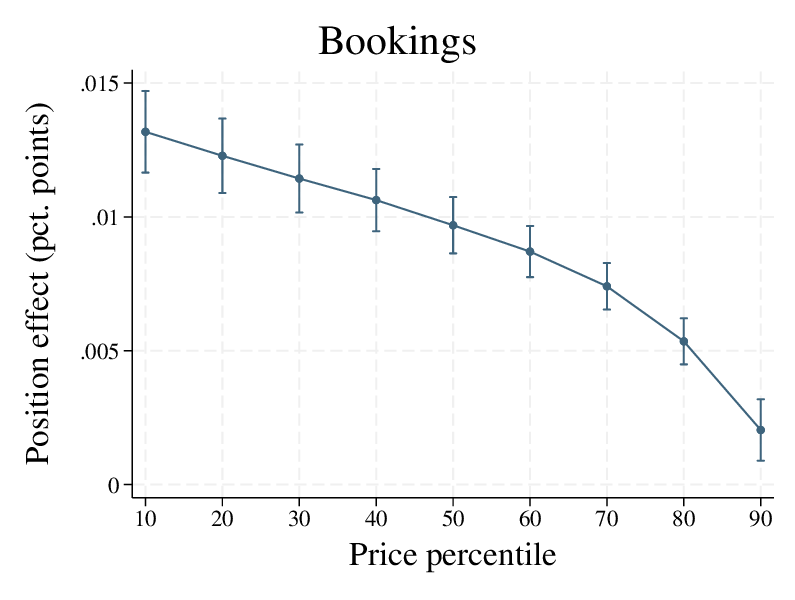}}}
  {Estimated Position Effects at Different Prices  \label{fig:Heterogeneous-position-effects-pricestar}}
  {Estimated position effects at different price percentiles, conditional on other attributes at their average. Bars show 95\% confidence intervals using session-clustered standard errors.}
\end{figure}

Differences in hotel attributes determine this heterogeneity. Figure \ref{fig:Heterogeneous-position-effects-pricestar} shows the estimated average position effect at different prices, with other attributes kept at their average.\footnote{The magnitude of the overall position effect depends on all hotel attributes, but the differences in position effects across price levels do not depend on other attributes' values (see equation \eqref{eq:def-position-effect}).} Conditional on other attributes, lower-priced hotels have larger position effects. For clicks, the position effect falls from 0.19 to below 0.07 as the price increases from the 10th to the 90th percentile, a reduction of more than 60\%. For bookings, the position effect drops from 0.013 to about 0.0025 percentage points, an even larger 80\% reduction.

\begin{figure}[t]
  \FIGURE
  {\subfloat{\hspace{-2.2em}\includegraphics[width=0.5\textwidth,trim=0 1em 0 0.5em,clip]{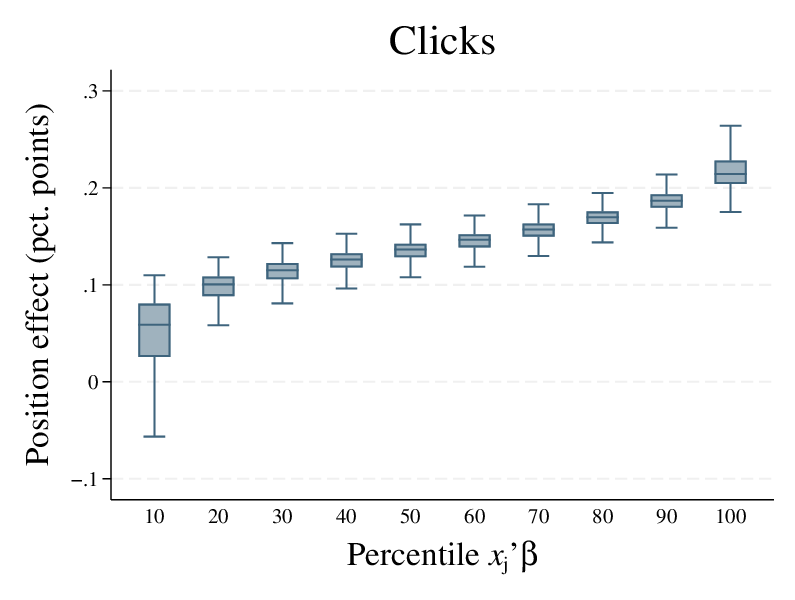}}
  \subfloat{\centering{}\includegraphics[width=0.5\textwidth,trim=0 1em 0 0.5em,clip]{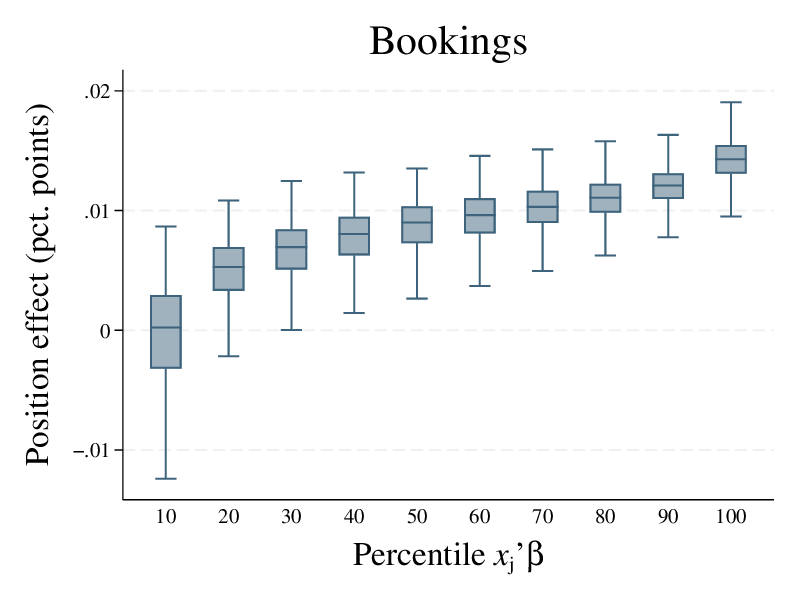}}}
  {Estimated Position Effects for Different Direct Effects \label{fig:Heterogeneous-position-effects}}
  {Boxplots for the estimated position effects for hotels in the data, grouped into percentiles of their direct effect $x_j'\beta$. Values outside the 1.5 interquartile-range are omitted. }
\end{figure}

Figure \ref{fig:Heterogeneous-position-effects} generalizes the analysis to other attributes. The figure shows the estimated position effects for hotels in the data, grouped by their direct effects $x_j'\beta$. This direct effect determines the average number of clicks or bookings a hotel receives independent of its position, which I use as a data-driven measure of hotel utility. The figure reveals a clear pattern: hotels with larger direct effects tend to have larger position effects. For example, the median position effect for clicks ranges from 0.05 percentage points among hotels in the lowest percentile of direct effects to above 0.2 for hotels in the highest percentile.

Table \ref{tab:reduced-form} in Online Appendix \ref{subsec:Robustness:-Position-effects} reports the corresponding coefficient estimates. The interaction terms in $\theta$ are negative for price and positive for all other attributes, confirming that lower-priced and high-utility alternatives---those that on average are more likely to be searched and chosen---have larger position effects on average.

\subsection{Robustness of Results\label{subsec:Robustness-of-Results}}

The randomized ranking ensures that a hotel's position within a search session is exogenous so that the position effects can be estimated consistently \citep[][]{Ursu2018}. However, a hotel's price may not be exogenous if it is correlated with unobservable hotel attributes. Table \ref{tab:reduced-form} in Online Appendix \ref{subsec:Robustness:-Position-effects} addresses this concern by additionally reporting estimates for a specification that adds hotel fixed effects, effectively controlling for such unobservable hotel attributes. The results confirm that lower-priced and high-utility alternatives have larger position effects on average. Online Appendix \ref{subsec:Robustness:-Position-effects} further shows that the results are also robust to using a Probit model or alternative specifications that treat position more flexibly.

\subsection{Implications} \label{subsec:Implications-Descr-Evidence}

The descriptive evidence shows that position effects differ substantially across alternatives. This heterogeneity has important implications for the trade-offs between ranking objectives.

Because high-utility alternatives have larger position effects, ranking them higher maximizes transactions. By helping consumers find these high-utility options early in the list, such a ranking also benefits consumers, suggesting limited trade-offs between maximizing transactions and consumer welfare.

Whether these objectives also align with maximizing platform revenues depends on the relative size of two effects. The results imply that moving higher-priced alternatives higher on the list decreases transactions (the \emph{demand effect}) but steers demand toward these alternatives (the \emph{price effect}). Ranking higher-priced alternatives higher thus parallels a price increase in a monopolist pricing problem: it reduces quantity sold while increasing revenue per unit sold.

Maximizing revenues by ranking alternatives requires balancing these two effects. If the demand effect dominates, ranking lower-priced alternatives higher can increase revenues through more transactions, aligning this objective with maximizing transactions and consumer welfare. Conversely, if the price effect dominates, ranking higher-priced alternatives higher increases revenues despite reducing transactions, creating trade-offs between the objectives.

While the descriptive evidence reveals the opposing demand and price effects, it does not reveal their relative size or the magnitude of the trade-offs they imply. Quantifying these trade-offs requires an empirical framework that can predict how different rankings affect platform revenues and consumer welfare. I now introduce such a framework, estimate it, and use it to compare rankings targeting the different objectives.

\section{Model and Estimation\label{sec:Model-and-estimation}}

I build on the \emph{search and discovery} framework of \citet{Greminger2021} to model how consumers interact with a ranked product list. In the model, consumers sequentially decide between discovering products, searching among already-discovered alternatives, and ending their search by taking the best option found so far or the outside option. This decision process closely matches how consumers interact with ranked product lists: they discover alternatives by scrolling down the list, reveal more information about an alternative by viewing its detail page, and end their search by choosing an alternative or leaving the list.  Moreover, it extends Weitzman models \citep[e.g.,][]{Ursu2018, Compiani2023}, which assume consumers immediately see the entire list and only choose between searching and stopping.

\cite{Greminger2021} defines the decision process and solves for the optimal policy but does not show how to estimate the model. I now extend that paper by introducing an empirical specification and developing an estimation procedure.

\subsection{The Empirical Search and Discovery Model\label{subsec:SD-model}}

Each consumer $i$ faces a ranked product list displaying a finite number of hotels \(j \in J\) in positions \(h \in \{1, \dots, |J|\}\). To simplify notation, I suppress consumer subscripts throughout.

\paragraph*{Utility Specification.} When booking hotel $j$, a consumer receives utility
\begin{align}
u_j & = u_j^l + u_j^d = x_j^{l}{}'\beta + \nu_j + x_j^{d}{}'\kappa + \delta_j + \varepsilon_j\  ,\label{eq:utility-spec}
\end{align}
where the \emph{list utility} \(u_j^l = x_j^l{}' \beta + \nu_j\) is the part of utility already revealed on the product list, and the \emph{detail page utility} $u_j^d = x_j^{d}{}'\kappa + \delta_j + \varepsilon_j$ is revealed only on the detail page. $x_j^l$ and $x_j^d$ are vectors of hotel attributes observed in the data, and $\beta$ and $\kappa$ are the corresponding preference weights.\footnote{I do not include consumer heterogeneity in preferences and search costs because it would be difficult to separately identify the two without observing the search order \citep[see][]{Morozov2021, Yavorsky2021}.} $\delta_j$ is a product-specific fixed effect that I estimate for a subset of products and normalize to zero for others.\footnote{These fixed effects add flexibility, but estimating them for all products is not computationally feasible. Hence, I follow \cite{Compiani2023} and estimate them only for the subset of products observed most frequently.} \(\nu_j\) and \(\varepsilon_j\) are idiosyncratic taste shocks revealed on the list page and the detail page, respectively.

This utility specification incorporates search over vertical attributes $x_j^d$ and $\delta_j$. The model can therefore capture that products differ in how often they are chosen after being searched, which \cite{Compiani2023} show matters for optimal rankings. In my empirical application, $x_j^d$ contains only a hotel's location score, the only attribute consumers cannot observe on the list page \citep[][]{Abaluck2025}.\footnote{The results in Online Appendix Table \ref{tab:reduced-form} show that the location score has a statistically significant effect on clicks. This result does not contradict the fact that consumers cannot observe this attribute on the list page. Instead, it suggests that consumers' beliefs about a hotel's location score are sufficiently accurate that their search decisions are correlated with hotels' actual location scores. In the model, this is captured through the belief parameters $\gamma$ and $\tilde{\gamma}_j$.}

The utility of the outside option of not booking any hotel is given by \(u_0 = \beta_0 + \eta\), where $\eta$ is another idiosyncratic taste shock and $\beta_0$ captures the mean willingness to pay. This specification follows \cite{Ursu2018} and adds $\beta_0$ to the outside option rather than introducing a common brand intercept into $u_j^d$ for all hotels. With $\delta_j$ included in $u_j^d$ for some alternatives and normalized to a constant (zero) for others, both specifications are equivalent when the fixed effects are appropriately redefined.

The taste shocks $\nu_j$ and $\varepsilon_j$ are i.i.d. normal across consumers and alternatives, with respective means $\mu_\nu$ and $\mu_\varepsilon$, and variances $\sigma^2_\nu$ and $\sigma^2_\varepsilon$. In estimation, I set \(\mu_\nu = \mu_\varepsilon= 0\) and \(\sigma_\nu^2 = \sigma_\varepsilon^2 = 1\). The shock for the outside option $\eta$ follows a standard uniform distribution. I discuss these distributional assumptions in more detail in Section \ref{subsec:Identification-and-normalization}.

\paragraph{Search Process.}
Initially, consumers know their preference weights and the value of the outside option but not the attributes and taste shocks of the hotels on the list. Consumers reveal these values by sequentially deciding between scrolling down the list (discovering) and clicking on hotels to view their detail pages (searching). By scrolling down the product list, consumers reveal the list utility \(u_j^l\) for the next hotel on the list.\footnote{Expedia's website currently reveals one alternative at a time and also did so during the sample period \citep[see online appendix C of][]{Ursu2018}. Moreover, Online Appendix \ref{sec:Comparison-of-specifications} provides results for a specification where consumers discover three alternatives with each discovery, showing that such a specification fits the data worse.} By searching alternative \(j\), consumers reveal its detail page utility $u_j^d$.

I further impose two precedence constraints: consumers cannot search a hotel before discovering it, and they cannot book a hotel before searching it. These constraints are necessary for the optimal policy to remain tractable \citep{Greminger2021}. They also apply naturally in this setting: consumers cannot click on hotels not yet on the screen and can only book hotels from the detail page.

Both clicking and scrolling are costly actions. When scrolling to discover additional hotels, consumers incur discovery costs \(c_d\). Because I do not observe consumers who do not visit Expedia and only observe consumers who discover at least the first hotel on the list, I assume that the first discovery is free.\footnote{Expedia initially reveals up to three alternatives depending on the screen size. However, assuming that consumers discover the first three alternatives for free leads to worse model fit (see Online Appendix \ref{sec:Comparison-of-specifications}).} When clicking on a listing to search some alternative $j$, consumers incur search costs $c_s$, which do not differ across products. I further assume free recall: consumers can return to search a previously discovered listing or book a previously searched hotel without incurring additional costs.

\paragraph{Beliefs About Products.}
Consumers have beliefs about the alternatives they will discover. To model these beliefs, I assume that consumers believe alternatives are independent draws from the joint distribution of hotel attributes and taste shocks. More formally, consumers believe that $x_j^{l}$, $x_j^{d}$, and $\delta_j$ are realizations of random variables that may be correlated but are independent of the shocks. Moreover, I assume that consumers know the distributions of these random variables and of the taste shocks.

Consumers form beliefs about an alternative's detail page utility from its list page attributes $x_j^l$. Given $x_j^l$, consumers believe the detail page utility is a random variable $U_j^d| x_j^l = x_j^l{}'\gamma + \tilde{\gamma}_j + \tilde{\varepsilon}_j$, where $\tilde{\gamma}_j$ is a product-specific fixed effect that I include for a subset of products,\footnote{As for $\delta_j$, consumers have correct beliefs about the distribution of these fixed effects. I also follow \cite{Compiani2023} and estimate these fixed effects for the same subset of products for which I estimate $\delta_j$.} and $\tilde{\varepsilon}_j$ is i.i.d. conditional on $x_j^l$ and $\tilde{\gamma}_j$. Consumers know the distribution of $\tilde{\varepsilon}_j$, which is given by the demeaned empirical distribution of $x_j^d{}'\kappa + \delta_j$ and the distribution of $\varepsilon_j$.

 Similar to the first microfoundation in \cite{Compiani2023}, $\gamma$ captures the perceived correlation between $x_j^l$ and the detail page utility $u_j^d$, while $\tilde{\gamma}_j$ adds flexibility by capturing that consumers may believe some alternatives offer higher average detail page utility beyond this correlation. By estimating $\gamma$ and $\tilde{\gamma}_j$ from consumers' choices, the model recovers these beliefs from the data rather than imposing them through rational expectations. Nonetheless, assuming consumers know the distribution of product attributes and shocks constrains these beliefs and ties them to the data. Without additional data on consumers' beliefs, this assumption is necessary to recover search costs because high search costs are observationally equivalent to consumers expecting low detail page utilities.

\paragraph{Beliefs About the Ranking.} To model consumers' beliefs about the ranking, I assume that consumers form beliefs about the \emph{effective value} of alternatives they expect to discover in each position. Specifying beliefs about effective values is sufficient because the expected benefits of discovering alternatives depend only on the distribution of effective values, not on how this distribution is determined by the distributions of list and detail page utilities \citep[][]{Greminger2021}.\footnote{Effective values were introduced by \cite{Armstrong2017} and \cite{Choi2018} for the Weitzman model. \cite{Greminger2021} generalizes the idea to the search and discovery model with product discovery.} In my empirical specification, the effective value of an alternative is given by
\begin{align}
  \tilde{w}_j = u_j^l + \min\{x_j^l{}'\gamma +\tilde{\gamma}_j + \xi, x_j^d{}' \kappa + \delta_j + \varepsilon_j\} \ , \label{eq:def-effective-value}
\end{align}
where $\xi$ is implicitly defined by
\begin{equation}
  \int_{\xi}^{\infty}[1-F(\tilde{\varepsilon})]\text{d}\tilde{\varepsilon} = c_s  \label{eq:search-value}
\end{equation}
and $F$ is the CDF of $\tilde{\varepsilon}_j$.

I assume that consumers believe the expected value of $\tilde{W}_j(h)$---the random variable for the effective value of the alternative revealed in position $h$---is a function $\mu(h)$ that changes with position $h$. Substantively, this means that consumers have some uncertainty about the ranking and do not know exactly how the available alternatives are ranked. For example, when \(\mu(2) > \mu(3)\), consumers expect that the alternative they reveal in the second position will offer a larger list utility $u_j^l$, a larger detail page utility $u_j^d$, or some combination of both \emph{on average}, while also knowing that the alternatives they reveal can deviate from this expectation.

This belief specification has the advantage that $\mu(h)$ fully characterizes consumers' beliefs about the ranking. This allows me to estimate $\mu(h)$ rather than imposing rational expectations that would require consumers to know exactly how Expedia ranks alternatives based on hotel attributes. However, beliefs are not fully flexible. The assumption that consumers know the distribution of product attributes and shocks constrains them and ties them to the data. Because consumers know these distributions, beliefs shift $\mu(h)$ across positions, while the baseline distribution determines the average across positions $1/|J| \sum_h \mu(h)$. I therefore recover the average from the empirical attribute distribution across all positions, while I estimate how $\mu(h)$ changes with $h$ to capture consumers' beliefs about how the ranking algorithm shifts this distribution across positions. As for search costs, tying the mean across positions to the data is necessary to identify discovery costs. Without data on consumers' beliefs, it is not possible to determine whether high discovery costs or consumers expecting low effective values prevent them from scrolling down the list.

Finally, I impose the substantive assumption that \(\mu(h)\) weakly decreases in \(h\). This means that consumers expect alternatives further down the list to be less worth discovering. This assumption is required for the optimal policy to remain tractable \citep{Greminger2021} and simplifies estimation. However, Online Appendix \ref{subsec:proof-nonincreasing-mu-h} shows that this assumption does not restrict the model in the present setting. If I estimated the model without this restriction, I would necessarily obtain estimates satisfying this assumption: the Expedia data with the randomized ranking exhibits significant position effects in clicks for consumers who do not choose a hotel, a pattern inconsistent with $\mu(h)$ increasing (or remaining constant) in $h$.

To operationalize this assumption, I use the following functional form
\begin{equation}
  \mu(h)=\mu_1 + \rho \log(h) \  \label{eq:functional_form_beliefs}
\end{equation}
for positions $h=1, ..., |J|$ and impose the constraint $\rho \leq 0$. The shape of this function determines how clicks and bookings decrease across positions. With $\log(1)=0$, this functional form implies a non-linear decrease starting from $\mu_1$, allowing the model to fit the observed non-linear decrease in clicks across positions by estimating only a single parameter, $\rho$ (see Online Appendix \ref{sec:Comparison-of-specifications}).\footnote{Imposing such a functional form is similar to imposing a functional form on how search costs change with an alternative's position, as in prior work using the Weitzman model to capture position effects \citep[e.g.,][]{Chen2017, Ursu2018, Chung2023}.} $\mu_1$ is not a parameter to be estimated. Instead, it is implied by the assumption that the average $\mu(h)$ across positions is known to consumers and will be recovered from the empirical distribution of hotel attributes (see Online Appendix \ref{sec:Discovery-Values}).

\subsection{Optimal Policy \label{subsec:Optimal-policy-and}}

Consumers maximize their expected utility by sequentially choosing one of the available actions. Let \(A(t)\) be the set of alternatives discovered and \(S(t)\) the set of alternatives searched up to period \(t\), and let $h(t)$ denote the position to be discovered next. In each period $t$, the consumer chooses between discovering position \(h(t)\), searching an alternative from \(A(t)\), and ending search by choosing an alternative from \(S(t)\).

\citet{Greminger2021} proves that the optimal policy for this dynamic decision process is fully characterized by three types of reservation values, one for each type of action. An action's reservation value is the value of a hypothetical outside option that would make a myopic consumer indifferent between choosing the action and taking this outside option. Crucially, reservation values do not depend on other available alternatives or future discoveries, so they can be computed without considering other actions or future periods. In my empirical model, the reservation values for the three actions are:\footnote{Online Appendix \ref{sec:Discovery-Values} and online appendix EC.2 of \cite{Greminger2021} provide further details.}
\begin{itemize}
\item \textbf{Purchase value}: \(z_j^p = u_j = x_j^l{}'\beta + \nu_j +x_j^d{}'\kappa + \delta_j +  \varepsilon_j\).
\item \textbf{Search value}: \(z_j^s = x_j^l{}'\beta + \nu_j + x_j^l{}'\gamma + \tilde{\gamma}_j + \xi\), where $\xi$ is defined by \eqref{eq:search-value}.
\item \textbf{Discovery value}: \(z^d(h) = \mu(h) + \Xi\), where \(\Xi\) solves
\begin{equation}
\int_{\Xi}^{\infty}[1-G(w)]\text{d}w
  = c_d\label{eq:discovery-value}
\end{equation}
and \(G\) is the CDF of the demeaned effective value consumers expect to reveal in position $h$, given by $\tilde{W}_j(h) - \mu(h)$. Note that only the expected value of $\tilde{W}_j(h)$ changes across positions, so the CDF $G$ does not depend on $h$.
\end{itemize}

Given these reservation values, the optimal policy is straightforward: always choose the available action with the largest reservation value. For example, when the purchase value of an alternative in the consideration set $S(t)$ is the largest available reservation value, the consumer ends her search by choosing that alternative. If instead the search value of an alternative in the awareness set $A(t)$ is largest, the consumer searches that alternative. Finally, if the discovery value is largest, the consumer scrolls down to reveal the alternative in the next position.

More formally, the optimal policy is characterized by three rules. Let \(\tilde{u}(t) = \max_{j \in S(t)} u_j\) and \(\tilde{z}^s(t) = \max_{j \in A(t)} z_j^s\) denote the maximum utility and search value available in period \(t\). Then, in period \(t\):
\begin{itemize}
  \item \textbf{Stopping:} Choose \(j \in S(t)\) and end search whenever \(u_j = \tilde{u}(t) \geq \max\{\tilde{z}^s(t), z^d(h(t))\}\).
  \item \textbf{Search:} Search \(j \in A(t)\) whenever \(z_j^s = \tilde{z}^s(t) \geq \max\{\tilde{u}(t), z^d(h(t))\}\).
  \item \textbf{Discovery:} Discover the next position \(h(t)\) whenever \(z^d(h(t)) \geq \max\{\tilde{u}(t), \tilde{z}^s(t)\}\).
\end{itemize}

\subsection{Position Effect Mechanism \label{subsec:Heterogeneous-Position-Effects}}
The search and discovery model microfounds the observed position effects through a clear mechanism: consumers stop scrolling before discovering alternatives further down the list. The following proposition shows that this discovery mechanism also predicts the heterogeneity in position effects revealed by the descriptive evidence.

\begin{proposition}
\label{prop:position-effect-model} The position effect mechanism in the search and discovery model implies:
\begin{enumerate}
    \item Alternatives more likely to be searched conditional on their position have weakly larger position effects in searches.
    \item Alternatives more likely to be chosen conditional on their position have weakly larger position effects in purchases.
\end{enumerate}
\end{proposition}

The proposition predicts the pattern in Figure \ref{fig:Heterogeneous-position-effects-pricestar}: because a lower price leads to more clicks and bookings conditional on position, hotels with lower prices have larger position effects in clicks and bookings, respectively. Similarly, it predicts the pattern in Figure \ref{fig:Heterogeneous-position-effects}: hotels that are more likely to be clicked and booked also have larger position effects in clicks and bookings. As I further show in Section \ref{subsec:Fit-measures}, the model not only predicts these patterns but, when estimated on these data, also fits the respective magnitudes well. 

Intuitively, the proposition follows from consumers being more likely to search and choose high-utility alternatives when they discover them, which has two implications. First, high-utility alternatives on average are more likely to be searched and chosen conditional on their position. Second, these alternatives also gain disproportionately more searches and purchases when ranked higher and thus discovered more often. Combined, these implications yield the predictions in Proposition \ref{prop:position-effect-model}. Appendix \ref{subsec:proof-prop-position-effect-model} provides the formal proof, extending \citet{Greminger2021} by deriving expressions for position effects conditional on an alternative's observable attributes. Crucially, the result does not rely on specific distributional assumptions or parameter values: it follows from the discovery mechanism itself, not from any particular parameterization.

This mechanism differs from the one in models building on \cite{Weitzman1979}, an established framework for quantifying ranking effects \citep[see][for a recent overview]{Ursu2024}. In these models, consumers observe the entire list, and position effects follow from position-specific search costs. This alternative mechanism does not yield the same testable predictions as Proposition \ref{prop:position-effect-model} and thus can match any pattern of heterogeneous position effects, which may appear to make Weitzman models more flexible. However, as I show in Section \ref{subsec:Fit-measures}, generating position effects through position-specific search costs limits Weitzman models in another dimension: they underestimate position effects in purchases, an important moment governing the trade-offs between ranking objectives.

\subsection{Estimation Approach \label{subsec:Estimation-approach}}

I develop a simulated maximum likelihood approach to estimate the search and discovery model. The approach is available as Julia and Python packages, including detailed documentation and examples.\footnote{The Julia package is available at \url{https://github.com/rgreminger/StructuralSearchModels.jl} and the Python package at \url{https://github.com/rgreminger/structuralsearchmodels}.} Online Appendix \ref{subsec:software} provides further details on the software implementation. The approach and packages can also be used to estimate the Weitzman model (see Online Appendix \ref{sec:weitzman-appendix}).

The goal of the estimation is to estimate the model parameters $\theta = (\beta, \beta_0, c_s, c_d, \rho, \tilde{\gamma}_j, \delta_j)$ using data for $N$ consumers visiting the product list. For each consumer $i$, the data contain the chosen alternative $j$; the consideration set $S_i$ of alternatives the consumer clicked on; and the positions $h_{ij}$ and attributes $(x_j^l, x_j^d)$ for all alternatives in the set $J_i$ of alternatives available to consumer $i$.

Given these data, the estimation procedure solves
\begin{align}\label{eq:likelihood}
  \max_{\theta}\mathcal{L}(\theta) = & \sum_{i=1}^{N} \log \mathbb{P}(\text{observed choices of } i| (x_j^l, x_j^d,h_{ij})\forall j\in J_i;\theta) \nonumber \\
   = & \sum_{i=1}^{N}\log\mathbb{P}(\text{click all }k\in S_i\text{, choose j }\in S_i|(x_j^l, x_j^d,h_{ij})\forall j\in J_i;\theta)\ .
\end{align}
The inner probability depends on the parameters gathered in $\theta$ and conditions on the hotel attributes and ranking observed in the data. In what follows, I omit highlighting this dependence to simplify exposition and again suppress the consumer subscript $i$.

\paragraph{Likelihood Contributions.} The individual likelihood contribution of a consumer is given by the probability of her observed choices under the optimal policy. One approach would be to use the three rules of the optimal policy to construct inequalities and maximize the likelihood of these inequalities holding. However, as with the Weitzman model, this approach leads to likelihood contributions that are costly to compute and requires observing the sequence of actions, which is not available in the Expedia data.\footnote{Related work estimating the Weitzman model on the Expedia data assumes a top-down search sequence \citep[][]{Ursu2018, Kaye2024, Chung2023} or integrates over all possible search sequences \citep[][]{Compiani2023}.}

I develop an approach that circumvents both issues. Specifically, I derive several implications of the optimal policy that characterize the choices of a consumer who eventually chooses some (inside or outside) option $j$. By characterizing all other actions relative to the chosen option irrespective of the sequence in which they occurred, the implications substantially simplify the likelihood computation and eliminate the need to observe the search sequence.\footnote{By discarding the search sequence, the approach can also be used when it is observed.}

Proposition \ref{prop:implications} provides the four implications. It builds on effective values, which combine the different reservation values and fully characterize which alternative a consumer eventually chooses in the model \citep{Greminger2021}. Formally, the \emph{generalized effective value} of an option $j$ is defined as $w_j=\min\{ z^{d}(h_j),\tilde{w}_j\}$, where $\tilde{w}_j=\min\{ z_j^{s},u_j\}$ is the effective value defined in \eqref{eq:def-effective-value} and does not depend on the position. For the outside option, the effective values are $w_0=\tilde{w}_0=u_0$.

\begin{proposition}
\label{prop:implications}
  Suppose a consumer chooses some (inside or outside) option $j$. For a given generalized effective value $w_j$ and effective value $\tilde{w}_j$, the optimal policy implies the following:
  \begin{itemize}
  \item \textbf{Discovery:} The consumer discovers all alternatives in the awareness set $A(\tilde{w}_j)$, which is the set of alternatives up to position $ \bar{h}(\tilde{w}_j) =  \arg \min_{h_k > h_j } \tilde{w}_j \geq z^d(h_k)$.
  \item \textbf{Search and early discovery:} The consumer searches all alternatives in $ S^{-}=\{k:z_k\geq w_j,k\in A(\tilde{w}_j),h_k<h_j\}$ and no other alternatives discovered before $j$.
  \item \textbf{Search and late discovery:} The consumer searches all alternatives in $ S^{+}=\{k:z_k\geq\tilde{w}_j,k\in A(\tilde{w}_j),h_k\geq h_j\}$ and no other alternatives discovered at the same time or after $j$.
  \item \textbf{Choice:} Utilities of products other than $j$ satisfy $u_k < w_j \ \forall k \in S^{-} \cup \{0\}$ and $u_k \leq \tilde{w}_j \ \forall k \in S^{+} \setminus \{j\}$.
  \end{itemize}
\end{proposition}

The proposition fully characterizes a consumer's actions for a given effective value of the chosen alternative. For a consumer choosing $j$, it shows which alternatives the consumer discovers and searches and provides conditions guaranteeing that $j$ is searched and chosen. The four implications follow from the optimal policy and extend the generalized eventual purchase theorem of \cite{Greminger2021} by additionally characterizing actions beyond the eventual choice. For example, the discovery implication follows from the optimal discovery rule: a consumer chooses $j$ and stops scrolling before position $\bar{h} > h_j$ if and only if the discovery value at that position is sufficiently small so that $j$ is searched $(z_j^s>z^d(\bar{h}))$ and chosen $(u_j > z^d(\bar{h}))$. These conditions are combined through the effective value $\tilde{w}_j$. Appendix \ref{subsec:proof-prop-implications} provides the formal proof, showing that the optimal policy and the proposition's implications always prescribe the same actions conditional on choosing option $j$ with effective value $\tilde{w}_j$.

Because Proposition \ref{prop:implications} fully characterizes a consumer's actions, choices observed in the data are consistent with the optimal policy only if they satisfy the four implications. Consequently, the individual log-likelihood contribution $\mathcal{L}^i(\theta)$ can be calculated by taking expectations over the shocks determining the effective value $\tilde{w}_j$ of the chosen option $j$ and computing the probability of the inequalities implied by the four implications holding.

Formally, the individual log-likelihood contribution of a consumer who eventually chooses option $j$ is
\begin{multline}
\mathcal{L}^i(\theta)=\log\int \underbrace{1(h_k \leq \bar{h}(\tilde{w}_j)\forall k\in S)}_{\text{Stopping}} \\
\,\,\,\,\,\,\,\,\,\,\,\,\,\,\,\,\,\,\,\, \times\underbrace{\textstyle\prod_{k\in S^{-} \cup \{0\}}\mathbb{P}(Z_k^{s}\geq w_j\cap U_k\leq w_j)}_{\text{Search and early discovery \& choice}}\times\underbrace{\textstyle\prod_{k\in S^{+} \setminus \{j\}}\mathbb{P}(Z_k^{s}\geq\tilde{w}_j\cap U_k\leq\tilde{w}_j)}_{\text{Search and late discovery \& choice}}\\
\times
\underbrace{\textstyle\prod_{k\in A^-(\tilde{w}_j)\backslash S^-}\mathbb{P}(Z_k^{s}\leq w_j)}_{\text{Search and early  discovery }} \times
\underbrace{\textstyle\prod_{k\in A^+(\tilde{w}_j)\backslash S^+ \setminus \{j\}}\mathbb{P}(Z_k^{s}\leq \tilde{w}_j)}_{\text{Search and  late discovery }}\text{d}H(\eta,\nu_j,\varepsilon_j)\ ,\label{eq:individual-likelihood-contribution}
\end{multline}
where $H(\eta,\nu_j,\varepsilon_j)$ is the joint CDF of the shocks forming the effective values $\tilde{w}_j$ and $w_j$ and the utility $u_j$ of the chosen option.\footnote{Note that the outside option $k=0$ enters the likelihood as part of the search and early discovery implication because the outside option is always discovered, and its search value is given by $z_0 =\infty$.} Depending on whether the consumer chooses the outside or an inside option, either the outside option shock $\eta$ or the taste shocks $(\nu_j, \varepsilon_j)$ determine these values. The integral thus has at most two dimensions. $1(\cdot)$ is the indicator function that equals one only if the consumer has not stopped scrolling before discovering all alternatives she clicked on.\footnote{If it is observed how far consumers scroll, then the condition can be implemented as $h_k = \bar{h}(\tilde{w}_j)$.} $A^-(\tilde{w}_j)$ and $A^+(\tilde{w}_j)$ denote the awareness sets of alternatives discovered before or after $j$, respectively. These sets are determined by the effective value $\tilde{w}_j$ and are not observed. $S=S^- \cup S^+$ denotes the set of alternatives the consumer searched, divided into those discovered before $j$ ($S^-$) and with or after $j$ ($S^+$). Unlike $A(\tilde{w}_j)$, the consumer's consideration sets in $S$ and eventual choice of inside or outside option $j$ are observed in the data.

\paragraph{Smooth Likelihood Simulation.} The individual likelihood contribution \eqref{eq:individual-likelihood-contribution} has no closed-form. Computing it requires simulation techniques that pose two challenges. First, simulating the integral requires calculating probabilities of the form $\mathbb{P}(Z_k^{s}\geq q\cap U_k\leq q)$ for many different values of $q$. This probability is difficult to compute because $Z_j^{s}$ and $U_j$ are not independent; the pre-search shock $\nu_j$ enters both. In Online Appendix \ref{subsec:joint-probability-search-non-purchase}, I decompose this probability into functions that can be computed using standard numerical methods, eliminating the need for a computationally costly numerical integration routine.

Second, naive Monte Carlo integration does not yield a smooth likelihood function, making \eqref{eq:likelihood} difficult to solve numerically. The set $A(\tilde{w}_j)$ determines the inequalities entering the likelihood and depends on parameter values through $\tilde{w}_j$. With fixed simulation draws for the shocks, changing the parameter values can change $A(\tilde{w}_j)$, creating discontinuities in the likelihood when it is computed as the average across simulation draws. Moreover, the effective value $\tilde{w}_j$ has a kink because of the minimum function, creating further discontinuities. In Online Appendix \ref{subsec:Smooth-Monte-Carlo}, I show how to obtain a smooth likelihood by partitioning the probability space and computing conditional likelihoods for each region separately.


\subsection{Parameter Identification and Normalizations \label{subsec:Identification-and-normalization}}

Conditional on discovery, the search and discovery model reduces to a Weitzman model. Hence, the same identification arguments apply for the preference parameters $\beta_0$, $\beta$, and $\delta_j$, belief parameters $\gamma$ and $\tilde{\gamma}_j$, and search costs $c_s$. For example, the average number of clicks per session identifies the search cost $c_s$ because larger search costs make consumers less likely to search alternatives. As in \cite{Compiani2023}, belief parameters $\gamma$ are separately identified from $\beta$ through the difference in an attribute's effect on the likelihood of searching versus choosing. $\tilde{\gamma}_j$ is also separately identified from $\delta_j$ by product-specific differences in click-through and purchase rates not captured by observable attributes. For identification of the other search parameters, I refer to \cite{Ursu2024}, who provide a recent review of the empirical Weitzman model and the respective identification arguments. 

The discovery parameters $c_d$ and $\rho$ are identified without data on how far consumers scrolled, provided the ranking is randomized. The search and discovery model generates position effects through consumers not discovering all alternatives on the list. When the ranking is randomized, the model can thus match the position effects in the data only through the discovery parameters that govern when consumers stop scrolling. Intuitively, higher discovery costs $c_d$ generate larger position effects because fewer consumers discover alternatives in lower positions, while the belief parameter $\rho$ determines how position effects vary across the list. Online Appendix \ref{sec:appendix_identification} provides a formal discussion of this identification argument, including a derivation of the moments that separately identify the search and discovery parameters when the discovery process is latent.

The shock variances are not estimated. Instead, I set $\sigma_\nu = 1$, which is a true normalization \citep[see][]{Ursu2024}. As highlighted by \cite{Morozov2021} and \cite{Yavorsky2021}, the second standard deviation $\sigma_\varepsilon$ is not a true normalization but can be difficult to identify separately from search costs because both similarly affect search values.\footnote{\cite{Yavorsky2021} show that increasing $\sigma_\varepsilon$ and decreasing $c_s$ similarly increase search values. In principle, the two parameters are separately identified by parametric form, but in simulations, the authors find that estimating both is difficult without additional shifters for the search costs.} I find a similar result in my model: jointly estimating $\sigma_\varepsilon$ with the other parameters does not reliably yield consistent estimates in Monte Carlo simulations. Hence, I follow prior work and set $\sigma_\varepsilon = 1$ \citep[e.g.,][]{Morozov2021}. Moreover, Online Appendix \ref{sec:Comparison-of-specifications} shows that the trade-offs between ranking objectives change little with $\sigma_\varepsilon$, despite the estimates for the search and discovery costs being sensitive to the choice of $\sigma_\varepsilon$. Hence, the main conclusions of the counterfactual analysis are robust to this assumption.

The outside option shock $\eta$ follows a standard uniform distribution. While not required for estimation, this assumption ensures that the functional form of $\mu(h)$ rather than the shape of the distribution of $\eta$ determines differences in stopping probabilities across positions.\footnote{This follows from consumers never discovering beyond the position where $u_0 \geq z^d (h)$ holds.} Moreover, it reduces the computational burden of taking draws from the truncated distribution of $\eta$. Online Appendix \ref{sec:Comparison-of-specifications} shows that increasing the upper bound does not affect the relative size of consumer welfare effects across rankings. Hence, insights into trade-offs between ranking objectives are robust to alternative assumptions for this upper bound.\footnote{In settings where the share of consumers who do not search any alternative is observed, it should be possible to estimate the upper bound of the distribution of $\eta$. Alternatively, one could assume that $\eta$ follows a normal distribution with mean zero and estimate the variance of this distribution.}


Online Appendix \ref{subsec:Monte-Carlo-Simulation} presents results of a Monte Carlo simulation. They confirm that the proposed estimation approach successfully recovers all parameters, including the discovery and search costs that are not directly estimated.

\subsection{Parametrizing Search and Discovery Values \label{subsec:Parameterizing-Search-and}}

Rather than directly estimating the search and discovery costs $c_s$ and $c_d$, I estimate parameters $\xi$ and $\tilde{z}^d = z^{d}(1)$ and then back out the search and discovery costs from these and other parameter estimates.\footnote{Discovery values for positions $h>1$ can be easily computed from  $z^d(1)$ and $\rho$ through the function $z^d(h)=z^d(1) + \rho \log(h)$.} Direct estimation would require repeatedly computing $\xi$ and $z^d(1)$ during estimation. These computations are costly because computing $\xi$ requires numerically solving for the root in \eqref{eq:search-value}, and computing $z^d(1)$ requires numerically solving for the root in \eqref{eq:discovery-value}.

This approach works because search costs enter the likelihood \emph{only} through $\xi$, which, given consumers' beliefs and the other parameters, uniquely determines $c_s$ through \eqref{eq:search-value}. Similarly, discovery costs $c_d$ enter the likelihood \emph{only} through $z^d(1)$, which uniquely determines $c_d$ through \eqref{eq:discovery-value}.\footnote{With this parametrization, the observed position effects are the only variation that determines the estimate of $\tilde{z}^d$ and hence of $c_d$. In small samples with little variation, this could potentially lead to differences from estimates obtained when directly estimating costs, where preference parameters and costs jointly enter $z^d(1)$ and $\xi$ during estimation.} Online Appendix \ref{sec:Discovery-Values} provides further details on this approach, including how I obtain estimates for $c_s$ and $c_d$ from the other estimates.

\subsection{Sample Selection \label{subsec:Estimation-sample}}

The Expedia data contains only sessions with at least one click. I address this sample selection by following \citet{Compiani2023}: I condition the individual likelihood contributions on the event that the consumer searches at least one product. Formally, this is equivalent to dividing each likelihood contribution by the probability of making at least one click. When evaluating model fit, I use fit measures conditional on consumers searching at least one alternative to account for this sample selection.

\section{Estimation Results and Model Fit \label{sec:Estimation-results}}

I estimate the search and discovery model using search sessions under the randomized ranking. I use 100 simulation draws for each region of the partitioned probability space, yielding over 6,000 draws per search session (see Online Appendix \ref{subsec:Smooth-Monte-Carlo}). To limit the computational burden, I restrict the estimation sample to the fifty destinations with the most search sessions.\footnote{I also follow \citet{Ursu2018} by including only sessions with at least 30 hotel listings in the results and omitting opaque offers from this analysis.}

\subsection{Parameter Estimates}
Table \ref{tab:results_pars} reports the parameter estimates and asymptotic standard errors. The results show that all parameter estimates have the expected sign. For example, $\beta_{price} < 0$ indicates that consumers are price sensitive when choosing an alternative from the consideration set, whereas $ \gamma_{price} >0$ implies that consumers believe higher-priced hotels have a larger detail page utility, which is consistent with the hotel location score being positively correlated with price in the data.

The discovery cost estimate suggests that an average consumer is willing to pay 2\textcent{} to reveal another alternative. Although small, these discovery costs prevent consumers from discovering all options, creating scope for rankings to affect their choices. Specifically, the model correctly captures that the next hotel discovered is unlikely to be searched and booked. As a result, the expected benefits of discovering an additional hotel are small, and consumers are unwilling to pay even small discovery costs to keep scrolling.\footnote{The discovery cost estimate increases with $\sigma_{\varepsilon}$, but remains small relative to the average price of hotel stays in the data (see Online Appendix \ref{sec:Comparison-of-specifications}).}

\begin{table}[t] \centering \footnotesize
\caption{Parameter Estimates} \label{tab:results_pars}

\begin{threeparttable}
\begin{tabular}{lllc} 
\midrule

 && \multicolumn{1}{c}{Estimate} & \multicolumn{1}{c}{Standard error} \\
\midrule
\textbf{Preference parameters in $u_j^l = x_j^l{}'\beta + \nu_j^l$} \\
\hspace{1em}$\beta$: Price (in \$100)&&\hphantom{00}-0.348***&(0.032)\\
\hspace{1em}$\beta$: Star rating&&\hphantom{00}\hphantom{-}0.155***&(0.035)\\
\hspace{1em}$\beta$: Review score&&\hphantom{00}\hphantom{-}0.166***&(0.049)\\
\hspace{1em}$\beta$: No reviews&&\hphantom{00}\hphantom{-}0.314&(0.263)\\
\hspace{1em}$\beta$: Chain&&\hphantom{00}\hphantom{-}0.003&(0.047)\\
\hspace{1em}$\beta$: On promotion&&\hphantom{00}\hphantom{-}0.165***&(0.046)\\
\hspace{1em}$\beta$: Outside option&&\hphantom{00}\hphantom{-}5.198***&(0.172)\\
\midrule
\textbf{Preference parameters in $u_j^d = x_j^d{}'\kappa + \delta_j + \varepsilon_j$} \\
\hspace{1em}$\kappa$: Location score&&\hphantom{00}\hphantom{-}0.079***&(0.016)\\
\midrule
\textbf{Search value parameters in $z_j^s = u_j^l + x_j^l{}'\gamma +\tilde{\gamma}_j + \xi$} \\
\hspace{1em}$\gamma$: Price (in \$100)&&\hphantom{00}\hphantom{-}0.155***&(0.032)\\
\hspace{1em}$\gamma$: Star rating&&\hphantom{00}\hphantom{-}0.065&(0.034)\\
\hspace{1em}$\gamma$: Review score&&\hphantom{00}-0.167***&(0.049)\\
\hspace{1em}$\gamma$: No reviews&&\hphantom{00}-0.371&(0.262)\\
\hspace{1em}$\gamma$: Chain&&\hphantom{00}-0.079&(0.047)\\
\hspace{1em}$\gamma$: On promotion&&\hphantom{00}-0.030&(0.045)\\
\hspace{1em}$\xi$&&\hphantom{00}\hphantom{-}2.649***&(0.172)\\
\midrule
\textbf{Discovery value parameters in $z^d(h) = \tilde{z}^d + \rho \log(h)$} \\
\hspace{1em}$\tilde{z}^d$&&\hphantom{00}\hphantom{-}5.773***&(0.174)\\
\hspace{1em}$\rho$&&\hphantom{00}-0.130***&(0.007)\\
\midrule
\textbf{Cost parameters} \\
\hspace{1em}Discovery costs (\$)&&\hphantom{00}\hphantom{-}0.020& \\
\hspace{1em}$c_d \times 100$&&\hphantom{00}\hphantom{-}0.007& \\
\hspace{1em}$c_s$&&\hphantom{00}\hphantom{-}0.001& \\
\midrule
\hspace{1em}Log likelihood&&-48,802.356& \\
\hspace{1em}N sessions&&\hphantom{-}11,467& \\
\midrule 
\end{tabular}
\begin{tablenotes}
\item \footnotesize{\emph{Notes:} Parameter estimates obtained using 100 simulation draws per region of the partitioned probability space. Product fixed effects $\delta_j$ and $\tilde{\gamma}_j$ are included for a limited set of products in the model but not shown in the table. Asymptotic standard errors are shown in parentheses. Search and discovery costs are computed using 10M simulation draws.  Statistical significance is indicated by * p < 0.1, ** p < 0.05, *** p < 0.01.  }
\end{tablenotes} 
\end{threeparttable}
\end{table}

\subsection{Model Fit\label{subsec:Fit-measures}}

I evaluate the \emph{search and discovery} (SD) model's fit by comparing model-implied position-specific click and booking probabilities with those in the estimation sample.\footnote{Out-of-sample fit measures are qualitatively similar.} Moreover, I compare these fit measures to a \emph{Weitzman} (WM) model, an established model for studying ranking effects \citep[e.g.,][]{Ursu2024}. I specify and estimate the WM model so that the main difference from the SD model is the mechanism that captures position effects. Notably, this means that the WM model I estimate includes search over vertical attributes as in \cite{Compiani2023}.\footnote{Similar results can be obtained using the specification used by \cite{Compiani2023}.} Online Appendix \ref{sec:weitzman-appendix} provides details on the implementation, including how the estimation approach developed in Section \ref{sec:Model-and-estimation} can be used to estimate the WM model.

\begin{figure}[t]
  \FIGURE
  {\begin{minipage}{\textwidth} \centering
    \subfloat[Search and Discovery Model]{\centering{}\includegraphics[width=0.5\textwidth,trim=0 1em 0 0,clip]{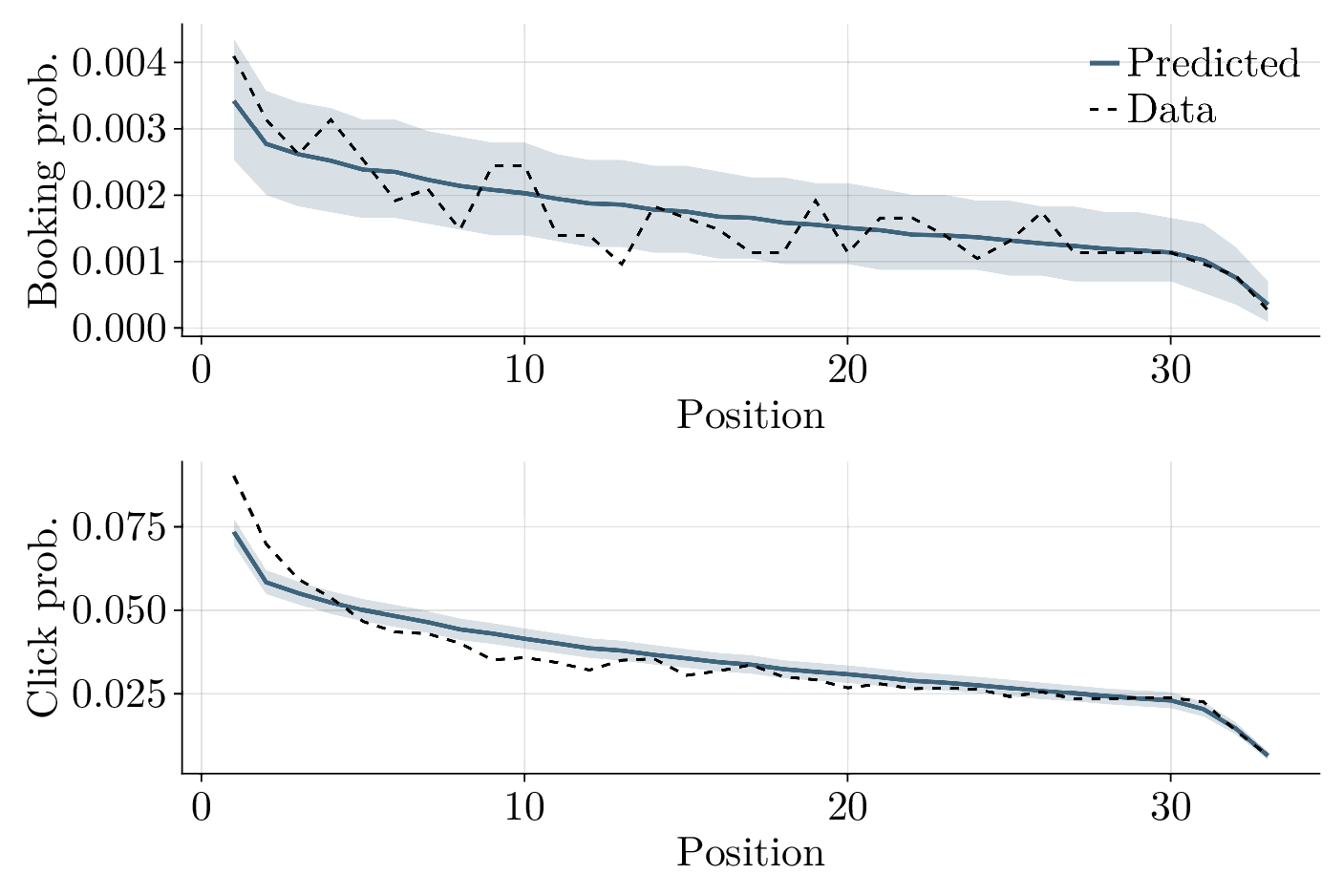}}%
    \subfloat[Weitzman Model]{\centering{}\includegraphics[width=0.5\textwidth,trim=0 1em 0 1em,clip]{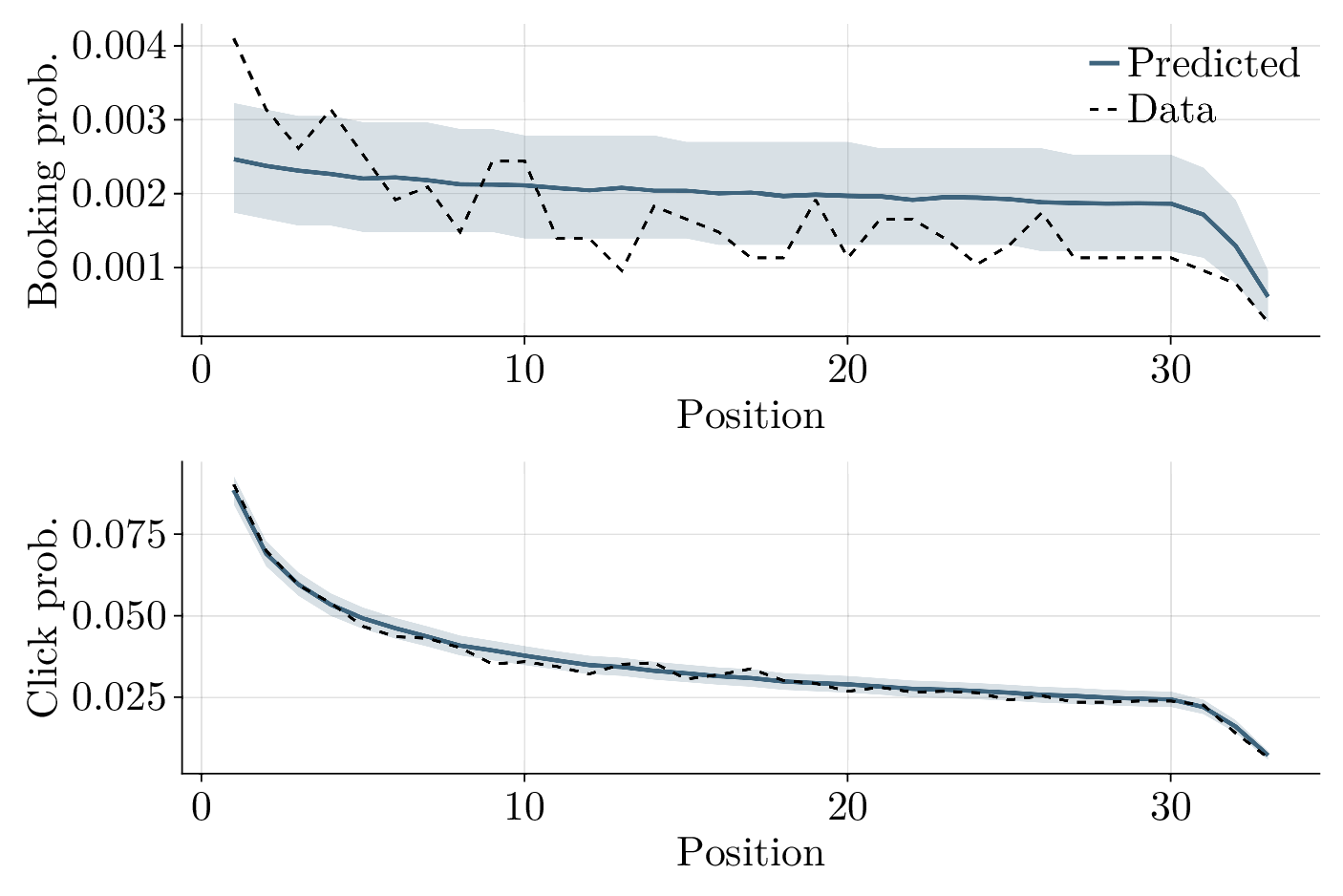}}
    \end{minipage}
  }
  {Booking and Click Probability by Position
\label{fig:fit}}
  {Click and booking probabilities in the data and predicted by the two models. The solid line presents the average across 10,000 simulation draws per consumer, conditional on consumers searching at least one hotel. The shaded areas represent the 95\% percentile of the minimum and maximum position-specific click or booking probability across simulations.}
\end{figure}

Figure \ref{fig:fit} reports the results. The left panel shows that the SD model fits the decrease in the booking probability across positions while also providing a reasonable fit for the decrease in the click probability. These results confirm that the small implied discovery costs are sufficient for the SD model to fit the position effects in the data. The right panel shows that the WM model fits the decrease in the click probability across positions better than the SD model. However, the WM model also substantially underpredicts the decrease in the booking probability, suggesting that it underestimates the position effects in bookings.

To effectively quantify the trade-offs between different rankings, the models should also fit how position effects in bookings vary with price, an important factor determining differences across rankings (see Section \ref{subsec:Implications-Descr-Evidence}). To analyze whether this is the case, I re-estimate the linear probability model from Section \ref{subsec:Descriptive-evidence-of} on data simulated from the two estimated models.

\begin{figure}[t]
  \FIGURE
  {\begin{minipage}{\textwidth} \centering
    \subfloat[Search and Discovery Model]{\centering{}\includegraphics[width=0.5\textwidth,trim=0 1em 0 2.7em,clip]{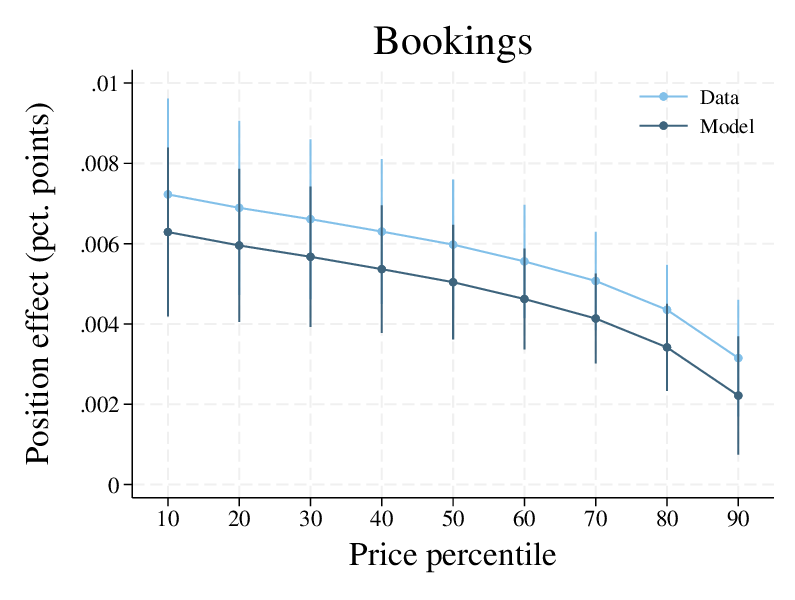}}%
    \subfloat[Weitzman Model]{\centering{}\includegraphics[width=0.5\textwidth,trim=0 1em 0 2.7em,clip]{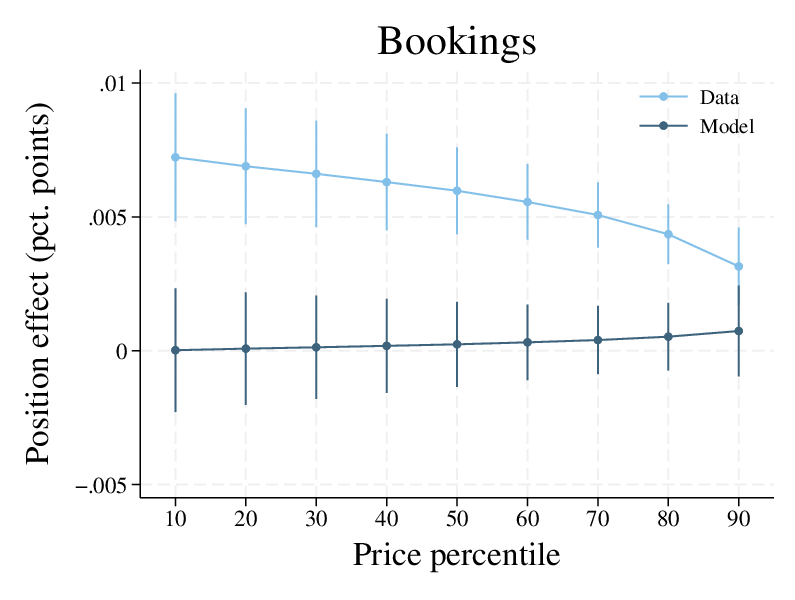}}
    \end{minipage}
  }
  {Position Effects by Price: Model Comparison
  \label{fig:fit-hetpos}}
  {Heterogeneous position effects for bookings as in Figure \ref{fig:Heterogeneous-position-effects}, estimated on the estimation sample with bookings either as observed or simulated from the model conditional on consumers searching at least one hotel.}
\end{figure}

Figure \ref{fig:fit-hetpos} presents the results. The left panel shows that the SD model fits the position effects across different price percentiles remarkably well.\footnote{The overall magnitude of position effects differs from those in Figure \ref{fig:Heterogeneous-position-effects} due to the different samples.} In contrast, the right panel shows that the WM model does not predict differences in position effects for differently priced hotels and, in line with Figure \ref{fig:fit}, underpredicts the position effects for all price percentiles.\footnote{For clicks, both models fit heterogeneous position effects by price reasonably well.}

\subsection{Discussion \label{subsec:comparison-weitzman}}

The difference in how the models fit position effects in bookings matters for the trade-offs between ranking objectives. These trade-offs depend on the relative strength of the price and demand effects, which is determined by heterogeneity in position effects in bookings (see Section \ref{subsec:Implications-Descr-Evidence}). Because the SD model better captures this heterogeneity, it also better captures the relative strength of the two effects and the resulting trade-offs.

This difference in model fit stems from how the two models generate position effects. In the WM model, consumers observe detail page utilities for all alternatives at the beginning of search. Without product discovery, the WM model generates position effects through position-specific search costs. With this mechanism, consumers search lower-ranked alternatives only if they are sufficiently promising to justify paying the higher search cost. Lower-ranked alternatives therefore need to offer larger list utility ($u_j^l$) to be searched, making it more likely they are chosen conditional on search. However, this prediction contradicts the data: \cite{Ursu2018} shows that the booking probability conditional on searching is constant across positions in the Expedia data.\footnote{Without a randomized ranking, \cite{Choi2019} also find that, conditional on clicking, position has no effect on choices.} Consequently, it is difficult for the WM model to fit how both clicks and bookings decrease across positions. In contrast, the SD model generates position effects through product discovery. This mechanism does not require lower-ranked alternatives to offer higher utility, allowing the SD model to better capture the position effects in bookings and their heterogeneity. Online Appendix \ref{subsec:Weitzman-discussion} provides a more formal discussion and compares other fit measures.

\section{Rankings and Their Trade-Offs \label{sec:Counterfactual-results}}

I now quantify the trade-offs between different ranking objectives by comparing ranking algorithms. Each algorithm is designed to maximize one of the three outcomes---transactions, revenues, or consumer welfare---by ordering the available alternatives on the list.\footnote{By iterating over the possible number of alternatives, the proposed algorithms can be generalized to the case where the assortment is not fixed.} Following prior work \citep[e.g.,][]{Compiani2023}, I assume the platform knows all alternatives' attributes, consumers' preferences, and consumers' search conduct from the estimated model. I also assume that the platform earns revenues through a commission proportional to the transaction price, so that its revenues are maximized when the total revenues across alternatives are maximized.\footnote{The ranking algorithms I propose can be readily adjusted to the case where a platform earns more by selling certain products. The trade-offs between ranking objectives in this case will depend on the utility those alternatives offer.}

I further focus on the case where consumers learn about the ranking change and update their beliefs accordingly. Such belief updating could occur through repeated interaction with the platform. Online Appendix \ref{subsec:ranking-effects-no-belief-adjustment} additionally provides results for the case where consumers do not update their beliefs, showing that the qualitative results remain unchanged. Hence, independent of whether consumers update their beliefs, the trade-offs between the different rankings are similar.

\subsection{Ranking Algorithms \label{subsec:The-six-rankings}}

To evaluate the trade-offs between the different ranking objectives, I develop ranking algorithms for each of the three objectives. However, developing provably optimal algorithms is challenging because switching the positions of two alternatives affects consumers' decisions with respect to all alternatives, leading to a complex combinatorial optimization problem. Algorithms from prior work also need not perform well in the search and discovery model because they were developed under different assumptions about the search process \citep[e.g.,][]{Derakhshan2022, Kosilova2025}.

One approach would be to simplify the model to the point where it becomes feasible to obtain provably optimal algorithms. However, this would require abstracting from important features of the empirical model. To retain all features, I instead derive heuristic algorithms that approximate optimal rankings. I then validate these heuristics by showing they outperform several alternatives. Beyond informing the trade-offs between objectives, these algorithms can help platforms implement rankings in practice, a challenging problem that brute-force algorithms or experimentation cannot easily solve unless there are very few alternatives to rank.

\paragraph{Maximizing Consumer Welfare.} I define consumer welfare as the expected utility consumers receive from a search session given a ranking, net of the discovery and search costs they incur. Formally, consumer welfare for a consumer $i$ is
\begin{equation} \label{eq:consumer-welfare}
  \text{CW}^i = \mathbb{E}[\max_{j \in S} U_j - N_s c_s - N_d c_d | (x_j^l, \tilde{\gamma}_j,x_j^d, \delta_j,h_{ij}) \forall j\in J_i]\, ,
\end{equation}
where the expectation is taken over the idiosyncratic preference shocks. These shocks determine the consideration set $S$, the number of searches $N_s$, and the number of discoveries $N_d$ through the optimal search policy.

To maximize consumer welfare, a ranking should help consumers find high-utility alternatives, reduce the search and discovery costs they pay, or both. Intuitively, this can be achieved by ranking high-utility alternatives higher, which makes more consumers discover them while also reducing the discovery costs they pay. However, consumers benefit from discovering high-utility alternatives only if they also search them, and a large utility $u_j$ alone does not guarantee that an alternative also has high search value $z_j^s$. Hence, ranking only by utility may rank higher alternatives that consumers do not search and thus cannot benefit from. Maximizing consumer welfare therefore requires ranking alternatives based on both their utility and search value.

I propose the \emph{Discovery-Value Ranking (DVR)}, which ranks alternatives in decreasing order of an index value $r_j^d$. This index is implicitly defined by
\begin{align}
  \int_{r_j^d}^{\infty} 1 - G(w; x_j^l, \tilde{\gamma}_j, x_j^d, \delta_j)\text{d}w = c_d \ ,\label{eq:discovery-value-ranking}
\end{align}
where $G(w; x_j^l, \tilde{\gamma}_j, x_j^d, \delta_j)$ denotes the CDF of the distribution of the effective value $\tilde{w}_j$, given the alternative's attributes and fixed effects. The index $r_j^d$ is equivalent to the discovery value $z^d$ for the hypothetical case where consumers already know that the next alternative they discover has attributes and fixed effects $(x_j^l, \tilde{\gamma}_j, x_j^d, \delta_j)$, so they reveal only $\nu_j$ when discovering it and $\varepsilon_j$ when searching it. Hence, $r_j^d$ balances the utility and search value of an alternative by capturing the expected gain from discovering $j$ when its attributes are already known.

Proposition \ref{prop:consumer-welfare-maximizing} in Online Appendix \ref{sec:comparison_heuristics} proves that the DVR maximizes consumer welfare in two cases: when consumers search only over idiosyncratic shocks and when the detail page reveals no vertical attributes. The empirical model, however, fits neither case. Consumers reveal attributes $x_j^l$ when discovering a product and the location score in $x_j^d$ when visiting the detail page. Nonetheless, the DVR provides a good heuristic. Online Appendix \ref{sec:comparison_heuristics} shows it outperforms several alternatives and approximates the optimum well: with a reduced set of alternatives where the true optimum can be computed, the DVR reduces welfare by only 0.16\textcent{}. Notably, the DVR outperforms a utility-based ranking, corroborating \cite{Compiani2023} who find that utility-based rankings fail to maximize consumer welfare when consumers search over vertical attributes. The DVR is therefore a useful heuristic for maximizing consumer welfare, both for the present analysis and for platforms in practice.

\paragraph{Maximizing Transactions.} Maximizing transactions---the probability that consumers choose any alternative from the list---requires ranking alternatives with larger position effects higher. Proposition \ref{prop:position-effect-model} shows that these are the high-utility alternatives consumers are more likely to search and choose upon discovering them. Maximizing transactions is therefore closely related to maximizing consumer welfare, such that the DVR may also maximize transactions.

Proposition \ref{prop:conversion-max-special} in Online Appendix \ref{sec:comparison_heuristics} confirms that the DVR maximizes both consumer welfare and transactions when there are no detail page attributes. With detail page attributes, however, maximizing transactions may differ from maximizing consumer welfare. Nonetheless, the DVR performs well in the estimated model: it lowers transactions by less than 1\% relative to the true optimum (in the case of a reduced set of alternatives) and consistently outperforms other heuristics. Hence, I conclude that the DVR is also a useful heuristic for maximizing transactions, and that the trade-offs between maximizing welfare and transactions are limited.

\paragraph{Maximizing Platform Revenues.} Maximizing total revenues requires balancing the demand and price effects highlighted in Section \ref{subsec:Implications-Descr-Evidence}. No single index captures this balance. Instead, I propose a heuristic algorithm that iteratively ranks alternatives from the bottom up, constructing the \emph{Bottom-Up Ranking (BUR)}. I compute the BUR assuming consumers do not learn about the ranking change, which simplifies the algorithm and follows prior work \citep[e.g.,][]{Derakhshan2022,Compiani2023}. However, I report the main effects for the case where consumers do update their beliefs, confirming that the BUR performs well even with belief updating.

Ranking from the bottom up is motivated by a prediction of the model: the demand for an alternative in a given position depends on which products are displayed in higher positions but not on how those products are ordered. Consequently, iteratively ranking from the bottom up avoids some of the interference from how alternatives in other positions are ordered, unlike when ranking from the top down.

For example, suppose product $A$ is a high-utility product that most consumers search and choose when they discover it, and product $B$ is a low-utility product that consumers never choose. Whether $A$ is in the first and $B$ in the second position, or vice versa, does not affect demand for product $C$ in the third position; consumers who choose $A$ when discovering it will not discover $C$ in either case. In contrast, when ranking from the top down, the order of alternatives in lower positions matters for demand in higher positions. Suppose consumers always stop scrolling before revealing the third position. Whether $B$ or $C$ is in the second position then determines which alternative $A$ in the first position is compared to. Hence, when choosing the top alternative first, how other alternatives are ordered affects demand for the alternative in the first position and, consequently, which product should be ranked first to maximize revenues.

The BUR algorithm works as follows: First, it initializes the ranking using the DVR. It then computes total revenues when each available alternative is moved to the last position. The alternative that maximizes these revenues is ranked last. The algorithm then moves to the second-to-last position and chooses the alternative that maximizes total revenues in that position, and so on. This iterative procedure is feasible to implement as it requires computing total revenues only $\sum_{k=1}^{N}(N-k)$ times when there are $N$ alternatives.\footnote{For example, with 20 alternatives, this requires 190 revenue comparisons, substantially fewer than the quintillions required for a brute-force algorithm. It is also fewer than the algorithms proposed by \cite{Compiani2023}, which would require computing revenues $20\times19\times18 = 6,840$ times to find the revenue-maximizing ranking for the first three positions.}

Online Appendix \ref{sec:comparison_heuristics} shows that the BUR effectively approximates the revenue-maximizing ranking. With only five alternatives, it lowers total revenues by only 0.62\% from the true optimum computed via brute-force algorithm. With more alternatives, it outperforms several alternative heuristics. Notably, it also substantially outperforms the ranking that would maximize revenues if there were no demand effect, highlighting the demand effect's importance. Hence, the BUR is a useful heuristic for the present analysis and for platforms wanting to maximize revenues in practice.

\subsection{Trade-Offs Between Objectives\label{subsec:ranking-effects}}

Using these algorithms, I now quantify the trade-offs between ranking objectives. These trade-offs matter for platforms deciding which objective to target. For example, when the revenue-maximizing ranking substantially harms consumers, platforms may be reluctant to implement it due to concerns about competition and long-term customer churn. In contrast, when trade-offs are limited, platforms can implement rankings that achieve the intended objective without detrimental effects on other outcomes.

I quantify how the proposed ranking algorithms perform relative to two benchmarks. The first benchmark is a fully randomized ranking, a neutral benchmark that does not favor any objective.\footnote{For the randomized ranking, I compute consumer welfare and revenues as the average over 1,000 random rankings.} The second benchmark is \emph{Expedia's ranking} (ER), which reveals how the proposed rankings could improve on the status quo during the sample period.\footnote{Expedia's algorithm was not trained on the demand system I impose with my model. Nonetheless, the estimated model closely fits the data, suggesting that Expedia's ranking should continue to work as intended within the model-implied demand system.}

To compute each ranking's effects, I simulate search paths for consumers given the ranking. Specifically, I take 100,000 draws from the distributions of the idiosyncratic preference shocks, apply the optimal search policy for each draw, and compute average outcomes across all draws and consumers.

Consumers update their beliefs following the ranking change. In the model, these beliefs are determined by $\mu(h)$ defined in \eqref{eq:functional_form_beliefs}, where the parameter $\rho$ governs the decline of the mean effective value across positions. In the neutral benchmark, consumers know the ranking is randomized, so $\rho=0$. For the other rankings, I estimate the updated $\rho$ via linear regression on data with the respective ranking. Online Appendix \ref{subsec:belief-updating} provides details. Online Appendix \ref{subsec:ranking-effects-no-belief-adjustment} reports results for the case where consumers do not learn of the ranking change, which are qualitatively similar.

Table \ref{tab:cf_ranking_effects} reports the effects of the different rankings relative to the neutral benchmark. As intended, the Bottom-Up Ranking (BUR) increases revenues the most (10.54\%), clearly outperforming Expedia's own ranking (ER). The BUR achieves this by increasing transactions and the average price per transaction. Despite being designed to maximize consumer welfare, the Discovery-Value Ranking (DVR) also substantially increases revenues (5.88\%), more so than the ER (1.40\%), albeit not as much as the BUR. The DVR achieves this revenue increase by increasing transactions, which increases revenues despite lowering the average price per transaction.

Turning to consumer welfare effects, the DVR increases average consumer welfare by a modest 11\textcent. This increase is small because most consumers are not predicted to book a hotel and, hence, benefit little from changes in the ranking. For consumers who eventually book a hotel, the DVR increases consumer welfare by a more substantial \$12.20. Despite being designed to maximize the platform's revenues, the BUR also benefits consumers by increasing average consumer welfare by 8\textcent{}, and by $\$9.30$ for consumers who eventually book a hotel. While these welfare gains are smaller, the BUR still achieves more than 70\% of the welfare increase of the DVR that targets consumer welfare directly, and clearly outperforms the ER.

The last row of Table \ref{tab:cf_ranking_effects} reveals how the rankings affect the discovery costs consumers pay. All rankings lower discovery costs, ranging from a reduction of 6\textcent{} for the ER to 17\textcent{} for the DVR. These small cost changes align with the small discovery cost estimates. Moreover, they make up only a small part of the overall welfare change, highlighting that rankings mainly affect consumers by determining which hotels they discover and eventually book rather than by reducing discovery costs.

\begin{table}[htb] \centering \footnotesize
\caption{Ranking Effects over Random Ranking} \label{tab:cf_ranking_effects}

\begin{threeparttable}
\begin{tabular}{llcccc} 
\midrule

 && \multicolumn{1}{c}{ER} & \multicolumn{1}{c}{DVR} & \multicolumn{1}{c}{BUR} \\
\midrule
\bfseries{Platform} \\
\hspace{1em}Total revenues (\%)&&\hphantom{-}1.40&\hphantom{0}\hphantom{-}5.88&\hphantom{-}10.54\\
\hspace{1em}Number of transactions (\%)&&\hphantom{-}1.83&\hphantom{-}12.09&\hphantom{0}\hphantom{-}9.78\\
\hspace{1em}Avg. price of booking (\%)&&-0.43&\hphantom{0}-5.54&\hphantom{0}\hphantom{-}0.70\\
\midrule
\bfseries{Consumers} \\
\hspace{1em}Consumer welfare (\$, average)&&\hphantom{-}0.02&\hphantom{0}\hphantom{-}0.11&\hphantom{0}\hphantom{-}0.08\\
\hspace{1em}Consumer welfare (\$, cond. on booking)&&\hphantom{-}2.15&\hphantom{-}12.20&\hphantom{0}\hphantom{-}9.30\\
\hspace{1em}Discovery costs (\$,  cond. on booking)&&-0.06&\hphantom{0}-0.17&\hphantom{0}-0.16\\
\midrule 
\end{tabular}
\begin{tablenotes}
\item \footnotesize{\emph{Notes:} Effects of different rankings relative to a randomized ranking when consumers adjust their beliefs. ER: Expedia's own ranking. DVR: Discover-Value Ranking (heuristic to maximize consumer welfare). BUR: Bottom-Up Ranking (heuristic to maximize revenues). }
\end{tablenotes} 
\end{threeparttable}
\end{table}

\subsection{Discussion and Implications \label{subsec:discussion-ranking-effects}}

The proposed Discovery Value and Bottom-Up rankings increase both consumer welfare and revenues over the neutral benchmark and the status quo. This alignment stems from the demand effect highlighted in Section \ref{subsec:Implications-Descr-Evidence}: ranking higher the alternatives that consumers are likely to search and choose increases the platform's revenues by increasing transactions while also benefiting consumers. Conversely, increasing consumer welfare requires ranking higher the high-utility alternatives that consumers will also search, which increases revenues by increasing transactions.

These limited trade-offs between the proposed rankings have important implications for platforms and regulators. First, the choice of ranking objective matters less than one might expect. Moreover, improvements to ranking algorithms can benefit both consumers and platforms, regardless of which objective the algorithm targets. The ranking comparison in Online Appendix \ref{sec:comparison_heuristics} confirms this: the proposed heuristics outperform others along all three outcomes, not just the one they are designed to maximize. Second, because maximizing consumer welfare is closely aligned with maximizing transactions, platforms aiming to increase consumer welfare can do so by maximizing transactions, an objective that is measurable and thus easier to target. Finally, the BUR also benefits consumers even when they do not update their beliefs following the ranking change (see Online Appendix \ref{subsec:ranking-effects-no-belief-adjustment}), suggesting that regulators' concerns about exploiting uninformed consumers may not be warranted \citep[e.g.,][]{CMA2021}.

These results differ from prior work using the same Expedia data. \cite{Ursu2018} estimates a Weitzman model and finds that a utility-based ranking substantially increases consumer welfare relative to Expedia's ranking but decreases revenues in three out of four destinations. With an extended Weitzman model, \cite{Compiani2023} consider the converse and find that a revenue-maximizing ranking substantially lowers consumer welfare. In contrast, I find that both the consumer-welfare-targeting DVR and the revenue-targeting BUR increase consumer welfare and revenues relative to Expedia's ranking.

The difference in results stems from the different models used to analyze the trade-offs. As Section \ref{sec:Estimation-results} revealed, the search and discovery model captures position effects in purchases, including their heterogeneity along the price dimension. In contrast, the Weitzman model underestimates these effects. It thus also underestimates the demand effect that aligns the different objectives, leading to larger trade-offs.\footnote{Other differences cannot explain the different results. Unlike \cite{Compiani2023} and \cite{Ursu2018}, I consider belief updating following ranking changes. Moreover, unlike \cite{Compiani2023} but like \cite{Ursu2018}, my consumer welfare measure accounts for search costs. Given the small changes in discovery costs across rankings and the similar results when consumers do not update their beliefs, neither of these differences would change my results. \cite{Compiani2023} also allow rankings to change the assortment by hiding some alternatives. However, the ranking algorithms in the authors' analysis tend to show all alternatives because of the low purchase probability, so this factor also cannot explain the difference.}

\section{Conclusion \label{sec:conclusion}}

This paper quantifies the trade-offs between different ranking objectives. I provide descriptive evidence for heterogeneity in position effects and show how this heterogeneity determines the trade-offs between ranking objectives. I further develop and estimate a structural model that microfounds this heterogeneity through product discovery. Using the model, I conduct a counterfactual analysis and find that the heuristic ranking algorithms I introduce benefit both the platform and consumers, suggesting that trade-offs between ranking objectives are limited. 

While I focus on the underlying demand effect and implications for the trade-offs between ranking objectives, I see two promising avenues for future research. First, studying rankings when individual sellers make strategic choices may reveal new insights into markets where sellers compete on rankings. Second, extending the empirical framework to accommodate multiple discovery technologies could provide novel insights into ranking effects when consumers can also discover products through more channels.

\clearpage
\begin{singlespace}
\bibliographystyle{chicago}
\bibliography{library}
\end{singlespace}

\clearpage

\appendix

\part*{Appendix }

\setcounter{section}{0}
\setcounter{equation}{0}
\setcounter{figure}{0}
\setcounter{table}{0}

\counterwithout{equation}{section}
\counterwithout{figure}{section}
\counterwithout{table}{section}

\renewcommand{\theequation}{A.\arabic{equation}}
\renewcommand{\thefigure}{A.\arabic{figure}}
\renewcommand{\thetable}{A.\arabic{table}}

This appendix provides proofs for the propositions presented in the main text.

\section{Proof of Proposition \ref{prop:position-effect-model}} \label{subsec:proof-prop-position-effect-model}

To prove Proposition \ref{prop:position-effect-model}, I first derive a lemma that provides general expressions for position effects in searches and purchases given an alternative's observable attributes, extending \citet{Greminger2021} who focused on unconditional position effects.
\begin{lemma}
\label{lem:position-effect} Let $\bar{W}_h=\max_{k\in\{k:h_k\leq h\}}\tilde{W}_k\geq z^d(h)$ denote the maximal effective value discovered up to position $h$. Conditional on its observable attributes, the position effect for the demand for some alternative $j$ in position $h$ is given by
\begin{multline}
\Delta d_j(h)=\mathbb{P}(\bar{W}_{h-1}\in(z^d(h),z^d(h-1)])\mathbb{P}(\tilde{W}_j>\bar{W}_{h-1}|\bar{W}_{h-1}\in(z^d(h),z^d(h-1)])\\
+ \mathbb{P}(\bar{W}_{h-1}\leq z^d(h))\mathbb{P}(\tilde{W}_j>z^d(h))\mathbb{P}(\tilde{W}_k>z^d(h))\ ,\label{eq:position-effect-prop}
\end{multline}
where $\tilde{W}_k$ is the effective value of the alternative $k$ shown beside $j$ in either position $h$ or $h+1$.
Similarly, the position effect in searches is given by
\begin{multline}
\Delta s_j(h)=\mathbb{P}(\bar{W}_{h-1}\in(z^d(h),z^d(h-1)])\mathbb{P}(Z_j^s>\bar{W}_{h-1}|\bar{W}_{h-1}\in(z^d(h),z^d(h-1)])\\
+ \mathbb{P}(\bar{W}_{h-1}\leq z^d(h))\mathbb{P}(Z_j^s>z^d(h))\mathbb{P}(\tilde{W}_k>z^d(h))\ .\label{eq:position-effect-prop-1}
\end{multline}
\end{lemma}
\begin{proof}
The stopping and search rules imply that the consumer stops in position $h$ whenever $\bar{w}_h \geq z^d(h)$, i.e., whenever the maximal effective value discovered so far exceeds the discovery value. Hence, by partitioning the probability space, the position effect in purchases can be written as
\begin{align}
\Delta d_j= & d_j(h)-d_j(h+1)\nonumber \\
= & \mathbb{P}(\bar{W}_{h-1}\in[z^d(h),z^d(h-1)))\mathbb{P}(\tilde{W}_j>\bar{W}_{h-1}|\bar{W}_{h-1}\in[z^d(h),z^d(h-1)))+\nonumber \\
 & \,\,\,\,\,\,\,\,\,\,\mathbb{P}(\bar{W}_{h-1}\leq z^d(h))[\mathbb{P}(\tilde{W}_j>z^d(h))(1-\mathbb{P}(\tilde{W}_k\leq z^d(h)))+C] \ ,\label{eq:appendix_proof_lemma}
\end{align}
where
\begin{align}
C= & \mathbb{P}(\tilde{W}_j\leq z^d(h))[\mathbb{P}(\tilde{W}_j\geq\bar{W}_{h+1}|\tilde{W}_j\leq z^d(h), \bar{W}_h \leq z^d(h))-\nonumber \\
 & \,\,\,\,\,\,\,\,\,\,\mathbb{P}(\tilde{W}_k\leq z^d(h))\mathbb{P}(\tilde{W}_j\geq\bar{W}_{h+1}|\tilde{W}_j\leq z^d(h),\tilde{W}_h\leq z^d(h), \bar{W}_k \leq z^d(h))]\ .\label{eq:appendix_proof_lemma2}
\end{align}
Because $\mathbb{P}(\tilde{W}_j\geq\bar{W}_{h+1}|\tilde{W}_j\leq z^d(h),\tilde{W}_k>z^d(h))=0$, we get $C=0$ and obtain equation \eqref{eq:position-effect-prop}. As consumers always click a discovered alternative $j$ whenever $z_{ij}^s\geq\bar{w}_h$, the expression for position effects in clicks can be derived the same way.
\end{proof}

The expressions in Lemma \ref{lem:position-effect} immediately imply that position effects in purchases increase in $j$'s effective value $\tilde{w}_j$. With a larger realization $\tilde{w}_j$, the probability in \eqref{eq:appendix_proof_lemma} increases, so it also increases as larger realizations become more likely. Hence, alternatives for which larger realizations of $\tilde{w}_j$ are more likely have larger position effects in purchases. The same logic applies to position effects in searches, though now the search value $z_j^s$ is used instead of the effective value $\tilde{w}_j$. 

To conclude the proof, it remains to show that (i) alternatives that are more likely to be searched conditional on their position also have larger search values, and (ii) alternatives that are more likely to be chosen conditional on their position also have larger effective values. 

Both conditions follow from the search and stopping rules of the optimal policy. The search rule implies that consumers are more likely to search alternatives with larger search values conditional on being discovered, so (i) holds. The stopping rule implies that alternatives with larger utility are more likely to be chosen conditional on being discovered and searched. A larger utility also implies a weakly larger effective value, so (ii) holds. 

Combined, Lemma \ref{lem:position-effect} and these two conditions imply that alternatives with larger effective values are more likely to be chosen conditional on their position and also have larger position effects in purchases. Similarly, alternatives with larger search values are more likely to be searched conditional on their position and also have larger position effects in searches. Hence, the two statements in Proposition \ref{prop:position-effect-model} follow.

\clearpage
\section{Proof of Proposition \ref{prop:implications}} \label{subsec:proof-prop-implications}

The generalized purchase theorem of \cite{Greminger2021} establishes that a consumer following the optimal policy eventually chooses the alternative $j$ with the largest $w_j = \min\{\tilde{w}_j, z^d(h_j)\}$, where $\tilde{w}_j=\min\{ z_j^s,u_j\}$. Given some $\tilde{w}_j$ for the chosen alternative $j$, the generalized eventual purchase theorem provides the necessary conditions on other alternatives that guarantee $j$ is discovered, searched, and chosen. To prove the proposition, I extend these conditions to also prescribe whether alternatives other than $j$ are discovered and searched, and show they are equivalent to the implications in the proposition.

Which alternatives the consumer searches given $w_j$ and conditional on being discovered depends on whether the alternative is discovered before or after $j$. 

\emph{$k$ discovered after $j$}: The optimal search rule prescribes searching alternative $k$ before choosing $j$ if and only if $z_k^s \geq z_k^j$ (search $k$ before searching $j$) or $z_k^s \geq u_j$ (search $k$ after searching $j$) applies, i.e., if and only if $z_k^s \geq \tilde{w}_j$. In this case, $u_k < \tilde{w}_j$ is necessary for $\tilde{w}_j \geq \tilde{w}_k$ to hold so that $j$ is eventually chosen according to the generalized eventual purchase theorem. Note also that search ends when the consumer chooses $j$, so $k$ is never searched after choosing $j$.

\emph{$k$ discovered before $j$}: The optimal search rule additionally prescribes searching $k$ before discovering $j$ if and only if $z_k^s \geq z^d(h_j)$. Combined with the same two conditions that apply for $k$ to be searched after discovering but before choosing $j$, this implies that the optimal policy prescribes searching $k$ if and only if $z_k^s \geq w_j$. Note that here we have $h_k \leq h_j$, so $z^d(h_k) \geq z^d(h_j)$ by the assumption that $z^d(h)$ is weakly decreasing in $h$. Hence, $w_k < w_j$ now requires that $u_k < w_j$, so $j$ is chosen according to the generalized eventual purchase theorem.

Combined, these conditions are sufficient to characterize which alternatives the consumer searches conditional on discovering them, and guarantee that $j$ is discovered, searched, and chosen. The conditions are also equivalent to the two search implications and the choice implication. 

The remaining discovery implication follows from the optimal stopping and discovery rules. First, the consumer has to discover up to at least position $h > h_j$ because $j$ has to be discovered to be chosen. Second, the optimal stopping rule implies that the consumer ends search by choosing $j$ if and only if $u_j \geq z^d(h)$ and $j$ has been searched. At position $h$, $j$ is searched before continuing to discover if and only if $z_j^s \geq z^d(h)$. A consumer therefore ends search at some position $\bar{h}$ if and only if $\bar{h} > h_j$ and $\tilde{w}_j \geq z^d(\bar{h})$. By the same logic, the optimal discovery rule always prescribes revealing a position $\underline{h}$ when $\tilde{w}_j < z^d(\underline{h})$. Hence, the consumer always ends search when the next position to be revealed is $\bar{h}\geq h_j$ such that $z^d(\bar{h}-1) > \tilde{w}_j \geq z^d(\bar{h})$. As $z^d(h)$ decreases in $h$, this condition is equivalent to the discovery implication, concluding the proof.

\newpage

\part*{Online Appendices }

\setcounter{section}{0}
\setcounter{equation}{0}
\setcounter{figure}{0}
\setcounter{table}{0}

\counterwithout{equation}{section}
\counterwithout{figure}{section}
\counterwithout{table}{section}

\renewcommand{\theequation}{OA.\arabic{equation}}
\renewcommand{\thefigure}{OA.\arabic{figure}}
\renewcommand{\thetable}{OA.\arabic{table}}

\section{Further Evidence of Position Effects\label{subsec:Robustness:-Position-effects}}

Table \ref{tab:reduced-form} presents the parameter estimates for the main regression \eqref{eq:lpm}. As in the main text, the coefficient estimates are scaled to represent changes in percentage points.\footnote{Standard errors are clustered at the consumer query level, capturing that search behavior can induce correlation in the error terms. A large draw for one alternative can mean that consumers are less likely to click on or book another alternative, suggesting a potential negative correlation in the error terms.} Columns 1 and 4 show the results of a baseline model that does not include the interaction term and thus assumes that position effects are homogeneous. Columns 2 and 5 show the main results. Columns 3 and 6 show the results for the specification that additionally includes hotel fixed effects.\footnote{Several hotels are displayed only to a single consumer and are therefore excluded from this specification. Only price and whether the hotel is on promotion vary across different search sessions, so the coefficients in $\beta$ cannot be estimated for other characteristics. The interaction between hotel characteristics and position can still be estimated because it is identified by within-hotel variation in position.}

\begin{figure}
  \FIGURE  {\subfloat{\centering{}\hspace{-2.2em}\includegraphics[width=0.41\textwidth,trim=0 1em 0 0.5em,clip]{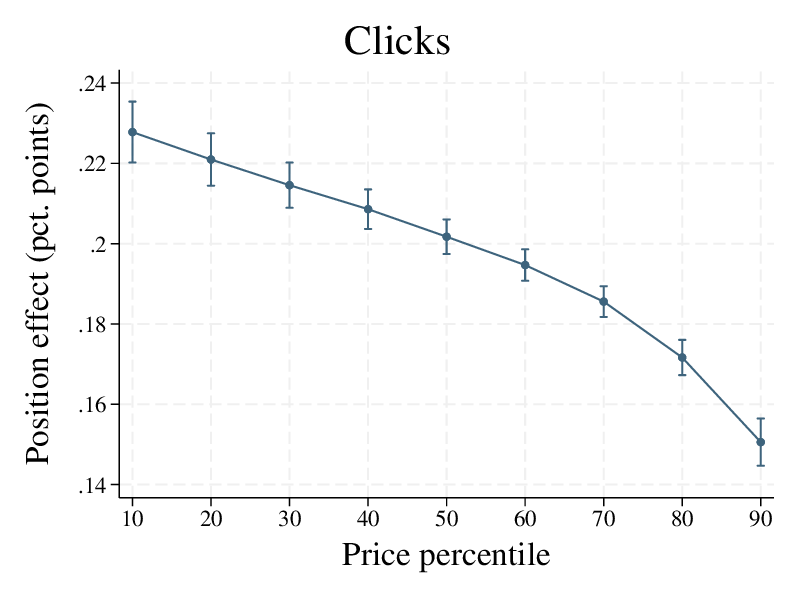}}
  \subfloat{\centering{}\includegraphics[width=0.41\textwidth,trim=0 1em 0 0.5em,clip]{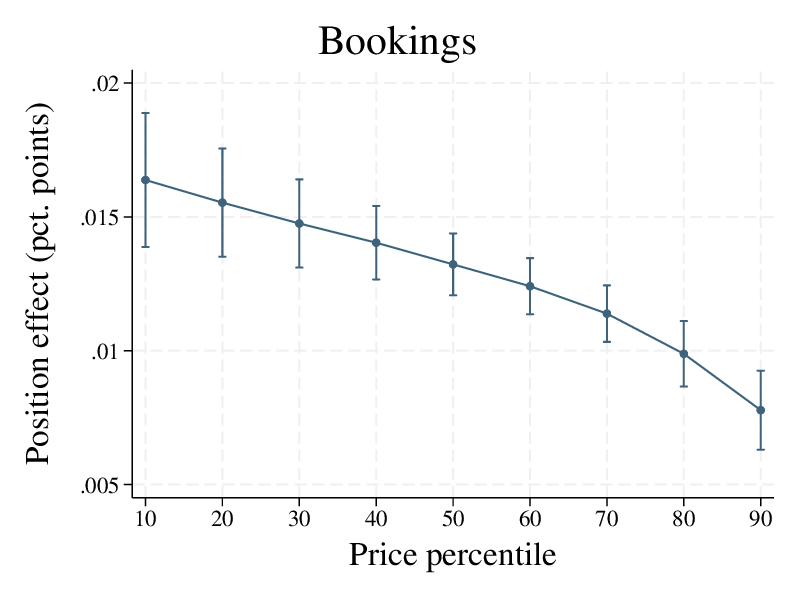}}}
  {Heterogeneous Position Effects: Probit \label{fig:Heterogeneous-position-effects-pricestar-probit}}
  {This figure replicates Figure \ref{fig:Heterogeneous-position-effects-pricestar} for a Probit model.}
\end{figure}

\begin{figure}
  \FIGURE {\subfloat{\centering{}\hspace{-2.2em}\includegraphics[width=0.41\textwidth,trim=0 1em 0 0.5em,clip]{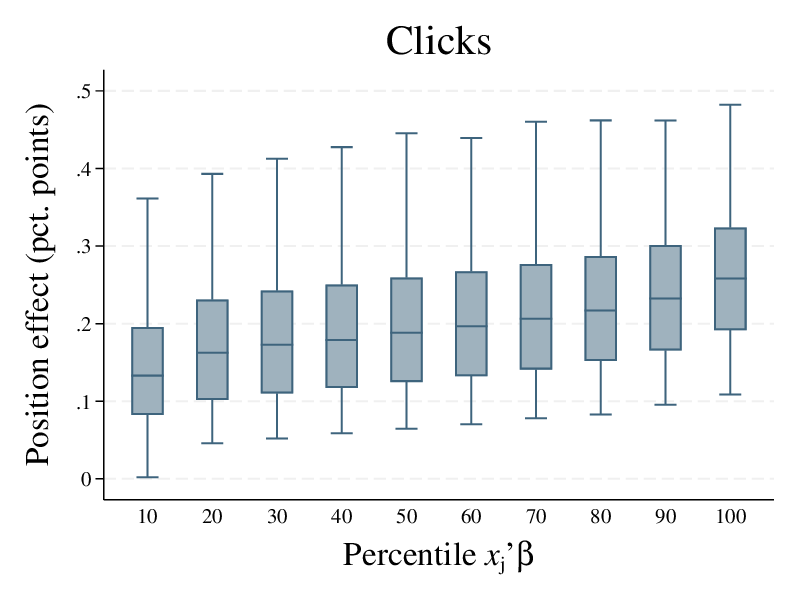}}
  \subfloat{\centering{}\includegraphics[width=0.41\textwidth,trim=0 1em 0 0.5em,clip]{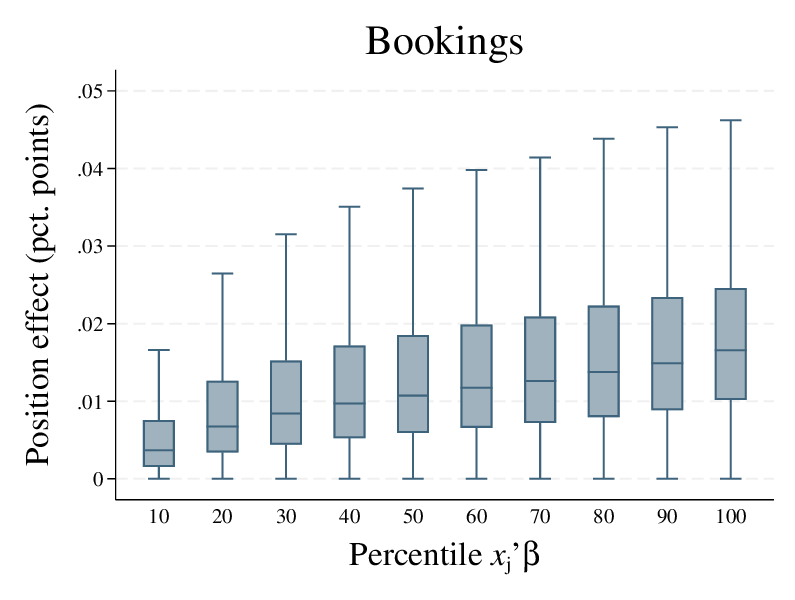}}}
  {Heterogeneous Position Effects: Probit \label{fig:Heterogeneous-position-effects-probit}}
  {\footnotesize{}This figure replicates Figure \ref{fig:Heterogeneous-position-effects} with estimates for a Probit model. }
\end{figure}

\begin{figure}
  \FIGURE
  {\subfloat{\centering{}\hspace{-2.2em}\includegraphics[width=0.41\textwidth,trim=0 1em 0 0.5em,clip]{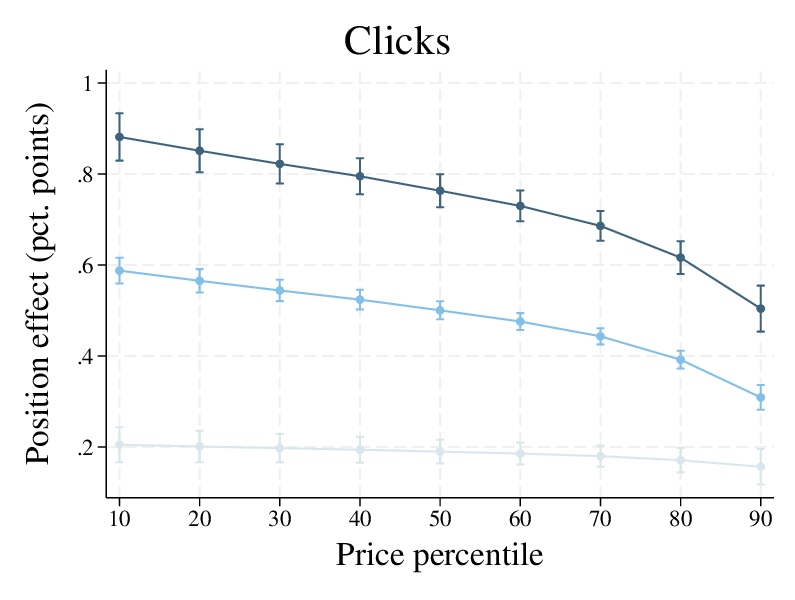}}
  \subfloat{\centering{}\includegraphics[width=0.41\textwidth,trim=0 1em 0 0.5em,clip]{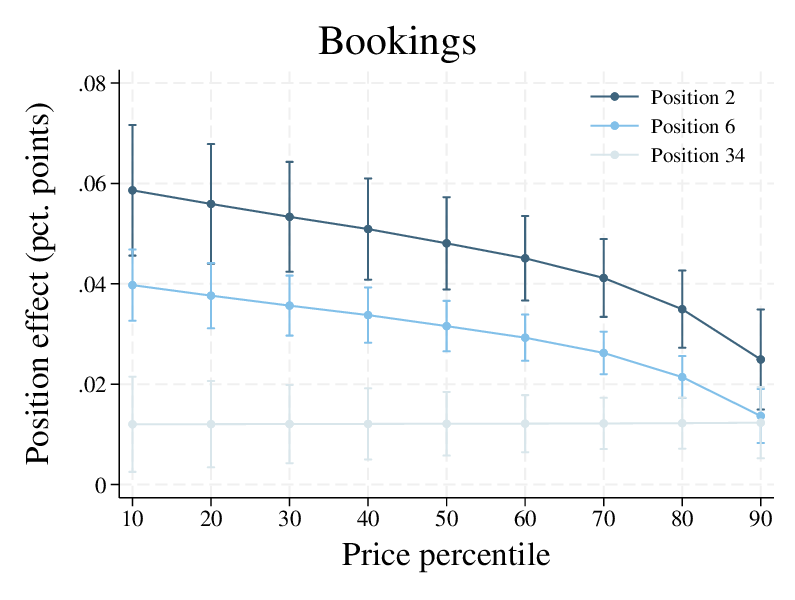}}}
  {Heterogeneous Position Effects: Cubic \label{fig:Heterogeneous-position-effects-pricestar-cubic}}
  {This figure replicates Figure \ref{fig:Heterogeneous-position-effects-pricestar} with estimates from a linear probability model that adds $pos_{ij}^2$, $pos_{ij}^2x_j$, $pos_{ij}^3$ and $pos_{ij}^3x_j$ to the baseline specification.}
\end{figure}

\begin{figure}
  \FIGURE
  {\subfloat{\centering{}\hspace{-2.2em}\includegraphics[width=0.41\textwidth,trim=0 1em 0 0.5em,clip]{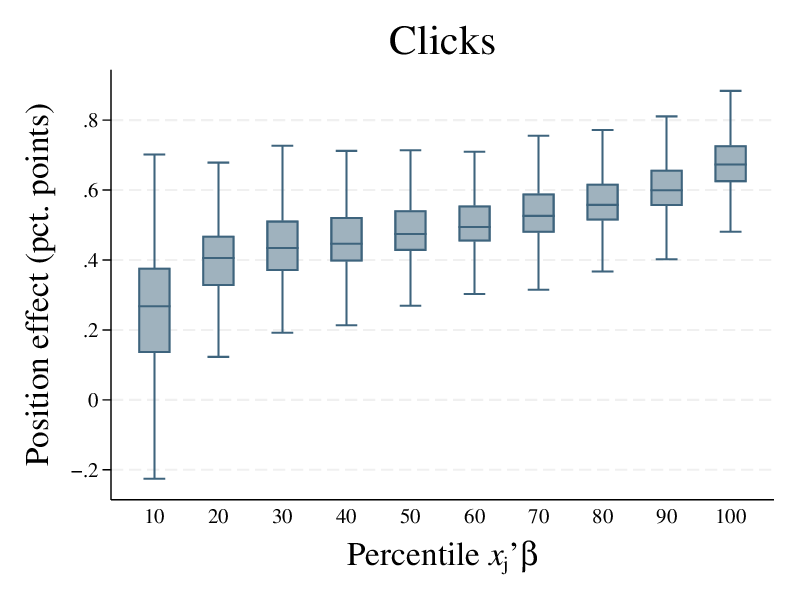}}
  \subfloat{\centering{}\includegraphics[width=0.41\textwidth,trim=0 1em 0 0.5em,clip]{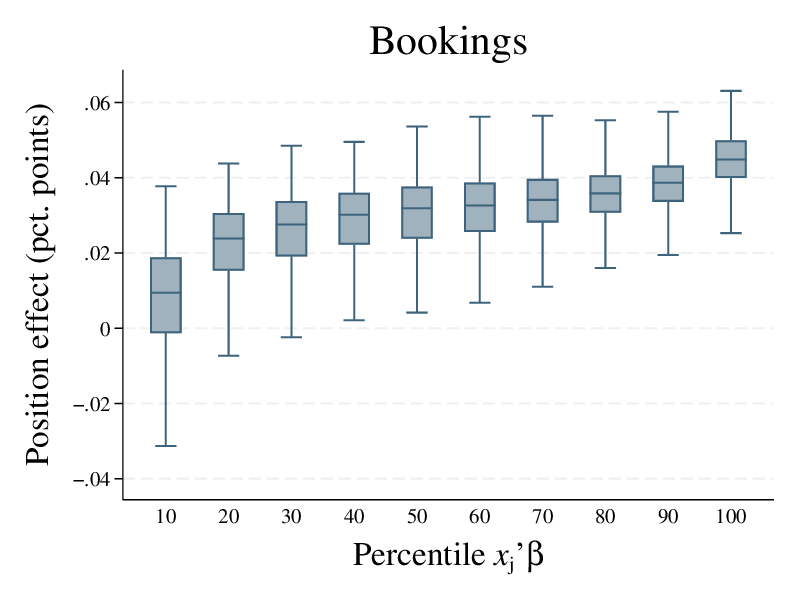}}}
  {Heterogeneous Position Effects: Cubic \label{fig:Heterogeneous-position-effects-cubic}}
  {This figure replicates Figure \ref{fig:Heterogeneous-position-effects} with estimates from a linear probability model that adds $pos_{ij}^2$, $pos_{ij}^2x_j$, $pos_{ij}^3$ and $pos_{ij}^3x_j$ to the baseline specification. Position effects are computed at $pos_{ij}=6$.}
\end{figure}

The remainder of this appendix presents results for alternative specifications. Throughout, the results indicate that the main findings are robust to these alternative specifications. First, I replicate the analysis of Figures \ref{fig:Heterogeneous-position-effects-pricestar} and \ref{fig:Heterogeneous-position-effects} for a Probit model and a linear probability model that parametrizes position effects more flexibly. To make estimation feasible, I include observable query characteristics for the Probit model rather than individual fixed effects. The results are shown in Figures \ref{fig:Heterogeneous-position-effects-pricestar-probit} to \ref{fig:Heterogeneous-position-effects-cubic}.

Table \ref{tab:reduced-form-flex} presents coefficient estimates for another linear probability model where the position effect is specific to the first three positions and then follows a linear specification for the remaining hotels. This specification is given by
\begin{multline}
\mathbb{P}(Y_{ij}=1|z_{ij}, pos_{ij})=x_{j}'\beta + \sum_{h=1}^{3}1(pos_{ij}\leq h)(\gamma_{h}+x_{j}'\theta_{h}) - \\
1(pos_{ij}\geq 4)(pos_{ij}-3)(\gamma_{4}+x_{j}'\theta_4)+\tau_{i}\ . \label{eq:reduced-form-flex1}
\end{multline}
In this specification, $\gamma_{h}$ and $\theta_h$ capture the position effect on position $h$. To see this, consider, for example, the effect of decreasing $pos_{ij}$ from $h=2$ to $h=1$. \eqref{eq:reduced-form-flex1} and the definition of the position effect in \eqref{eq:def-position-effect} implies that this effect is given by
\begin{equation}
  \text{Position effect}_{ij} = (\gamma_1 + \gamma_2) + x_j'(\theta_1+\theta_2) - \gamma_1 - x_j'\theta_1 = \gamma_2 + x_j'\theta_2 \, .
\end{equation}
Hence, $\gamma_2$ and $\theta_2$ capture the expected change in $Y_{ij}$ when moving from the second to the first position. Similarly, $\gamma_4$ and $\theta_4$ capture the expected change in $Y_{ij}$ when moving across positions beyond $h = 3$. For example, changing $pos_{ij}$ from $h=5$ to $h=4$ implies
\begin{equation}
  \text{Position effect}_{ij} = - (4 - 5 - 3)(\gamma_4 + x_j'\theta_4) = \gamma_4 + x_j'\theta_4 \, .
\end{equation}

Finally, I also estimate the following specification, which allows position effects to be fully flexible across all positions:
\begin{equation}
\mathbb{P}(Y_{ij}=1|z_{ij}, pos_{ij})=x_{j}'\beta + \sum_{h=1}^{33}1(pos_{ij}=h)(\gamma_{h}+x_{j}'\theta_{h})+\tau_{i}\ .\label{eq:full-flexible-spec}
\end{equation}
In this specification, $\gamma_{h}$ and $\theta_h$ capture the position effect on position $h$ relative to the bottom position $h=34$.\footnote{I drop sessions with more than 34 hotels to avoid estimating position effects for positions that are rarely observed.}

\begin{figure}[h]
  \FIGURE
  {\subfloat{\centering{}\hspace{-2.2em}\includegraphics[width=0.5\textwidth,trim=0 1.2em 0 1em,clip]{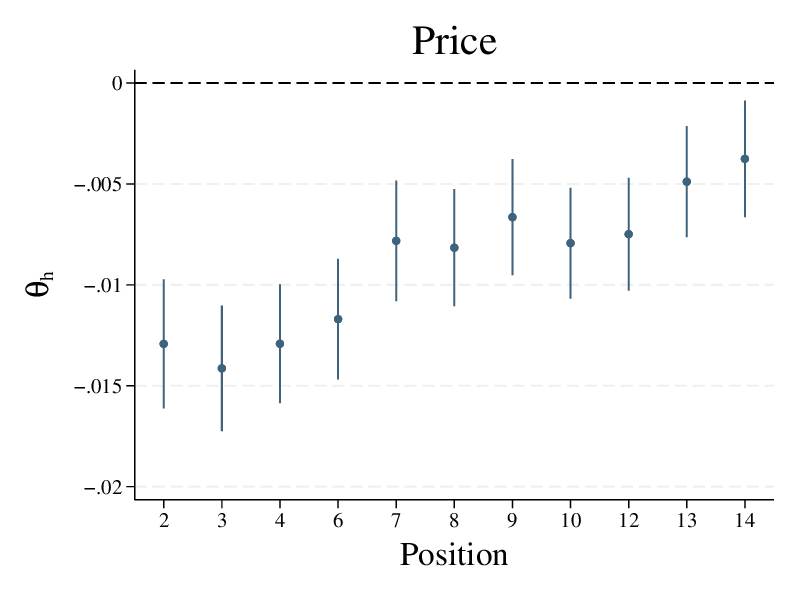}}
  \subfloat{\centering{}\includegraphics[width=0.5\textwidth,trim=0 1.2em 0 1em,clip]{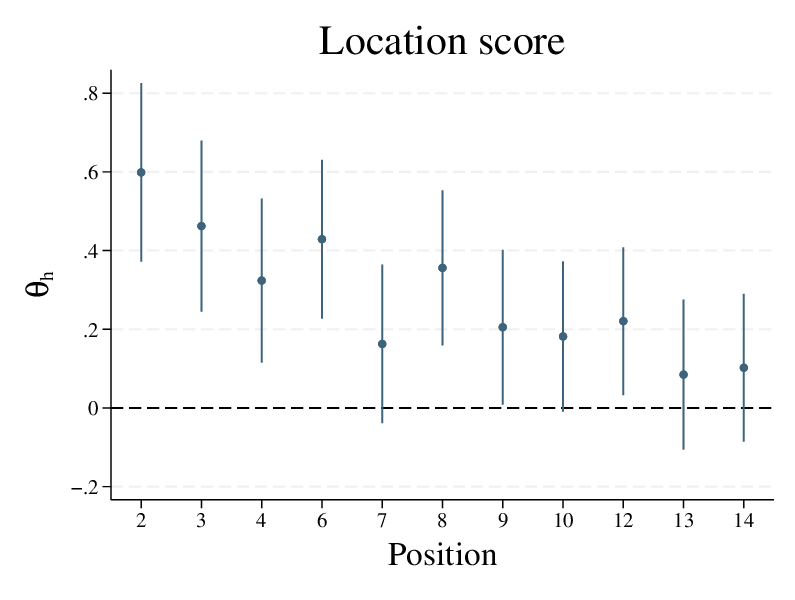}}}
  {Flexible Heterogeneous Position Effects for Clicks \label{fig:Flexible-Heterogeneous-Position-clicks}}
  {This figure shows the parameter estimates $\theta_h$ in specification \eqref{eq:full-flexible-spec} for the first 15 positions. A positive estimate implies that the attribute amplifies the position effect.}
  \end{figure}

\begin{figure}[h]
  \FIGURE
  {\subfloat{\centering{}\hspace{-2.2em}\includegraphics[width=0.5\textwidth,trim=0 1.2em 0 1em,clip]{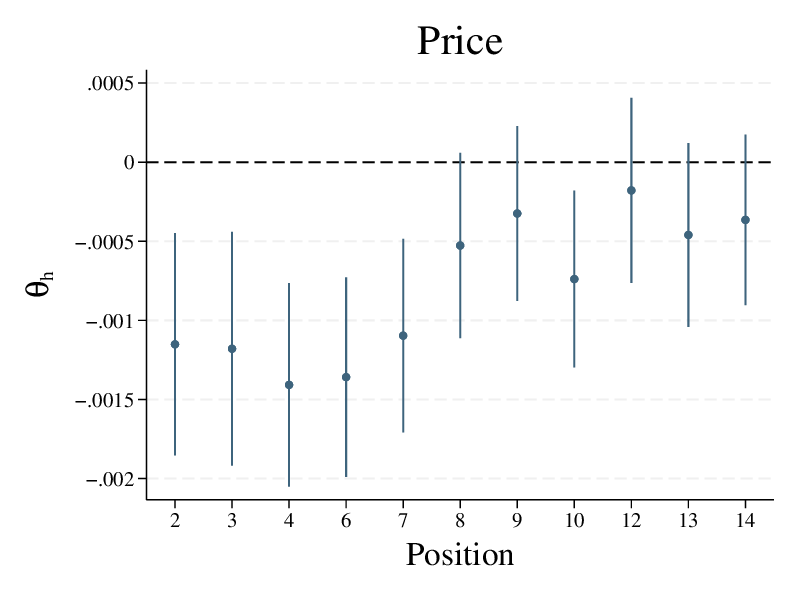}}
  \subfloat{\centering{}\includegraphics[width=0.5\textwidth,trim=0 1.2em 0 1em,clip]{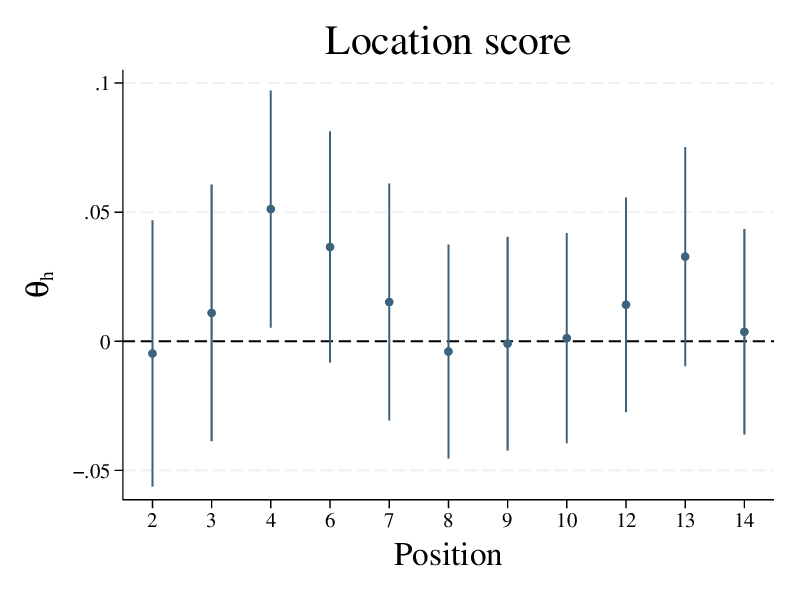}}}
  {Flexible Heterogeneous Position Effects for Bookings \label{fig:Flexible-Heterogeneous-Position-book}}
  {This figure shows parameter estimates  $\theta_h$ in specification \eqref{eq:full-flexible-spec} for the first 15 positions. A positive estimate implies that the attribute amplifies the position effect.}
\end{figure}

Figures \ref{fig:Flexible-Heterogeneous-Position-clicks} and \ref{fig:Flexible-Heterogeneous-Position-book} present the estimates for the main parameters $\theta_h$ for positions $h\leq15$. For clicks, the results are consistent with the parsimonious specification from the main text: price mutes position effects, while location score---a desirable attribute that also has a strong effect in the parsimonious specification---amplifies position effects. For the other variables, the effects are not statistically significant, indicating insufficient variation in these attributes in the data.

Given the small booking probabilities and many parameters to estimate, differences in position-specific position effects across different hotels are difficult to identify, as evidenced by the coefficients $\gamma_h$ not being statistically significant for the less flexible specification in Table \ref{tab:reduced-form-flex}. Nonetheless, the results from the extremely flexible specification continue to imply that lower-priced hotels have larger position effects conditional on other attributes.

\clearpage

\begin{table} \centering \footnotesize
  \caption{Coefficient Estimates (LPM, Random Ranking)} \label{tab:reduced-form}
  \scalebox{0.95}{
  \begin{threeparttable}
  {\renewcommand{\arraystretch}{0.8}
  \begin{tabular}{llllllllllll}
  \midrule
                      &\multicolumn{3}{c}{Clicks}                     &\multicolumn{3}{c}{Bookings}                   \\\cmidrule(lr){2-4}\cmidrule(lr){5-7}
                    &\multicolumn{1}{c}{(1)}   &\multicolumn{1}{c}{(2)}   &\multicolumn{1}{c}{(3)}   &\multicolumn{1}{c}{(4)}   &\multicolumn{1}{c}{(5)}   &\multicolumn{1}{c}{(6)}   \\
\midrule
Position            &      0.1398***&      0.0838***&      0.0809***&      0.0085***&     -0.0021   &     -0.0034   \\
                    &    (0.0019)   &    (0.0126)   &    (0.0128)   &    (0.0005)   &    (0.0028)   &    (0.0029)   \\
\addlinespace
Price               &     -0.0184***&     -0.0262***&     -0.0369***&     -0.0016***&     -0.0024***&     -0.0032***\\
                    &    (0.0003)   &    (0.0005)   &    (0.0006)   &    (0.0001)   &    (0.0001)   &    (0.0001)   \\
\addlinespace
On promotion        &      1.2784***&      1.7224***&      1.6935***&      0.1443***&      0.1692***&      0.1881***\\
                    &    (0.0522)   &    (0.1150)   &    (0.1291)   &    (0.0134)   &    (0.0290)   &    (0.0338)   \\
\addlinespace
Star rating         &      2.0188***&      2.6052***&               &      0.1367***&      0.1653***&               \\
                    &    (0.0354)   &    (0.0711)   &               &    (0.0088)   &    (0.0172)   &               \\
\addlinespace
Review score        &      0.2722***&      0.3090***&               &      0.0575***&      0.1049***&               \\
                    &    (0.0364)   &    (0.0846)   &               &    (0.0083)   &    (0.0195)   &               \\
\addlinespace
No reviews          &      0.4459***&      0.2226   &               &      0.1786***&      0.2467***&               \\
                    &    (0.1682)   &    (0.3984)   &               &    (0.0351)   &    (0.0838)   &               \\
\addlinespace
Chain               &      0.1993***&      0.2526** &               &      0.0196*  &      0.0611** &               \\
                    &    (0.0478)   &    (0.1004)   &               &    (0.0117)   &    (0.0246)   &               \\
\addlinespace
Location score      &      0.6157***&      0.8238***&               &      0.0637***&      0.0741***&               \\
                    &    (0.0186)   &    (0.0344)   &               &    (0.0040)   &    (0.0074)   &               \\
\addlinespace
Position $\times$ Price&               &     -0.0005***&     -0.0005***&               &     -0.0000***&     -0.0000***\\
                    &               &    (0.0000)   &    (0.0000)   &               &    (0.0000)   &    (0.0000)   \\
\addlinespace
Position $\times$ Star rating&               &      0.0342***&      0.0316***&               &      0.0017** &      0.0012*  \\
                    &               &    (0.0030)   &    (0.0030)   &               &    (0.0007)   &    (0.0007)   \\
\addlinespace
Position $\times$ Review score&               &      0.0024   &      0.0009   &               &      0.0027***&      0.0028***\\
                    &               &    (0.0036)   &    (0.0036)   &               &    (0.0008)   &    (0.0009)   \\
\addlinespace
Position $\times$ No reviews&               &     -0.0116   &     -0.0140   &               &      0.0039   &      0.0038   \\
                    &               &    (0.0170)   &    (0.0172)   &               &    (0.0036)   &    (0.0036)   \\
\addlinespace
Position $\times$ Chain&               &      0.0031   &     -0.0002   &               &      0.0024** &      0.0017   \\
                    &               &    (0.0042)   &    (0.0043)   &               &    (0.0010)   &    (0.0011)   \\
\addlinespace
Position $\times$ Location score&               &      0.0119***&      0.0124***&               &      0.0006** &      0.0010***\\
                    &               &    (0.0014)   &    (0.0014)   &               &    (0.0003)   &    (0.0003)   \\
\addlinespace
Position $\times$ On promotion&               &      0.0262***&      0.0274***&               &      0.0015   &      0.0014   \\
                    &               &    (0.0050)   &    (0.0051)   &               &    (0.0013)   &    (0.0013)   \\
\midrule
FE Hotel            &          no   &          no   &         yes   &          no   &          no   &         yes   \\
N                   &   1,220,909   &   1,220,909   &   1,219,242   &   1,220,909   &   1,220,909   &   1,219,242   \\
R2 (adj.)           &      0.0020   &      0.0026   &      0.0154   &      0.0098   &      0.0099   &      0.0054   \\
 \\ \vspace{-2em} \\ \midrule
  \end{tabular}
  }
  \vspace{-0.5em}
  \begin{tablenotes}
  \item \emph{Notes:} Coefficient estimates are scaled to represent changes in terms of percentage points.  All specifications include session fixed effects. Standard errors are shown in parentheses and are clustered at the query level. Star ratings are adjusted so that the position effect from the first row is for a hotel with 1 star and other characteristics equal to the minimum observed value. Statistical significance is indicated by *** $p<0.01$, ** $p<0.05$, * $p<0.1$.
  \end{tablenotes}
  \end{threeparttable}
  }
  \end{table}

\clearpage

\clearpage

\begin{table}[htb] \centering
\caption{Coefficient Estimates (Flexible LPM, Random Ranking)} \label{tab:reduced-form-flex}
\scalebox{0.65}{
\begin{threeparttable} \footnotesize
\begin{tabular}{llllllllllll}
\midrule
                    &\multicolumn{3}{c}{Clicks}                     &\multicolumn{3}{c}{Bookings}                   \\\cmidrule(lr){2-4}\cmidrule(lr){5-7}
                    &\multicolumn{1}{c}{(1)}   &\multicolumn{1}{c}{(2)}   &\multicolumn{1}{c}{(3)}   &\multicolumn{1}{c}{(4)}   &\multicolumn{1}{c}{(5)}   &\multicolumn{1}{c}{(6)}   \\
\midrule
1(Position $\leq$ 1) &      2.4021***&      5.5333***&      5.4790***&     -0.0072   &      0.1169   &      0.1317   \\
                    &    (0.2031)   &    (1.3558)   &    (1.3366)   &    (0.0505)   &    (0.2986)   &    (0.2986)   \\
1(Position $\leq$ 2) &      1.2725***&      1.0553   &      0.8896   &      0.0813*  &      0.0758   &      0.0016   \\
                    &    (0.1846)   &    (1.2205)   &    (1.2071)   &    (0.0490)   &    (0.2945)   &    (0.2965)   \\
1(Position $\leq$ 3) &      1.6571***&      2.1390** &      1.8724** &      0.1398***&     -0.0189   &     -0.0060   \\
                    &    (0.1333)   &    (0.8925)   &    (0.8834)   &    (0.0352)   &    (0.2201)   &    (0.2229)   \\
1(Position $\geq$ 4) $\times$ (Position - 3)&      0.0858***&     -0.0012   &      0.0024   &      0.0055***&     -0.0036   &     -0.0044   \\
                    &    (0.0021)   &    (0.0131)   &    (0.0136)   &    (0.0005)   &    (0.0030)   &    (0.0032)   \\
Price               &     -0.0184***&     -0.0228***&     -0.0334***&     -0.0016***&     -0.0021***&     -0.0030***\\
                    &    (0.0003)   &    (0.0005)   &    (0.0006)   &    (0.0001)   &    (0.0001)   &    (0.0001)   \\
On promotion        &      1.2637***&      1.4620***&      1.4306***&      0.1435***&      0.1368***&      0.1547***\\
                    &    (0.0521)   &    (0.1124)   &    (0.1270)   &    (0.0134)   &    (0.0280)   &    (0.0328)   \\
Star rating         &      2.0131***&      2.3959***&               &      0.1364***&      0.1573***&               \\
                    &    (0.0354)   &    (0.0688)   &               &    (0.0088)   &    (0.0169)   &               \\
Review score        &      0.2838***&      0.5136***&               &      0.0581***&      0.0848***&               \\
                    &    (0.0363)   &    (0.0797)   &               &    (0.0083)   &    (0.0187)   &               \\
No reviews          &      0.4901***&      1.0610***&               &      0.1810***&      0.2186***&               \\
                    &    (0.1677)   &    (0.3707)   &               &    (0.0351)   &    (0.0792)   &               \\
Chain               &      0.2090***&      0.4151***&               &      0.0201*  &      0.0252   &               \\
                    &    (0.0477)   &    (0.0968)   &               &    (0.0117)   &    (0.0241)   &               \\
Location score      &      0.6200***&      0.6854***&               &      0.0639***&      0.0794***&               \\
                    &    (0.0185)   &    (0.0335)   &               &    (0.0040)   &    (0.0073)   &               \\
1(Position $\leq$ 1) =1 $\times$ Price&               &     -0.0083***&     -0.0082***&               &     -0.0005   &     -0.0004   \\
                    &               &    (0.0018)   &    (0.0018)   &               &    (0.0004)   &    (0.0004)   \\
1(Position $\leq$ 2) =1 $\times$ Price&               &      0.0012   &      0.0011   &               &      0.0000   &     -0.0000   \\
                    &               &    (0.0017)   &    (0.0017)   &               &    (0.0004)   &    (0.0004)   \\
1(Position $\leq$ 3) =1 $\times$ Price&               &     -0.0042***&     -0.0040***&               &     -0.0002   &     -0.0002   \\
                    &               &    (0.0013)   &    (0.0013)   &               &    (0.0003)   &    (0.0003)   \\
1(Position $\geq$ 4) $\times$ (Position - 3) $\times$ Price&               &     -0.0004***&     -0.0004***&               &     -0.0000***&     -0.0000***\\
                    &               &    (0.0000)   &    (0.0000)   &               &    (0.0000)   &    (0.0000)   \\
1(Position $\leq$ 1) =1 $\times$ Star rating&               &      0.1464   &               &               &      0.0188   &               \\
                    &               &    (0.3004)   &               &               &    (0.0712)   &               \\
1(Position $\leq$ 2) =1 $\times$ Star rating&               &      0.1571   &               &               &      0.0446   &               \\
                    &               &    (0.2732)   &               &               &    (0.0684)   &               \\
1(Position $\leq$ 3) =1 $\times$ Star rating&               &      0.0949   &               &               &     -0.0337   &               \\
                    &               &    (0.1962)   &               &               &    (0.0489)   &               \\
1(Position $\geq$ 4) $\times$ (Position - 3) $\times$ Star rating&               &      0.0287***&      0.0267***&               &      0.0015** &      0.0013*  \\
                    &               &    (0.0031)   &    (0.0032)   &               &    (0.0008)   &    (0.0008)   \\
1(Position $\leq$ 1) =1 $\times$ Review score&               &     -0.5659   &               &               &     -0.0075   &               \\
                    &               &    (0.3882)   &               &               &    (0.0890)   &               \\
1(Position $\leq$ 2) =1 $\times$ Review score&               &     -0.0553   &               &               &     -0.0226   &               \\
                    &               &    (0.3481)   &               &               &    (0.0852)   &               \\
1(Position $\leq$ 3) =1 $\times$ Review score&               &     -0.2348   &               &               &      0.0619   &               \\
                    &               &    (0.2533)   &               &               &    (0.0617)   &               \\
1(Position $\geq$ 4) $\times$ (Position - 3) $\times$ Review score&               &      0.0124***&      0.0101***&               &      0.0022** &      0.0020** \\
                    &               &    (0.0037)   &    (0.0039)   &               &    (0.0009)   &    (0.0009)   \\
1(Position $\leq$ 1) =1 $\times$ No reviews&               &     -2.5397   &               &               &     -0.0922   &               \\
                    &               &    (1.8275)   &               &               &    (0.3962)   &               \\
1(Position $\leq$ 2) =1 $\times$ No reviews&               &     -0.2186   &               &               &     -0.0344   &               \\
                    &               &    (1.6511)   &               &               &    (0.3689)   &               \\
1(Position $\leq$ 3) =1 $\times$ No reviews&               &     -0.7859   &               &               &      0.1320   &               \\
                    &               &    (1.1897)   &               &               &    (0.2554)   &               \\
1(Position $\geq$ 4) $\times$ (Position - 3) $\times$ No reviews&               &      0.0265   &      0.0190   &               &      0.0032   &      0.0021   \\
                    &               &    (0.0174)   &    (0.0180)   &               &    (0.0037)   &    (0.0039)   \\
1(Position $\leq$ 1) =1 $\times$ Chain&               &     -0.2427   &               &               &     -0.0415   &               \\
                    &               &    (0.4426)   &               &               &    (0.1081)   &               \\
1(Position $\leq$ 2) =1 $\times$ Chain&               &     -1.2089***&               &               &     -0.0408   &               \\
                    &               &    (0.4012)   &               &               &    (0.1041)   &               \\
1(Position $\leq$ 3) =1 $\times$ Chain&               &      0.5719** &               &               &      0.1339*  &               \\
                    &               &    (0.2857)   &               &               &    (0.0749)   &               \\
1(Position $\geq$ 4) $\times$ (Position - 3) $\times$ Chain&               &      0.0118***&      0.0062   &               &      0.0011   &      0.0002   \\
                    &               &    (0.0045)   &    (0.0046)   &               &    (0.0011)   &    (0.0011)   \\
1(Position $\leq$ 1) =1 $\times$ Location score&               &      0.1312   &               &               &      0.0081   &               \\
                    &               &    (0.1428)   &               &               &    (0.0314)   &               \\
1(Position $\leq$ 2) =1 $\times$ Location score&               &      0.1355   &               &               &     -0.0154   &               \\
                    &               &    (0.1291)   &               &               &    (0.0311)   &               \\
1(Position $\leq$ 3) =1 $\times$ Location score&               &      0.1904** &               &               &     -0.0102   &               \\
                    &               &    (0.0931)   &               &               &    (0.0225)   &               \\
1(Position $\geq$ 4) $\times$ (Position - 3) $\times$ Location score&               &      0.0070***&      0.0076***&               &      0.0009***&      0.0013***\\
                    &               &    (0.0015)   &    (0.0015)   &               &    (0.0003)   &    (0.0003)   \\
1(Position $\leq$ 1) =1 $\times$ On promotion&               &     -0.0650   &     -0.0916   &               &     -0.1640   &     -0.1604   \\
                    &               &    (0.5072)   &    (0.5053)   &               &    (0.1294)   &    (0.1300)   \\
1(Position $\leq$ 2) =1 $\times$ On promotion&               &      0.5020   &      0.5675   &               &      0.2206*  &      0.2295*  \\
                    &               &    (0.4674)   &    (0.4671)   &               &    (0.1246)   &    (0.1254)   \\
1(Position $\leq$ 3) =1 $\times$ On promotion&               &      0.0370   &      0.0086   &               &     -0.0238   &     -0.0270   \\
                    &               &    (0.3366)   &    (0.3369)   &               &    (0.0872)   &    (0.0870)   \\
1(Position $\geq$ 4) $\times$ (Position - 3) $\times$ On promotion&               &      0.0174***&      0.0186***&               &      0.0001   &      0.0000   \\
                    &               &    (0.0054)   &    (0.0055)   &               &    (0.0013)   &    (0.0014)   \\
\midrule
FE Hotel            &          no   &          no   &         yes   &          no   &          no   &         yes   \\
N                   &   1,220,909   &   1,220,909   &   1,219,242   &   1,220,909   &   1,220,909   &   1,219,242   \\
R2 (adj.)           &      0.0044   &      0.0052   &      0.0178   &      0.0099   &      0.0100   &      0.0055   \\
 \\ \vspace{-5.8ex} \\ \midrule
\end{tabular}
\vspace{-0.5em}
\begin{tablenotes}
\item \emph{Notes:} Coefficient estimates are scaled to represent changes in terms of percentage points.  Estimates for query characteristics are omitted. Standard errors are shown in parentheses and are clustered at the query level. Star ratings are adjusted so that the position effect from the first row is for a hotel with 1 star and other characteristics equal to the minimum observed value. Statistical significance is indicated by *** $p<0.01$, ** $p<0.05$, * $p<0.1$.
\end{tablenotes}
\end{threeparttable}
}
\end{table}

\newpage
\clearpage

\newpage

\section{Comparing Ranking Algorithms} \label{sec:comparison_heuristics}

This appendix presents additional results for the comparison of rankings and their trade-offs. First, I provide the procedure used to obtain consumers' updated beliefs under alternative ranking algorithms and show the counterfactual results for the case where consumers' beliefs are fixed. Second, I provide results showing that the proposed Discovery-Value Ranking maximizes consumer welfare and transactions in some cases. Third, I compare different ranking algorithms with a brute-force algorithm that is guaranteed to find the optimal ranking. Because the number of possible rankings grows faster than exponentially in the number of alternatives, this comparison is only feasible for a small number of alternatives. Finally, I compare different ranking algorithms with each other to assess their relative performance for the case considered in the main counterfactual analysis.

\subsection{Consumers' Updated Beliefs under Alternative Rankings} \label{subsec:belief-updating}

In the model, consumers expect to discover alternatives with effective values that have a mean given by equation \eqref{eq:functional_form_beliefs}. When they adjust beliefs to the new rankings, $\mu_1$ and $\rho$ are updated to reflect the new ranking of alternatives and the resulting distribution of effective values across positions. To recover these parameters, I use the following procedure.

First, I take draws $\nu_{ij}$ and $\varepsilon_{ij}$ for each alternative $j$ and consumer $i$. Then, I use these draws to compute the effective values $\tilde{w}_{ij} = x_j^l{}' \hat{\beta} + \nu_{ij} + x_j^l{}' \hat{\gamma} + \hat{\tilde{\gamma}}_j+\min\{ \hat{\xi}, x_j^d{}' \hat{\kappa} - \hat{\mu}_{x\kappa} + \varepsilon_{ij}\}$ using the estimated parameter values.

Next, I estimate the following linear regression on the computed effective values:
\begin{equation}
\tilde{w}_{ij}=\pi_0 + \pi_1 \log(pos_{ij})+\epsilon_{ij}\ ,\label{eq:regression_beliefs}
\end{equation}
where $pos_{ij}$ is the position of alternative $j$ in the new ranking for consumer $i$. Moreover, I compute the mean $\mu_{\tilde{w}}$ as the average of $\tilde{w}_{ij}$ across all alternatives and consumers.

The estimated coefficient $\hat{\pi}_1$ provides the belief parameter $\rho^*$ for the new ranking. Under the assumption that consumers know the distribution of product attributes and shocks, the expected effective value across positions is the same as the unconditional expected effective value. Hence, given the mean $\mu_{\tilde{w}}$ and the functional form in \eqref{eq:functional_form_beliefs}, $\mu_1$ and its updated version $\mu_1^*$ can be recovered from the following relationship:
\begin{align}
\mu_{\tilde{w}} & = \frac{1}{|J|} \sum_{h=1}^{|J|} \mu(h) = \mu_1 + \frac{1}{|J|} \sum_{h=1}^{|J|} \rho\log(h) 
\end{align}
Using $\mu_1$ and $\mu_1^*$, I obtain the updated discovery value in the first position as $\tilde{z}^{d*} = \tilde{z}^d + \mu_1^* - \mu_1$. $\rho^*$ and $\tilde{z}^{d*}$ obtained this way fully characterize consumers' updated beliefs under the new ranking, as the functional form in \eqref{eq:functional_form_beliefs} remains unchanged.

Table \ref{tab:estimates_belief_adjustment} reports the estimates for $\pi_0$ and $\pi_1$ of the regression \eqref{eq:regression_beliefs} for the different rankings. Standard errors are small, so beliefs are estimated precisely. The reported estimates for $\pi_1 = \rho$ suggest that consumers expect alternatives to get worse at a faster rate under the DVR---the heuristic designed to maximize consumer welfare---than under Expedia's own ranking, which is in line with the DVR increasing consumer welfare by ranking high-utility alternatives higher. 

\begin{table}[th] \centering \normalsize
\caption{Estimates of Adjusted Beliefs} \label{tab:estimates_belief_adjustment}

\begin{threeparttable}
\begin{tabular}{llcccc} 
\midrule

 && \multicolumn{1}{c}{ER} & \multicolumn{1}{c}{DVR} & \multicolumn{1}{c}{BUR} \\
\midrule
$ \pi_0 $&&\hphantom{-}5.581&\hphantom{-}5.903&\hphantom{-}5.889\\
&&\hphantom{-}(0.026)&\hphantom{-}(0.026)&\hphantom{-}(0.025)\\
$ \pi_1$&&-0.056&-0.181&-0.175\\
&&\hphantom{-}(0.010)&\hphantom{-}(0.009)&\hphantom{-}(0.009)\\
\midrule 
\end{tabular}
\begin{tablenotes}
\item \footnotesize{\emph{Notes:} Standard errors in parentheses.  }
\end{tablenotes} 
\end{threeparttable}
\end{table}



\subsection{Ranking Effects With Fixed Beliefs} \label{subsec:ranking-effects-no-belief-adjustment}

Table \ref{tab:cf_ranking_effects_nobeliefadjust} shows the effects of the two main heuristics for the case where consumers do not update their beliefs after the ranking change. The effects are comparable to those reported in Table \ref{tab:cf_ranking_effects}. Importantly, differences between the rankings remain small, so the main conclusions about the trade-offs between rankings depend little on whether consumers update their beliefs. Also note that consumer welfare effects can be smaller when consumers update to the correct beliefs because the baseline comparison is different: with belief updating, the baseline ranking is the randomized ranking with consumers knowing that the ranking is randomized, whereas with fixed beliefs, the baseline ranking is the randomized ranking with consumers continuing to have the beliefs as estimated. 

\begin{table}[htb] \centering \footnotesize
\caption{Ranking Effects over Random Ranking} \label{tab:cf_ranking_effects_nobeliefadjust}

\begin{threeparttable}
\begin{tabular}{llcccc} 
\midrule

 && \multicolumn{1}{c}{ER} & \multicolumn{1}{c}{DVR} & \multicolumn{1}{c}{BUR} \\
\midrule
\bfseries{Platform} \\
\hspace{1em}Total revenues (\%)&&\hphantom{-}4.80&\hphantom{-}12.03&\hphantom{-}16.32\\
\hspace{1em}Number of transactions (\%)&&\hphantom{-}5.38&\hphantom{-}17.33&\hphantom{-}15.46\\
\hspace{1em}Avg. price of booking (\%)&&-0.55&\hphantom{0}-4.52&\hphantom{0}\hphantom{-}0.75\\
\midrule
\bfseries{Consumers} \\
\hspace{1em}Consumer welfare (\$, average)&&\hphantom{-}0.04&\hphantom{0}\hphantom{-}0.14&\hphantom{0}\hphantom{-}0.11\\
\hspace{1em}Consumer welfare (\$, cond. on booking)&&\hphantom{-}2.70&\hphantom{0}\hphantom{-}8.15&\hphantom{0}\hphantom{-}5.65\\
\hspace{1em}Discovery costs (\$,  cond. on booking)&&-0.02&\hphantom{0}-0.07&\hphantom{0}-0.06\\
\midrule 
\end{tabular}
\begin{tablenotes}
\item \footnotesize{\emph{Notes:} Effects of different rankings relative to a randomized ranking when consumers do not adjust their beliefs. ER: Expedia's own ranking. DVR: Discover-Value Ranking (heuristic to maximize consumer welfare). BUR: Bottom-Up Ranking (heuristic to maximize revenues). }
\end{tablenotes} 
\end{threeparttable}
\end{table}

\subsection{Consumer Welfare Maximization and the DVR} \label{subsec:consumer-welfare-maximization-dvr}

The Discovery-Value Ranking (DVR) balances the utility and the search value of an alternative to maximize consumer welfare (see Section \ref{subsec:The-six-rankings}). Proposition \ref{prop:consumer-welfare-maximizing} provides two cases in which the DVR is guaranteed to maximize consumer welfare.
\begin{proposition} \label{prop:consumer-welfare-maximizing}
  The Discovery-Value Ranking maximizes consumer welfare when consumers (i) know all alternatives' attributes and search costs at the beginning of search, or (ii) searching an alternative only reveals the shock but no other attributes.
\end{proposition}
\begin{proof}
    To prove this proposition, I use the following logic. Suppose a consumer could choose the order in which to discover products herself, rather than the order being determined by the ranking. In this case, the order that the consumer would optimally choose to discover products in by definition maximizes consumer welfare. Hence, to prove that the DVR maximizes consumer welfare, I show that the consumer's optimal policy would prescribe to discover products in the order of decreasing $r_j^d$, if that choice were available.

    The consumer's choice of the order in which to discover products can be formulated as a search and discovery problem with multiple discovery technologies. Specifically, each product $j$ is a discovery technology that only reveals the respective product on the list, and the consumer sequentially chooses which discovery technology to use next. As \cite{Greminger2021} notes, the optimal policy in such a setting is still characterized by the optimal discovery, search, and stopping rules presented in Section \ref{subsec:Optimal-policy-and}, with the extension that the optimal discovery rule now prescribes to choose the discovery technology with the largest discovery value $z_j^d$ next.

    For the case where the consumer knows the realizations of the product attributes ($x_j^l, x_j^d$) and search costs ($c_s$), the definition of the index $r_j^d$ means that it equals the discovery value $z_j^d$, so that the consumer would optimally choose to discover products in the order of decreasing $r_j^d$. Hence, the DVR maximizes consumer welfare in this case. 

    In the case where searching an alternative only reveals its shock and no vertical attributes ($x_j^l{}'\gamma = \tilde{\gamma}_j=x_j^d{}'\kappa=0$), the index $r_j^d$ sorts alternatives in order of first-order stochastic dominance of the effective value $\tilde{W}_j = x_j^l{}'\beta + \min\{\xi, \varepsilon_j\}$.\footnote{When there are vertical attributes revealed on the detail page, $r_j^d$ does not necessarily sort alternatives in order of first-order stochastic dominance of the effective value $\tilde{W}_j$.} Now note that the definition of the discovery value guarantees that when the distribution of some $\tilde{W}_j$ first-order stochastically dominates the distribution of some $\tilde{W}_k$, then $z_j^d \geq z_k^d$. Hence, the consumer would optimally choose to discover $j$ before $k$ and, as a result, that the consumer would optimally choose to discover products in the order of decreasing $r_j^d$ such that the DVR maximizes consumer welfare. Note that this rationale applies independently of whether consumers learn of the change in ranking or not because the optimal policy based on correct beliefs by definition cannot be improved upon, so that the ranking maximizing consumer welfare is the same in both cases.

\end{proof}

\subsection{Additional Ranking Algorithms} \label{subsec:additional-ranking-algorithms}

Table \ref{tab:list_algos} lists the ranking algorithms that I compare with the heuristic algorithms used for the main counterfactual analysis. The first three algorithms are brute-force algorithms that compare all possible rankings and select the best one according to the respective objective. To make these algorithms feasible, they are computed under the assumption that consumers do not update their beliefs about the ranking. However, they are still evaluated for the case when consumers update beliefs. 

\begin{table}[htb]  \footnotesize
\begin{centering}
\caption{Ranking Algorithms \label{tab:list_algos}}
\par\end{centering}
\centering{}\begin{tabular}{llll}  
\midrule

 & Name & Objective & Description \\
\midrule 
\bfseries{WBF} & Welfare Brute-Force & Welfare & Compares all possible rankings. \\
\bfseries{CBF} & Transactions Brute-Force & Transactions & Compares all possible rankings. \\
\bfseries{RBF} & Revenues Brute-Force & Revenues & Compares all possible rankings. \\
\midrule
\bfseries{DVR} & Discovery-Value Ranking & Welfare & Rank by $r_j^d$ defined in \eqref{eq:discovery-value-ranking}. \\
\bfseries{UBR} & Utility-Based Ranking & Welfare & Rank by $\mathbb{E}[U_j^l]=x_j^l{}'\beta$.  \\ 
\bfseries{WT3} & Welfare Top-3 Ranking & Welfare & OPT3G-CS algorithm of \cite{Compiani2023}. \\
\midrule
\bfseries{DVR} & Discovery-Value Ranking & Transactions & Rank by $r_j^d$ defined in \eqref{eq:discovery-value-ranking}. \\
\bfseries{CT3} & Transactions Top-3 Ranking & Transactions & Same as WT3, but for maximizing transactions. \\
\midrule
\bfseries{BUR}& Bottom-Up Ranking & Revenue & Start ranking from the bottom (see main text). \\ 
\bfseries{RT3}& Revenues Top-3 Ranking & Revenue & OPT3G-Rev algorithm of \cite{Compiani2023}. \\
\bfseries{PDR} & Price-Decreasing Ranking& Revenue & Decreasing order of $p_j$.  \\
\midrule 
\end{tabular}
\end{table}

\paragraph{Maximizing Consumer Welfare.} The next three algorithms are heuristic algorithms designed to maximize welfare. The DVR balances the utility and search costs of an alternative through the alternative-specific discovery value $r_j^d$ as described in Section \ref{subsec:The-six-rankings}. The WT3 is the OPT3G-CS heuristic algorithm for maximizing consumer welfare proposed by \cite{Compiani2023}. This algorithm works by first optimally ranking alternatives in the top three positions using a brute-force algorithm. Then, it fills the remaining positions iteratively by selecting the alternative that increases consumer welfare the most when added in the next position. As \cite{Compiani2023} show, this algorithm performs well in a Weitzman model when the number of positions is small or when consumers mostly search and choose alternatives just from the top of the ranking. Albeit without formal proof, the same logic likely applies in the SD model, so the algorithm provides a potentially good heuristic to maximize welfare in the SD model. Finally, the UBR ranks alternatives by their expected utility ($x_j^l{}'\beta + x_j^d{}'\kappa$). Proposition \ref{prop:consumer-welfare-maximizing} shows that when search costs are homogeneous and there are no attributes revealed on the list page, this ranking maximizes consumer welfare. Hence, depending on the estimated model, it provides another potentially good approximation of the consumer-welfare-maximizing ranking.

\paragraph{Maximizing Transactions.} The next two algorithms are heuristic algorithms designed to maximize the number of transactions. Maximizing transactions is similar to maximizing consumer welfare because consumers are more likely to search and buy alternatives that provide higher utility. Proposition \ref{prop:conversion-max-special} proves that when consumers do not reveal vertical attributes on the detail page and do not learn of the ranking change, the consumer-welfare-maximizing ranking also maximizes transactions. Hence, provided the DVR approximates the consumer-welfare-maximizing ranking well, it should also provide a good heuristic to maximize transactions.

\begin{proposition}
\label{prop:conversion-max-special}In the SD model without search over vertical attributes ($x_j^d{}'\kappa = 0$), the consumer-welfare-maximizing ranking also maximizes transactions when consumers do not update their beliefs about the ranking.
\end{proposition}
\begin{proof}
    The generalized eventual purchase theorem of \cite{Greminger2021} implies that the probability that a consumer purchases an alternative is given by $\mathbb{P}(U_0 \leq \max_{j \in J} W_j)$, where $W_j = \min\{z^d(h_j), \tilde{W}_j\}$. Hence, if the distribution of $\tilde{W}_j$ improves in higher positions in a first-order stochastic dominance sense, then this probability weakly increases, provided consumers do not update their beliefs so that the discovery values remain the same. When search costs are homogeneous and no attributes are revealed on the detail page, then the DVR orders alternatives based on first-order stochastic dominance of their effective values $\tilde{W}_j$, and by Proposition \ref{prop:consumer-welfare-maximizing}, the DVR also maximizes consumer welfare. Hence, the consumer-welfare-maximizing ranking also maximizes transactions in this case.
\end{proof}

The CT3 is another heuristic to maximize transactions. It works in the same way as the WT3, but targets transactions rather than consumer welfare. 

\paragraph{Maximizing Platform Revenues.} The final three ranking algorithms are heuristic algorithms designed to maximize platform revenues. The BUR starts ranking from the bottom position and is described in detail in Section \ref{subsec:The-six-rankings}. The RT3 is the same as the OPT3G-Rev heuristic algorithm for maximizing revenues proposed by \cite{Compiani2023}. It works with the same steps as WT3, but targets revenues rather than consumer welfare. Finally, the PDR ranks alternatives in decreasing order of their prices. This heuristic is motivated by the price effect highlighted in Section \ref{subsec:Implications-Descr-Evidence}. As Proposition \ref{prop:price-decreasing} below shows, the PDR maximizes revenues when position effects are homogeneous. Hence, how well the PDR performs relative to other algorithms provides a measure of how important heterogeneous position effects are for revenue maximization.

\begin{proposition}
\label{prop:price-decreasing}The Price-Decreasing Ranking (PDR) maximizes revenues when position effects are homogeneous.
\end{proposition}
\begin{proof}
    The proposition follows from the fact that position effects are homogeneous in the SD model only when alternatives are homogeneous in the expected list utilities ($u_j^e = x_j^l{}'\beta + x_j^d{}'\kappa$). Hence, if an alternative has a higher price, other attributes in $x_j^l$ have to adjust so that $x_j^l{}'\beta$ remains unchanged. As $x_j^l{}'\beta$ and $u_j^e$ are the same for all alternatives, switching two alternatives will only shift demand from the alternative moved down to the one moved up, but does not affect the demand for other alternatives. Hence, whenever a higher-priced alternative can be moved up, moving it up will shift demand to that alternative so that revenues increase. This implies that any ranking that does not order in decreasing order of price cannot be optimal when position effects are homogeneous, which proves the proposition because an optimal ranking always exists with a finite number of alternatives.
\end{proof}

\subsection{Ranking Comparison with Reduced Set of Alternatives} \label{subsec:comparison-reduced-set-alternatives}

To compare the rankings obtained from the different heuristics to the optimal one, I randomly sample 1,000 consumers and reduce the number of alternatives by randomly selecting five for each consumer. For each of the ranking algorithms, I simulate 800,000 search paths and compute the average consumer welfare, revenues, and transactions. I consider two settings: one where consumers update their beliefs following the change in ranking and one where they keep the beliefs as estimated. For the former, I use the procedure described in Online Appendix \ref{subsec:belief-updating} to obtain consumers' beliefs under the new ranking.

Table \ref{tab:heuristic-comparison} reports the results. Relative to the consumer-welfare-maximizing ranking (WBF), the proposed DVR heuristic performs best in both scenarios, outperforming even the WT3, which ranks the first three positions using brute-force. The DVR also performs substantially better than the utility-based ranking (UBR). This result is in line with \cite{Compiani2023}, who showed that the UBR is not consumer-welfare-maximizing when consumers search over vertical attributes. Overall consumer welfare differences are small because there are only five alternatives. 

Among the heuristic algorithms, the DVR also performs best in terms of transactions, lowering transactions by less than 1\% relative to the conversion-maximizing ranking (CBF). Hence, I conclude that the DVR provides a good heuristic to maximize consumer welfare and transactions. Nonetheless, comparing the WBF to the CBF shows that the two objectives are not perfectly aligned, and that the DVR seems to balance the two objectives.

For revenues, the proposed BUR heuristic performs best among the heuristics. Depending on whether consumers adjust beliefs or not, the BUR reduces revenues only by about 0.62\% and 0.59\% relative to the revenue-maximizing ranking (RBF). The PDR performs by far the worst (-13.09\%) because it ignores the underlying demand effects, highlighting the importance of heterogeneous position effects when designing rankings to maximize revenues.

\subsection{Ranking Comparison with All Alternatives} \label{subsec:comparison-all-alternatives}

Computing the optimal ranking with a brute-force algorithm is not feasible beyond a few alternatives. However, I can still compare the different heuristic algorithms with each other following the same procedure as for the main counterfactual analysis. 

Table \ref{tab:heuristic-comparison-allalternatives} reports the effects of the different rankings relative to Expedia's own ranking. Throughout, the results are in line with the five-alternative comparison. Importantly, the proposed heuristics DVR and BUR continue to outperform all other heuristics for all objectives, confirming they are well-suited to evaluate the trade-offs between ranking objectives.

\vspace{3em} 

\begin{table}[htb] \centering \footnotesize
\caption{Comparison of Rankings (5 alternatives)} \label{tab:heuristic-comparison}

\begin{threeparttable}
\begin{tabular}{llccccccc} 
\midrule

 && \multicolumn{3}{c}{Fixed Beliefs}& \multicolumn{3}{c}{Updated Beliefs}\\
\cmidrule(rl){3-5} \cmidrule(rl){6-8} 
&& \multicolumn{1}{c}{$\Delta$ Wel. (\$)}& \multicolumn{1}{c}{$\Delta$ Trans. (\%)}& \multicolumn{1}{c}{$\Delta$ Rev. (\%)}& \multicolumn{1}{c}{$\Delta$ Wel. (\$)}& \multicolumn{1}{c}{$\Delta$ Trans. (\%)}& \multicolumn{1}{c}{$\Delta$ Rev. (\%)}\\
\midrule
WBF&&\hphantom{-}0.0000&\hphantom{0}-0.2950&\hphantom{0}-3.1847&\hphantom{-}0.0000&-0.2617&-3.2511\\
CBF&&-0.0005&\hphantom{0}\hphantom{-}0.0000&\hphantom{0}-2.4705&-0.0005&\hphantom{-}0.0000&-2.5661\\
RBF&&-0.0061&\hphantom{0}-2.4806&\hphantom{0}\hphantom{-}0.0000&-0.0062&-2.4253&\hphantom{-}0.0000\\
\midrule
DVR&&-0.0016&\hphantom{0}-0.8100&\hphantom{0}-3.1152&-0.0014&-0.8036&-3.2200\\
UBR&&-0.0106&\hphantom{0}-4.9227&\hphantom{0}-6.7839&-0.0114&-4.3390&-6.2589\\
WT3&&-0.0072&\hphantom{0}-3.8738&\hphantom{0}-6.3346&-0.0071&-3.7126&-6.2726\\
CT3&&-0.0077&\hphantom{0}-3.9172&\hphantom{0}-6.1223&-0.0076&-3.7624&-6.0659\\
\midrule
BUR&&-0.0068&\hphantom{0}-3.2485&\hphantom{0}-0.5852&-0.0067&-3.2370&-0.6230\\
RT3&&-0.0137&\hphantom{0}-6.5145&\hphantom{0}-4.0089&-0.0138&-6.3343&-3.8820\\
PDR&&-0.0407&-20.6330&-13.0901& & & \\
\midrule 
\end{tabular}
\begin{tablenotes}
\item \footnotesize{\emph{Notes:} Effects of different rankings for a reduced set of five alternatives. The effects are all relative to the brute-force algorithm for the respective objective. Table \ref{tab:list_algos} describes the different ranking algorithms. For the PDR, beliefs would imply that the discovery value increases across positions, so that it is not possible to compute the effects under the updated beliefs. }
\end{tablenotes} 
\end{threeparttable}
\end{table}

\begin{table}[htb] \centering \footnotesize
\caption{Comparison of Rankings (all alternatives)} \label{tab:heuristic-comparison-allalternatives}

\begin{threeparttable}
\begin{tabular}{llccccccc} 
\midrule

 && \multicolumn{3}{c}{Fixed Beliefs}& \multicolumn{3}{c}{Updated Beliefs}\\
\cmidrule(rl){3-5} \cmidrule(rl){6-8} 
&& \multicolumn{1}{c}{$\Delta$ Wel. (\$)}& \multicolumn{1}{c}{$\Delta$ Trans. (\%)}& \multicolumn{1}{c}{$\Delta$ Rev. (\%)}& \multicolumn{1}{c}{$\Delta$ Wel. (\$)}& \multicolumn{1}{c}{$\Delta$ Trans. (\%)}& \multicolumn{1}{c}{$\Delta$ Rev. (\%)}\\
\midrule
DVR&&\hphantom{-}0.1385&\hphantom{-}17.3350&12.0297&0.1116&12.0933&\hphantom{0}5.8780\\
UBR&&\hphantom{-}0.1010&\hphantom{-}13.2085&\hphantom{0}8.4352&0.0648&\hphantom{0}7.5064&\hphantom{0}3.3675\\
WT3&&\hphantom{-}0.0804&\hphantom{0}\hphantom{-}9.7571&\hphantom{0}6.7152&0.0510&\hphantom{0}5.2999&\hphantom{0}2.7185\\
CT3&&\hphantom{-}0.0792&\hphantom{0}\hphantom{-}9.9339&\hphantom{0}6.9932&0.0499&\hphantom{0}5.4531&\hphantom{0}2.9554\\
\midrule
BUR&&\hphantom{-}0.1143&\hphantom{-}15.4569&16.3182&0.0825&\hphantom{0}9.7802&10.5441\\
RT3&&\hphantom{-}0.0627&\hphantom{0}\hphantom{-}8.3979&\hphantom{0}8.2462&0.0354&\hphantom{0}4.2130&\hphantom{0}4.1640\\
PDR&&-0.0553&\hphantom{0}-6.8478&\hphantom{0}0.7675& & & \\
\midrule 
\end{tabular}
\begin{tablenotes}
\item \footnotesize{\emph{Notes:} Effects of different rankings relative to a randomized ranking. Table \ref{tab:list_algos} describes the different ranking algorithms. }
\end{tablenotes} 
\end{threeparttable}
\end{table}

\clearpage

\newpage

\section{Parameter Identification \label{sec:appendix_identification}}

This appendix formally shows that data without information on how far consumers scrolled is sufficient to identify the model parameters. A main concern is the separate identification of the search and discovery parameters. For example, an increase in either the search or the discovery costs will induce consumers to end discovery sooner and search fewer products. Throughout, I assume that the ranking is randomized, ensuring that the discovery parameters can be separately identified from the preference parameters. 

With data on how far consumers scrolled, separate identification follows from moments that condition on the discovered alternatives. For example, larger search costs imply that consumers are less likely to search the alternatives they discovered on the list, whereas larger discovery costs only change the probability where consumers stop scrolling. 

Without such data, these identification arguments no longer apply directly because it is unclear how the stopping probabilities and conditional moments could be recovered. However, I now provide two propositions that show how these moments can be recovered in data that does not track how far consumers scrolled.  As a result, the same identification arguments that rely on conditional moments still apply even without data on how far consumers scrolled. 

The first proposition establishes that the stopping probabilities are identified by the position-specific click probabilities. The second proposition shows that the relevant conditional moments are uniquely determined by the stopping probabilities and the unconditional moments. Hence, these moments are effectively present in the data, even without information on how many alternatives consumers discovered.\footnote{These results could also be used to develop a simulated methods of moments estimation approach even without having to observe the discovery process. } As a result, the same identification arguments still apply, even in the absence of this information. 

\subsection{Identification of Stopping Probabilities \label{subsec:proof-identification-stopping-prob}}

\begin{proposition} \label{prop:identification-stopping-prob}
Suppose there are $N$ alternatives randomly shown in $N$ positions, and let $q_h = \mathbb{P}(\text{Click in position }h)$ denote the probability of clicking in position $h$ across consumers. The stopping probability in the search and discovery model then is given by
  \begin{equation}
    \mathbb{P}(\text{Stop discovery on position }h) = m(q_1, \dots , q_N) \ ,\label{eq:identification-stopping-prob}
  \end{equation}
  where the function $m()$ has a known form.
\end{proposition}
\begin{proof} 
I first derive two conditions required for the stopping probabilities to be identified by the data. Intuitively, the two conditions guarantee that only the stopping probabilities produce position effects and that there are position effects whenever consumers do not discover all alternatives. After establishing the two conditions, I show that they apply in the model. 

To simplify exposition, I introduce the following notation: $q_h=\mathbb{P}(\text{Click in position }h)$; $\Delta q_h=q_h-q_{h-1}$; $D_h$ denotes the event of a consumer ending discovery on position $h$; $R_h$ denotes the event of a consumer discovering at least position $h$; $R_h^c$ denotes the complement event of a consumer stopping before reaching position $h$; and $S_{jh}$ denotes the event of a consumer searching alternative $j$ in position $h$.

With $N$ alternatives randomly shown across $N$ positions, the expected difference in clicks across two positions is the average across the click-probabilities for each alternative when it is shown in the respective positions. Formally, this is given by
\begin{equation}
\Delta q_h=\frac{1}{N}\left[\sum_{j}\mathbb{P}(S_{jh})-\mathbb{P}(S_{jh+1})\right],
\end{equation}
which is an average across the $N$ possible products that could be shown in the respective positions.

Consumers cannot search products they did not discover such that $\mathbb{P}(S_{jh}|R_h^c) = 0$. Hence, conditioning on reaching position $h$ and differentiating the two cases of stopping on position $h$ or reaching $h+1$, the expected difference in clicks can be written as
\begin{multline}
N\Delta q_h=\mathbb{P}(R_h) \left[ \mathbb{P}(D_h|R_h)[\sum_{j}\mathbb{P}(S_{jh}|D_h)] \right. \\
+ \left. \mathbb{P}(R_{h+1}|R_h)[\sum_{j}\mathbb{P}(S_{jh}|R_{h+1})-\sum_{j}\mathbb{P}(S_{jh+1}|R_{h+1})] \right] .\label{eq:diff2-conditional}
\end{multline}
Now suppose that the following two conditions hold for all positions $h$:
\begin{enumerate}
\item $\Delta q_h=q_h\iff\mathbb{P}(D_h|R_h)=1$: whenever there are no clicks in the next position ($q_{h+1}=0$), consumers who reach position $h$ never discover $h+1$.
\item $\Delta q_h=0\iff\mathbb{P}(D_h|R_h)=0$: whenever there is no difference in clicks across the two positions, consumers who reach position $h$ always continue to discover $h+1$.
\end{enumerate}

Applying the first condition and setting $\Delta q_h = q_h$ and  $\mathbb{P}(D_h|R_h)=1$  in \eqref{eq:diff2-conditional} yields $\sum_{j}\mathbb{P}(S_{jh}|D_h)=q_h$. This is because $\mathbb{P}(D_h|R_h)=1$ implies that $\mathbb{P}(R_{h+1}|R_h)=1-\mathbb{P}(D_h|R_h) = 0$. Similarly, applying the second condition and setting $\Delta q_h = 0$ and $\mathbb{P}(D_h|R_h) = 0$ in \eqref{eq:diff2-conditional} yields $\sum_{j}\mathbb{P}(S_{jh}|R_{h+1})-\sum_{j}\mathbb{P}(S_{jh+1}|R_{h+1})=0$. Substituting these two expressions in \eqref{eq:diff2-conditional} then yields
\begin{align}
N\Delta q_h & =\mathbb{P}(D_h|R_h)Nq_h=\mathbb{P}(D_h|R_h)N\frac{1}{N}\left[\sum_{j}\mathbb{P}(S_{jh}|D_h)\right]\\
\Rightarrow & \mathbb{P}(D_h|R_h)=\frac{N\Delta q_h}{\sum_{j}q_j}
\end{align}

This relation shows that the probability of stopping on a particular position $h$ conditional on reaching it is nonparametrically identified as long as the position-specific click probabilities can be estimated. The unconditional stopping probabilities then are given by the recursive relation
\begin{align}
  \mathbb{P}(D_h) & = \mathbb{P}(D_h|R_h)\mathbb{P}(R_h) + 0 \nonumber \\
  & = \frac{N\Delta q_h}{\sum_{j}q_j}\left(1-\sum_{q=1}^{h-1}\mathbb{P}(D_q)\right) \ , \label{eq:recursive-stopping}
\end{align}
where I use that the probability of consumers stopping on position $h$ conditional on not having reached is always zero. This recursive relation characterizes the unconditional stopping probabilities for all positions $h$ starting from $h=1$. At $h = 1$, $\mathbb{P}(D_0) = 0$ by the assumption that the first discovery is free. Given $\mathbb{P}(D_0)$, \eqref{eq:recursive-stopping} yields  $\mathbb{P}(D_1)$ from the conditional stopping probability. Given $\mathbb{P}(D_1)$, \eqref{eq:recursive-stopping} yields $\mathbb{P}(D_2)$ from the conditional stopping probability, and so on. Hence, once the conditional stopping probabilities are identified, so are the unconditional stopping probabilities. 

It remains to show that the two conditions hold in the search and discovery model. The first condition holds as long as consumers cannot search an alternative without discovering it, which is the case in the search and discovery model. This guarantees that $q_{h+1}=0$ whenever $h+1$ is not discovered. The random ranking then guarantees that there is a non-zero probability of clicking on an alternative in position $h+1$ when that position is revealed, which proves that $\Delta q_h=0$ if and only if $\mathbb{P}(D_h|R_h)=0$.

The second condition holds because the search and discovery model guarantees that an alternative is searched when shown in some position whenever it will also be searched when shown in the previous position. This follows immediately from the two search implications that provide the conditions under which an alternative is searched, given an effective value for the chosen option. Given that $w_j\leq\tilde{w}_j$, the requirement for an alternative to be searched in earlier positions is weaker than the requirement for it to be searched in later positions. As a result, the second condition also holds in the search and discovery model. 
\end{proof}

\subsection{Identification of Conditional Moments \label{subsec:proof-identification-conditional-moments}}

\begin{proposition} \label{prop:identification-conditional-moments}
Moments on clicks and purchases conditional on discovery are uniquely determined by the stopping probabilities and the unconditional moments.
\end{proposition}
\begin{proof} 
Let $M$ be an average quantity of interest determined by either clicks or purchases, and let $M_h$ be the respective position-specific counterparts. For example, $M$ could be the number of clicks consumers make on average and $M_h$ the clicks consumers make on average in position $h$. The unconditional moments then are $\mathbb{E}[M] = \sum_h \mathbb{E}[M_h]$ and the conditional moments are $\mathbb{E}[M_h|D_h]$, where $D_h$ again denotes the event of ending discovery on position $h$.  Throughout, I assume that sufficient data is available so that $\mathbb{E}[M]$, $\mathbb{E}[M_h]$ and $\mathbb{P}(D_h)$ (through Proposition \ref{prop:identification-stopping-prob}) are precisely estimated and, hence, can be treated as known. The conditional moments $\mathbb{E}[M_h|D_h]$ then are identified if  $\mathbb{E}[M_h|D_h]$ is uniquely determined by these known moments for all positions $h$. 

I now show that this is indeed the case through a recursive relation. The fact that consumers cannot click or choose alternatives from positions they have not discovered implies that $\mathbb{E}[M_h | D_k]=0$ for all $k < h$ . As a result, the law of total probability implies the following:
\begin{align} \label{eq:recursive-relation}
     \mathbb{E}[M_h] = \sum_{k\geq h}\mathbb{E}[M_{h} | D_k]\mathbb{P}(D_k)  \ .
\end{align}
In the last available position, denoted by $\tilde{h}$, this relation simplifies to $\mathbb{E}[M_{\tilde{h}}] = \mathbb{E}[M_{\tilde{h}} | D_{\tilde{h}}]\mathbb{P}(D_{\tilde{h}})$. Hence, $ \mathbb{E}[M_{\tilde{h}} | D_{\tilde{h}}]$ is fully determined by $ \mathbb{E}[M_{\tilde{h}}]$ and $\mathbb{P}(D_{\tilde{h}})$. Given $ \mathbb{E}[M_{\tilde{h}} | D_{\tilde{h}}]$,  $ \mathbb{E}[M_{\tilde{h}-1} | D_{\tilde{h}-1}]$ then is also determined through \eqref{eq:recursive-relation}, which then allows determining the conditional moment for $\tilde{h}-2$, and so on. Hence,  the recursive relation \eqref{eq:recursive-relation} uniquely determines all conditional moments from the stopping probabilities and the unconditional moments. 
\end{proof}

\newpage

\section{Estimation Details \label{sec:estimation-appendix}}

This appendix provides details on the estimation approach. First, it shows how to compute the probability of searching an alternative without choosing it. Second, it provides the partitioning of the probability space that guarantees a smooth simulated likelihood function. Finally, it reports results from a Monte Carlo simulation study.
  
\subsection{Probability of Search and Non-Purchase \label{subsec:joint-probability-search-non-purchase}}

The probability of searching an alternative $k$ without purchasing it enters the likelihood as $\mathbb{P}(Z_k^s\geq \tilde{w}_j \cap U_k\leq \tilde{w}_j)$, where $\tilde{w}_j$ is the effective value of the eventually chosen option $j$ and is not random. By expressing it as the CDF of a standard normal distribution and the CDF of a bivariate normal distribution, Proposition \ref{prop:calculating_probability} provides a way to calculate this probability using standard numerical methods, avoiding computationally costly numerical integration.\footnote{To calculate the bivariate normal CDF, I use the method by \citet{Drezner1990} as implemented in the Julia StatsFuns.jl package.}

\begin{proposition}
  \label{prop:calculating_probability} If the two shocks are independent and $\nu_j \sim N(0, 1)$ and $\varepsilon_j \sim N(0, \sigma^2_\varepsilon)$, then
  \begin{equation}
  \mathbb{P}(Z_j^s\geq q\cap U_j\leq q)=\mathbb{P}\left(Y\leq\frac{q-u_j^e}{\tilde{\sigma}}\right)-\mathbb{P}\left(Y_1\leq\frac{q-u_j^e}{\tilde{\sigma}}\cap Y_2\leq q-z_j^e\right)\  \label{eq:probability-cdf-bvn}
  \end{equation}
for some constant $q$, where $u_j^e = x_j^l{}'\beta + x_j^d{}'\kappa$, $z_j^e =x_j^l{}'\beta + x_j^l{}'\gamma+ \tilde{\gamma}_j + \xi$,  $Y$ follows a standard normal distribution, $[Y_1,Y_2]$ follows a bivariate normal distribution with correlation $-\frac{1}{\tilde{\sigma}\sigma_\varepsilon}$, and $\tilde{\sigma}=\sqrt{1+\frac{1}{\sigma_\varepsilon^2}}.$
\end{proposition}

\begin{proof}

Under the assumption that the shocks are normally distributed and independent, $\mathbb{P}(Z_j\geq q\cap U_j\leq q)$ can be written as
\begin{align}
\mathbb{P}(Z_j^s\geq q\cap U_j\leq q) & =\int_{q-z_j^e}\Phi\left(\frac{q-u_j^e-\nu}{\sigma_\varepsilon}\right)\phi(\nu)\text{d}\nu\label{eq:appendix_zwu}\\
& =\int_{-\infty}^{\infty}\Phi\left(\frac{q-u_j^e-\nu}{\sigma_\varepsilon}\right)\phi(\nu)\text{d}\nu- \notag \\
& \hspace{3em} \int^{q-z_j^e}_{-\infty}\Phi\left(\frac{q-u_j^e-\nu}{\sigma_\varepsilon}\right)\phi(\nu)\text{d}\nu\ ,
\end{align}
where $u_j^e$ and $z_j^e$ are defined in the proposition. Using results 10,010.8 and 10,010.1 for normal integrals from \citet{Owen1980}, the expression follows.

\end{proof}

\subsection{Partitioning the Probability Space \label{subsec:Smooth-Monte-Carlo}}

I calculate the likelihood contribution in \eqref{eq:likelihood} as
\begin{align}
  \mathbb{P}(\text{observed choices of } i) = \sum_{R_k \in R} \mathbb{P}(\text{observed choices of }i| R_k) \mathbb{P}(R_k) \ ,\label{eq:individual-likelihood-contribution-conditional}
\end{align}
where $R = \bigcup_{k} R_k$ is a partition of the probability space of $(\eta, \nu_j, \varepsilon_j)$, the shocks for the chosen option.  The partition is defined so that two conditions hold for any region $R_k$: (i) the number of alternatives the consumer discovered remains fixed, and (ii) $\varepsilon_j$ is either above or below the threshold $\xi_j = x_j^l{}'\beta + \tilde{\gamma}_j + \xi$.

Recall that the set of alternatives a consumer discovers, $A(\tilde{w}_j)$, is determined by the effective value of the chosen (inside or outside) option, $\tilde{w}_j$. Hence, the number of alternatives the consumer discovered can be fixed by constructing the following intervals:
\begin{equation}
B = (-\infty,z^d(|J|)]\cup(z^d(|J|),z^d(|J|-1)]\cup,\dots,\cup(z^d(1),\infty] \ .\label{eq:definition-intervals}
\end{equation}
Using these intervals and additionally partitioning on $\varepsilon_j > \xi_j$, I calculate the individual likelihood contributions as
\begin{multline}
\mathcal{L}^i(\theta)=\sum_{b \in B}\mathbb{P}(\tilde{W}_j\in b \cap\varepsilon_j\leq\xi_j)\mathbb{E}[p(\tilde{W}_j)|\tilde{W}_j\in b\cap\varepsilon_j\leq\xi_j]+\\
\sum_{b\in B}\mathbb{P}(\tilde{W}_j\in b\cap\varepsilon_j>\xi_j)\mathbb{E}[p(\tilde{W}_j)|\tilde{W}_j\in b\cap\varepsilon_j>\xi_j]\ ,\label{eq:likelihood-partitioned}
\end{multline}
where $p(\tilde{w}_j)$ denotes the product of the integral in \eqref{eq:individual-likelihood-contribution}. As it has no closed-form solution, I use Monte Carlo integration to compute $\mathbb{E}[p(\tilde{W}_j)|\tilde{W}_j\in b\cap\varepsilon_j\leq\xi_j]$, i.e., I take draws $r = 1, \dots, N_l$ from distributions truncated so that the draws satisfy the respective conditions. This produces a smooth likelihood because all other expressions in the likelihood are smooth functions.

If the outside option is chosen, $\tilde{w}_j=\beta_0+\eta$ so taking draws for $\tilde{w}_j \in b$ is the same as taking truncated draws $\beta_0 + \eta \in b$. If instead an alternative $j>0$ is chosen, the expression further depends on the condition $\varepsilon_j\leq\xi_j$, and I take draws for $\tilde{w}_j \in b$ as follows:
\begin{enumerate}
\item Take a draw $\varepsilon_j^r$ from its distribution truncated on $\varepsilon_j\leq\xi_j$.
\item Take a draw $\nu_j^r$ from its distribution truncated on $\tilde{w}_j(\varepsilon_j^r) = x_j^l{}'\beta + \nu_j + x_j^d{}'\kappa + \delta_j + \varepsilon_j^r \in b$.
\item Calculate $\tilde{w}_j^r$ using draws $\nu_j^r$ and $\varepsilon_j^r$ and compute the inner probability $p(\tilde{w}_j^r)$.
\end{enumerate}
Based on draws generated by this procedure, the expression can be calculated as the weighted average
\begin{equation}
\mathbb{P}(\varepsilon_j\leq\xi_j)\sum_{r}\mathbb{P}(\tilde{W}_j(\varepsilon_j^r)\in b)p(\tilde{w}_j^r)\ .
\end{equation}
I calculate $\mathbb{P}(\tilde{W}_j\in b\cap\varepsilon_j>\xi_j)\mathbb{E}[p(\tilde{W}_j)|\tilde{W}_j\in b\cap\varepsilon_j>\xi_j]$ using the same steps with the truncation on $\varepsilon_j>\xi_j$ in step 1 and the corresponding change of $\tilde{w}_j(\varepsilon_j^r)$ in step 2.

\clearpage

\subsection{Monte Carlo Simulation Study \label{subsec:Monte-Carlo-Simulation}}

I run a Monte Carlo simulation study to verify whether the estimation procedure recovers the parameters with the present data. To this end, I first simulate data for parameter values close to those obtained from the estimation. For the observable product characteristics and the ranking, I use the hotel characteristics from the estimation sample with the randomized ranking. This setup leads to simulated data comparable to the observed data in terms of variation in observable characteristics and consumers' choices. To reduce the computational burden, I randomly sample 5,000 search sessions rather than using the entire estimation sample.

Table \ref{tab:simulation} reports the results. The estimation procedure recovers the parameters well. Importantly, the search and discovery costs are also recovered well, confirming that the proposed approach to back these out after the estimation to save on computation time works as intended.  

\begin{table}[htb] \centering \footnotesize
\caption{Monte Carlo Simulation Results} \label{tab:simulation}

\begin{threeparttable}
\begin{tabular}{llcccc} 
\midrule

 && \multicolumn{1}{c}{True} & \multicolumn{1}{c}{Estimate} & \multicolumn{1}{c}{Std. Error} \\
\midrule
$\beta$: Price (in \$100)&&\hphantom{00}-0.350&\hphantom{00}-0.402&(0.044)\\
$\beta$: Star rating&&\hphantom{00}\hphantom{-}0.150&\hphantom{00}\hphantom{-}0.169&(0.045)\\
$\beta$: Review score&&\hphantom{00}\hphantom{-}0.150&\hphantom{00}\hphantom{-}0.179&(0.062)\\
$\beta$: No reviews&&\hphantom{00}\hphantom{-}0.250&\hphantom{00}\hphantom{-}0.498&(0.299)\\
$\beta$: Chain&&\hphantom{00}\hphantom{-}0.000&\hphantom{00}-0.086&(0.059)\\
$\beta$: On promotion&&\hphantom{00}\hphantom{-}0.200&\hphantom{00}\hphantom{-}0.287&(0.059)\\
$\beta$: Outside option&&\hphantom{00}\hphantom{-}5.100&\hphantom{00}\hphantom{-}5.189&(0.219)\\
$\kappa$: Location score&&\hphantom{00}\hphantom{-}0.100&\hphantom{00}\hphantom{-}0.117&(0.021)\\
$\gamma$: Price (in \$100)&&\hphantom{00}\hphantom{-}0.150&\hphantom{00}\hphantom{-}0.172&(0.044)\\
$\gamma$: Star rating&&\hphantom{00}\hphantom{-}0.050&\hphantom{00}\hphantom{-}0.045&(0.044)\\
$\gamma$: Review score&&\hphantom{00}-0.150&\hphantom{00}-0.174&(0.061)\\
$\gamma$: No reviews&&\hphantom{00}-0.300&\hphantom{00}-0.555&(0.297)\\
$\gamma$: Chain&&\hphantom{00}-0.100&\hphantom{00}-0.023&(0.058)\\
$\gamma$: On promotion&&\hphantom{00}-0.050&\hphantom{00}-0.125&(0.058)\\
$\xi$&&\hphantom{00}\hphantom{-}2.600&\hphantom{00}\hphantom{-}2.694&(0.217)\\
$\tilde{z}^d$&&\hphantom{-}565.000&\hphantom{00}\hphantom{-}576.897&(0.224)\\
$\rho$&&\hphantom{00}-0.100&\hphantom{00}\hphantom{-}0.115&(0.013)\\
\midrule
$ c_d \times 100 $&&\hphantom{00}\hphantom{-}0.007&\hphantom{00}\hphantom{-}0.007& \\
$ c_s $&&\hphantom{00}\hphantom{-}0.002&\hphantom{00}\hphantom{-}0.001& \\
\midrule
Log likelihood&& &-22748.501& \\
N consumers&& &\hphantom{0}\hphantom{-}5,000& \\
\midrule
No. clicks (average)&& &\hphantom{00}\hphantom{-}1.160& \\
No. bookings (average)&& &\hphantom{00}\hphantom{-}0.072& \\
\midrule 
\end{tabular}
\begin{tablenotes}
\item \footnotesize{\emph{Notes:}  Parameter estimates obtained using 100 simulation draws per region of the partitioned probability space. Asymptotic standard errors are shown in parentheses. Search and discovery costs are computed using 10M simulation draws.  Statistical significance is indicated by * p < 0.1, ** p < 0.05, *** p < 0.01.  }
\end{tablenotes} 
\end{threeparttable}
\end{table}

\section{Comparison of Specifications\label{sec:Comparison-of-specifications}}

This appendix compares different specifications of the search and discovery model to determine how they affect the results. To limit the computational burden, I estimate all specifications without fixed effects $\delta_j$ and $\tilde{\gamma}_j$.\footnote{As the fixed effects capture alternative-specific differences, omitting them does not affect how different specifications compare to each other in terms of relative fit and counterfactual effects.}

\subsection{Distributions of Idiosyncratic Shocks}

I evaluate the sensitivity of the results to using different values for $\sigma_\varepsilon$ (variance of $\varepsilon$) and the upper bound of the distribution of the outside option shock $\eta_i$. To this end, I estimate the model under different values for these parameters using the same sample as in the main analysis, simulate how many and where clicks and bookings occur, and predict the revenue and consumer welfare effects of implementing either the Discovery-Value Ranking (DVR) or the Bottom-Up Ranking (BUR) over Expedia's own ranking (ER). By comparing effects to the ER rather than a randomized ranking, I save on computation time, while the results remain qualitatively the same. Other than the different baseline ranking, I use the same sample and procedures as for the main counterfactual analysis.

The results of this comparison are presented in Table \ref{tab:comparison_specs}. They reveal that parameter estimates tend to scale with the variance of $\varepsilon_j$. This includes the search and discovery cost estimates, which increase as $\sigma_\varepsilon$ increases. Notably, however, the differences $\tilde{z}^d - \beta_0$ and $\xi - \beta_0$ do not change much, suggesting that the net benefits of the different search actions over taking the outside option change little across specifications. Similarly, search values $z_j^s=x_j^l{}'(\beta + \gamma) + \tilde{\gamma}_j + \xi$ change little with $\sigma_\varepsilon$ because both $\beta$ and $\gamma$ scale similarly.

Whereas the revenue and consumer welfare effects change with $\sigma_\varepsilon$, the qualitative results remain the same across all values. Both the DVR and the BUR increase revenues and consumer welfare relative to the ER in all specifications, with the DVR generating larger consumer welfare effects and the BUR generating larger revenue effects. Notably, the differences between the rankings remain quite stable, and the suggested trade-offs between ranking objectives become even smaller compared to the main specification. Hence, the qualitative results are not very sensitive to the assumed variance of the idiosyncratic shocks.

The last columns of Table \ref{tab:comparison_specs} provide results for specifications where the upper bound ($b$) of the distribution of the outside option shock $\eta$ differs from the baseline. The results from these specifications show that the search and discovery cost estimates change very little with $b$. Though the sizes of the effects of the different rankings depend on $b$, the results also show that both rankings continue to increase both consumer welfare and platform revenues, suggesting that the qualitative results are not very sensitive to this upper bound.

\begin{table}[htb] \centering \footnotesize
\caption{Comparison of Specifications} \label{tab:comparison_specs}
\resizebox{\columnwidth}{!}{%
\begin{threeparttable}
\begin{tabular}{llcccccc} 
\midrule

 && \multicolumn{1}{c}{Main} & \multicolumn{1}{c}{$\sigma_\varepsilon = 5$} & \multicolumn{1}{c}{$\sigma_\varepsilon = 10$} & \multicolumn{1}{c}{$b = 0.5$} & \multicolumn{1}{c}{$b = 5$} \\
\midrule
\bfseries{Parameter estimates} \\
\hspace{1em}$\beta$: Price (in \$100)&&\hphantom{00}-0.362&\hphantom{00}-1.005&\hphantom{00}-2.018&\hphantom{00}-0.364&\hphantom{00}-0.361\\
\hspace{1em}$\beta$: Star rating&&\hphantom{00}\hphantom{-}0.168&\hphantom{00}\hphantom{-}0.089&\hphantom{00}\hphantom{-}0.071&\hphantom{00}\hphantom{-}0.170&\hphantom{00}\hphantom{-}0.167\\
\hspace{1em}$\beta$: Review score&&\hphantom{00}\hphantom{-}0.135&\hphantom{00}\hphantom{-}0.671&\hphantom{00}\hphantom{-}1.306&\hphantom{00}\hphantom{-}0.145&\hphantom{00}\hphantom{-}0.130\\
\hspace{1em}$\beta$: No reviews&&\hphantom{00}\hphantom{-}0.196&\hphantom{00}\hphantom{-}1.201&\hphantom{00}\hphantom{-}2.439&\hphantom{00}\hphantom{-}0.226&\hphantom{00}\hphantom{-}0.178\\
\hspace{1em}$\beta$: Chain&&\hphantom{00}\hphantom{-}0.031&\hphantom{00}\hphantom{-}0.320&\hphantom{00}\hphantom{-}0.823&\hphantom{00}\hphantom{-}0.025&\hphantom{00}\hphantom{-}0.035\\
\hspace{1em}$\beta$: On promotion&&\hphantom{00}\hphantom{-}0.187&\hphantom{00}\hphantom{-}0.582&\hphantom{00}\hphantom{-}0.709&\hphantom{00}\hphantom{-}0.191&\hphantom{00}\hphantom{-}0.184\\
\hspace{1em}$\beta$: Outside option&&\hphantom{00}\hphantom{-}5.128&\hphantom{0}\hphantom{-}13.711&\hphantom{0}\hphantom{-}24.978&\hphantom{00}\hphantom{-}5.168&\hphantom{00}\hphantom{-}5.104\\
\hspace{1em}$\kappa$: Location score&&\hphantom{00}\hphantom{-}0.083&\hphantom{00}\hphantom{-}0.323&\hphantom{00}\hphantom{-}0.749&\hphantom{00}\hphantom{-}0.083&\hphantom{00}\hphantom{-}0.082\\
\hspace{1em}$\gamma$: Price (in \$100)&&\hphantom{00}\hphantom{-}0.164&\hphantom{00}\hphantom{-}0.808&\hphantom{00}\hphantom{-}1.812&\hphantom{00}\hphantom{-}0.163&\hphantom{00}\hphantom{-}0.164\\
\hspace{1em}$\gamma$: Star rating&&\hphantom{00}\hphantom{-}0.052&\hphantom{00}\hphantom{-}0.130&\hphantom{00}\hphantom{-}0.148&\hphantom{00}\hphantom{-}0.052&\hphantom{00}\hphantom{-}0.052\\
\hspace{1em}$\gamma$: Review score&&\hphantom{00}-0.135&\hphantom{00}-0.673&\hphantom{00}-1.288&\hphantom{00}-0.135&\hphantom{00}-0.133\\
\hspace{1em}$\gamma$: No reviews&&\hphantom{00}-0.247&\hphantom{00}-1.258&\hphantom{00}-2.407&\hphantom{00}-0.242&\hphantom{00}-0.239\\
\hspace{1em}$\gamma$: Chain&&\hphantom{00}-0.103&\hphantom{00}-0.390&\hphantom{00}-0.867&\hphantom{00}-0.104&\hphantom{00}-0.103\\
\hspace{1em}$\gamma$: On promotion&&\hphantom{00}-0.048&\hphantom{00}-0.443&\hphantom{00}-0.603&\hphantom{00}-0.046&\hphantom{00}-0.047\\
\hspace{1em}$\xi$&&\hphantom{00}\hphantom{-}5.695&\hphantom{0}\hphantom{-}14.272&\hphantom{0}\hphantom{-}25.387&\hphantom{00}\hphantom{-}5.650&\hphantom{00}\hphantom{-}5.701\\
\hspace{1em}$\tilde{z}^d$&&\hphantom{00}-0.128&\hphantom{00}-0.126&\hphantom{00}-0.091&\hphantom{00}-0.105&\hphantom{00}-0.136\\
\hspace{1em}$\rho$&&\hphantom{00}\hphantom{-}2.578&\hphantom{0}\hphantom{-}11.167&\hphantom{0}\hphantom{-}22.263&\hphantom{00}\hphantom{-}2.580&\hphantom{00}\hphantom{-}2.567\\
\hspace{1em}$ \tilde{z}^d - \beta_0 $&&\hphantom{00}\hphantom{-}0.568&\hphantom{00}\hphantom{-}0.560&\hphantom{00}\hphantom{-}0.408&\hphantom{00}\hphantom{-}0.482&\hphantom{00}\hphantom{-}0.597\\
\hspace{1em}$ \bar{\xi} - \beta_0 $&&\hphantom{00}-2.549&\hphantom{00}-2.544&\hphantom{00}-2.715&\hphantom{00}-2.587&\hphantom{00}-2.537\\
\midrule
\hspace{1em}Discovery costs (\$)&&\hphantom{00}\hphantom{-}0.023&\hphantom{00}\hphantom{-}0.164&\hphantom{00}\hphantom{-}0.451&\hphantom{00}\hphantom{-}0.025&\hphantom{00}\hphantom{-}0.023\\
\hspace{1em}Search costs (\$)&&\hphantom{00}\hphantom{-}0.471&\hphantom{00}\hphantom{-}2.302&\hphantom{00}\hphantom{-}2.369&\hphantom{00}\hphantom{-}0.464&\hphantom{00}\hphantom{-}0.488\\
\hspace{1em}Log likelihood&&-48,893.726&-48,860.509&-49,196.096&-48,951.917&-48,888.534\\
\midrule
\bfseries{Fit: bookings} \\
\hspace{1em}N, model (cond. on click)&&\hphantom{00}\hphantom{-}0.057&\hphantom{00}\hphantom{-}0.056&\hphantom{00}\hphantom{-}0.053&\hphantom{00}\hphantom{-}0.057&\hphantom{00}\hphantom{-}0.056\\
\hspace{1em}N, data&&\hphantom{00}\hphantom{-}0.055&\hphantom{00}\hphantom{-}0.055&\hphantom{00}\hphantom{-}0.055&\hphantom{00}\hphantom{-}0.055&\hphantom{00}\hphantom{-}0.055\\
\hspace{1em}Mean position, model&&\hphantom{0}\hphantom{-}13.758&\hphantom{0}\hphantom{-}13.683&\hphantom{0}\hphantom{-}13.362&\hphantom{0}\hphantom{-}13.947&\hphantom{0}\hphantom{-}13.721\\
\hspace{1em}Mean position, data&&\hphantom{0}\hphantom{-}13.468&\hphantom{0}\hphantom{-}13.468&\hphantom{0}\hphantom{-}13.468&\hphantom{0}\hphantom{-}13.468&\hphantom{0}\hphantom{-}13.468\\
\midrule
\bfseries{Fit: clicks} \\
\hspace{1em}N, model&&\hphantom{00}\hphantom{-}1.170&\hphantom{00}\hphantom{-}1.169&\hphantom{00}\hphantom{-}1.133&\hphantom{00}\hphantom{-}1.178&\hphantom{00}\hphantom{-}1.168\\
\hspace{1em}N, data&&\hphantom{00}\hphantom{-}1.134&\hphantom{00}\hphantom{-}1.134&\hphantom{00}\hphantom{-}1.134&\hphantom{00}\hphantom{-}1.134&\hphantom{00}\hphantom{-}1.134\\
\hspace{1em}Mean position, model&&\hphantom{0}\hphantom{-}13.622&\hphantom{0}\hphantom{-}13.653&\hphantom{0}\hphantom{-}13.339&\hphantom{0}\hphantom{-}13.841&\hphantom{0}\hphantom{-}13.577\\
\hspace{1em}Mean position, data&&\hphantom{0}\hphantom{-}13.305&\hphantom{0}\hphantom{-}13.305&\hphantom{0}\hphantom{-}13.305&\hphantom{0}\hphantom{-}13.305&\hphantom{0}\hphantom{-}13.305\\
\midrule
\bfseries{Effects of DVR} \\
\hspace{1em}$ \Delta $ total revenues (\%)&&\hphantom{00}\hphantom{-}3.770&\hphantom{00}\hphantom{-}4.411&\hphantom{00}\hphantom{-}4.953&\hphantom{00}\hphantom{-}3.939&\hphantom{00}\hphantom{-}3.826\\
\hspace{1em}$ \Delta $ consumer welfare (\$, average)&&\hphantom{00}\hphantom{-}0.061&\hphantom{00}\hphantom{-}0.120&\hphantom{00}\hphantom{-}0.154&\hphantom{00}\hphantom{-}0.120&\hphantom{00}\hphantom{-}0.013\\
\hspace{1em}$ \Delta $ consumer welfare (\$, cond. on click)&&\hphantom{00}\hphantom{-}7.189&\hphantom{00}\hphantom{-}3.288&\hphantom{00}\hphantom{-}4.886&\hphantom{00}\hphantom{-}7.001&\hphantom{00}\hphantom{-}7.582\\
\hspace{1em}$ \Delta $ consumer welfare (\$, cond. on booking)&&\hphantom{00}\hphantom{-}7.456&\hphantom{00}\hphantom{-}4.311&\hphantom{00}\hphantom{-}5.511&\hphantom{00}\hphantom{-}7.180&\hphantom{00}\hphantom{-}8.123\\
\midrule
\bfseries{Effects of BUR} \\
\hspace{1em}$ \Delta $ total revenues (\%)&&\hphantom{00}\hphantom{-}8.243&\hphantom{00}\hphantom{-}8.860&\hphantom{00}\hphantom{-}9.345&\hphantom{00}\hphantom{-}8.364&\hphantom{0}\hphantom{-}10.006\\
\hspace{1em}$ \Delta $ consumer welfare (\$, average)&&\hphantom{00}\hphantom{-}0.037&\hphantom{00}\hphantom{-}0.094&\hphantom{00}\hphantom{-}0.146&\hphantom{00}\hphantom{-}0.067&\hphantom{00}\hphantom{-}0.010\\
\hspace{1em}$ \Delta $ consumer welfare (\$, cond. on click)&&\hphantom{00}\hphantom{-}6.467&\hphantom{00}\hphantom{-}3.429&\hphantom{00}\hphantom{-}6.589&\hphantom{00}\hphantom{-}6.359&\hphantom{00}\hphantom{-}6.769\\
\hspace{1em}$ \Delta $ consumer welfare (\$, cond. on booking)&&\hphantom{00}\hphantom{-}5.003&\hphantom{00}\hphantom{-}2.709&\hphantom{00}\hphantom{-}5.106&\hphantom{00}\hphantom{-}5.256&\hphantom{00}\hphantom{-}3.599\\
\midrule 
\end{tabular}
\begin{tablenotes}
\item \footnotesize{\emph{Notes:} The main specification uses $\sigma_\varepsilon = 1$ and $ \eta_i \sim Uniform(0,1) $. The other specifications are the same, except for either $\sigma_\varepsilon$ or $ \eta_i \sim Uniform(0,b)$.  The welfare effects are all relative to Expedia's own ranking. All specifications are estimated without fixed effects. }
\end{tablenotes} 
\end{threeparttable}
}
\end{table}

\clearpage

\subsection{Functional Form for Beliefs and Number of Alternatives Discovered}

To evaluate the impact of the functional form for beliefs, specified in \eqref{eq:functional_form_beliefs}, I estimate the model with the following linear specification for $h>1$:

\begin{equation}
\mu_{u}(h)=\mu_{u}(1) + \rho h\ .\label{eq:functional_form_beliefs_linear}
\end{equation}

\begin{figure}[hbt]
  \FIGURE
  {\includegraphics[width=0.9\textwidth]{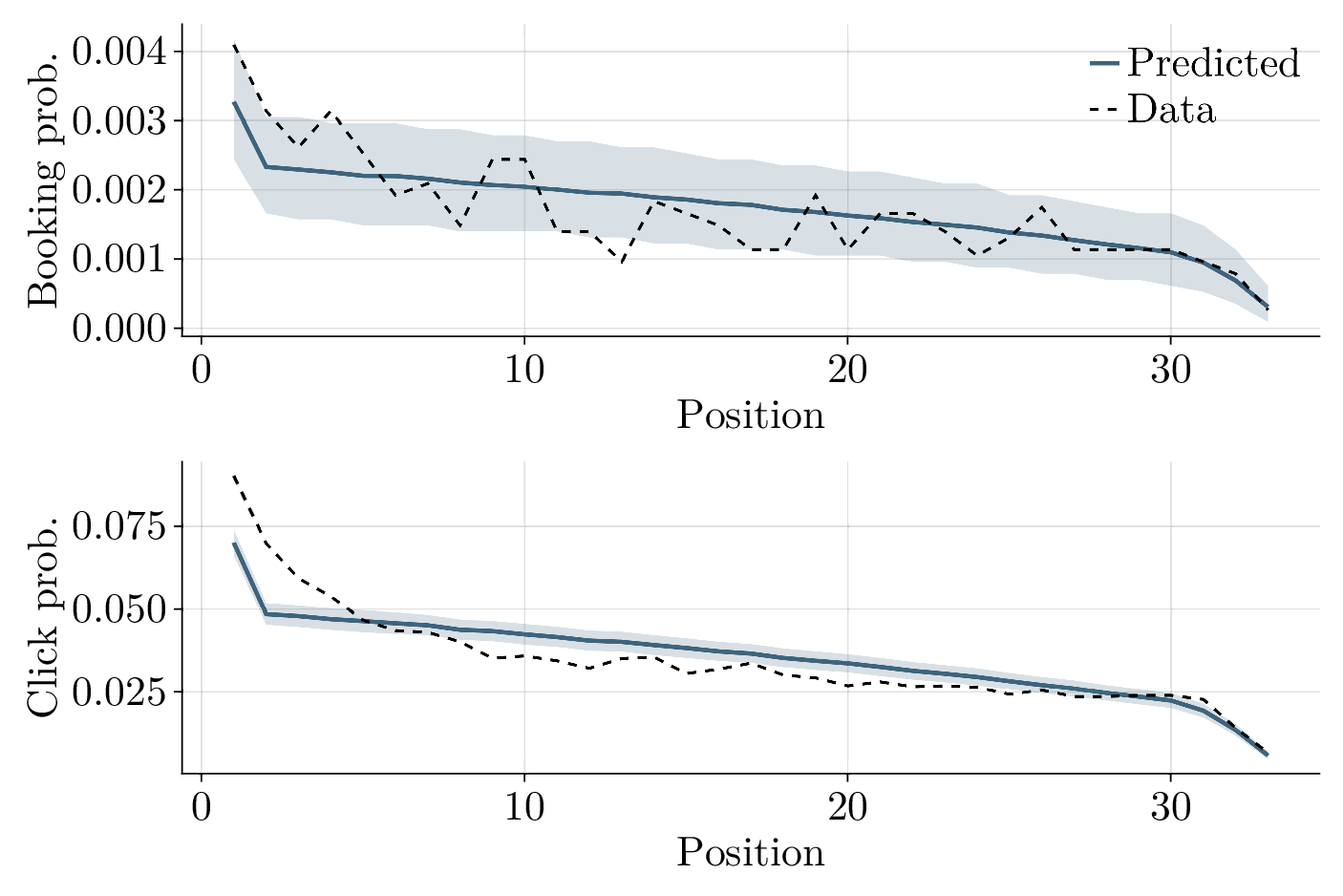}}
  {Model Fit Under the Linear Belief Specification
  \label{fig:fit-linear-spec}}
  {Click and booking probabilities averaged across 10,000 draws per consumer, conditional on consumers searching at least one hotel. The shaded area indicates the 95th percentile of the minimum and maximum number of clicks or bookings across draws and consumers.}
\end{figure}

Figure \ref{fig:fit-linear-spec} shows the position-specific booking and click probabilities predicted by the model under this linear specification. The results show that the linear specification does not capture that clicks and bookings decrease at a decreasing rate across positions.

\begin{figure}[t]
\FIGURE
{\includegraphics[width=0.9\textwidth]{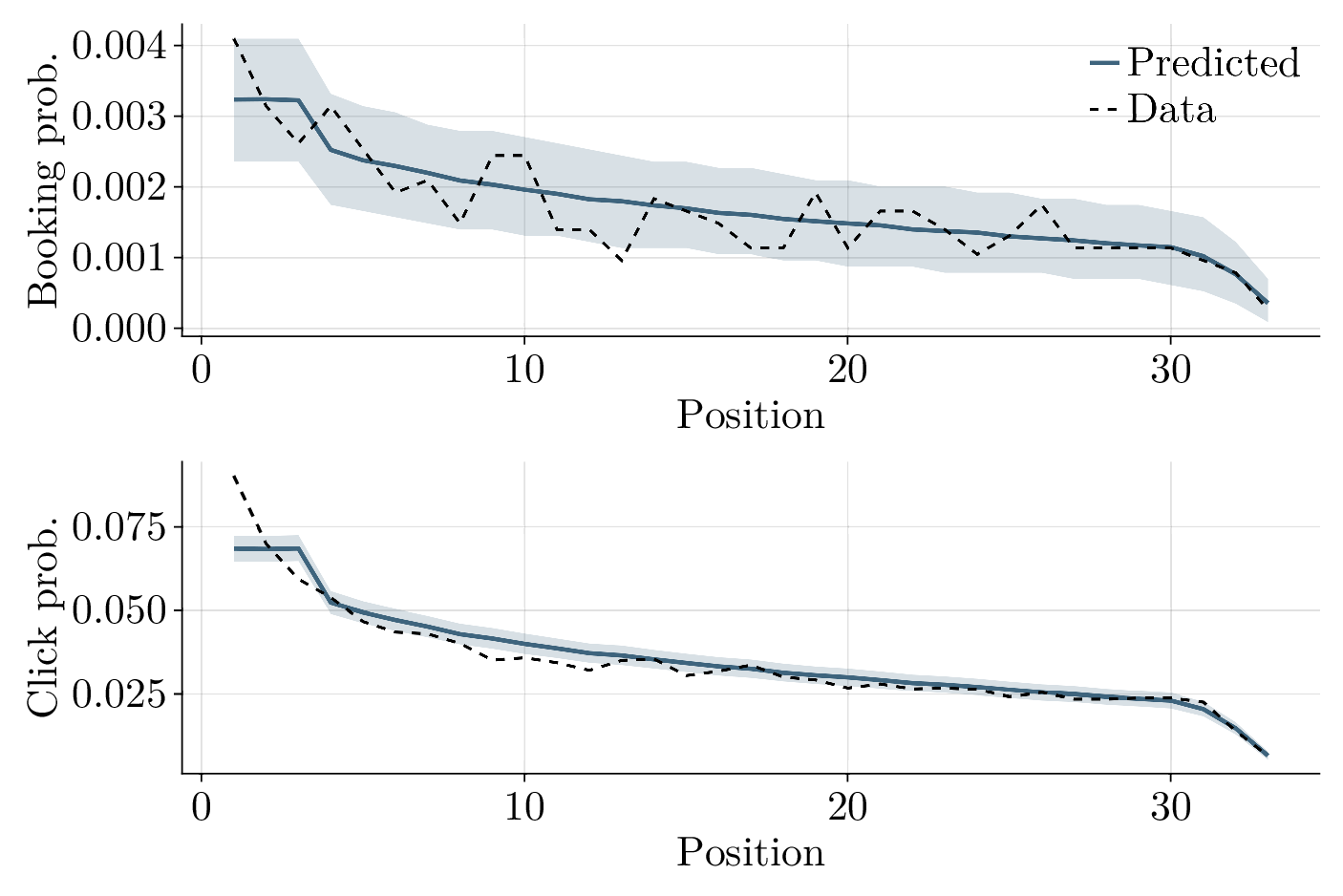}}
{Model Fit When Consumers Initially Discover 3 Alternatives \label{fig:fit-nA03-spec}}
{Click and booking probabilities averaged across 10,000 draws per consumer, conditional on consumers searching at least one hotel. The shaded area indicates the 95\% percentile of the minimum and maximum number of clicks or bookings across draws and consumers.}
\end{figure}

\begin{figure}[t]
\FIGURE
{\includegraphics[width=0.9\textwidth]{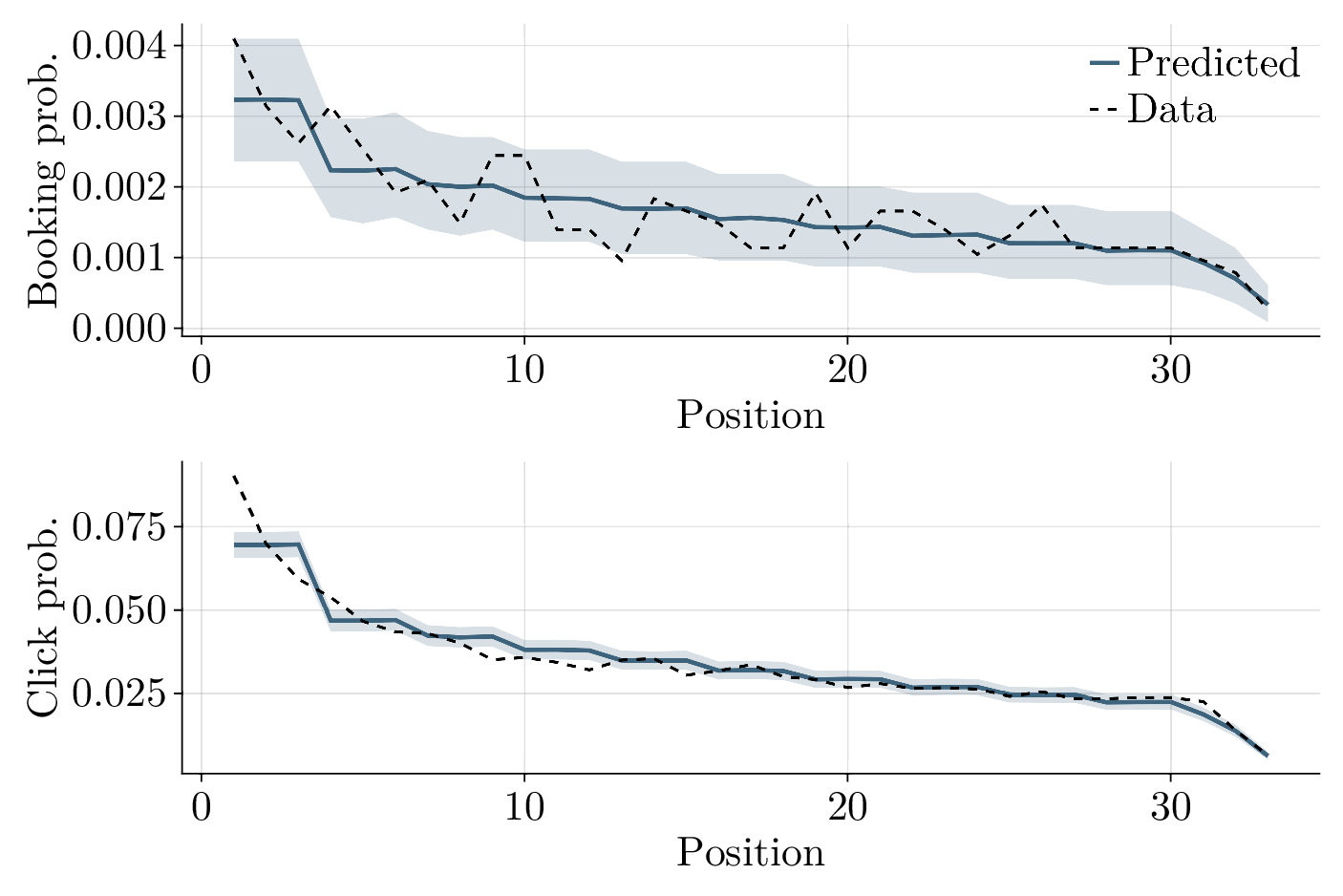}}
{Model Fit When Consumers Discover 3 Alternatives at a Time \label{fig:fit-nd3-spec}}
  {Click and booking probabilities averaged across 10,000 draws per consumer, conditional on consumers searching at least one hotel. The shaded area indicates the 95\% percentile of the minimum and maximum number of clicks or bookings across draws and consumers.}
\end{figure}

To evaluate the impact of the assumptions that consumers discover the first alternative for free and that they discover alternatives one by one, I also estimate two specifications that relax these assumptions. The results in Figures \ref{fig:fit-nA03-spec} and \ref{fig:fit-nd3-spec} show that under these alternative specifications, the model predicts click and booking probabilities that are constant among the positions discovered for free and among positions discovered at the same time, respectively. As the click and booking probabilities decrease across all positions, these results suggest that the main specification better captures the position effects in the data.

\newpage
\clearpage

\section{Comparison with the Weitzman Model\label{sec:weitzman-appendix}}

This appendix provides the specification and estimation approach used for the model comparison in Section \ref{subsec:Fit-measures}. It also discusses how the position effect mechanism in the Weitzman model limits its flexibility to simultaneously fit position effects in searches and purchases, and provides the full estimation results for this model.

\subsection{The Weitzman Model}

The \emph{Weitzman} (WM) model is an established approach to quantify the effects of rankings and other changes to the search environment \citep[see][]{Ursu2024}. Unlike in the \emph{search and discovery} (SD) model, consumers in the WM model know all product attributes and preference shocks revealed on the list page ($x_j^l$, $\tilde{\gamma}_j$, and $\nu_j$) at the beginning of search. Without product discovery, the WM model generates position effects through higher search costs for products in lower positions. Formally, search costs determine $\xi(h)$ defined in \eqref{eq:search-value}, which now depends on position $h$. As search costs increase, $\xi(h)$ decreases, leading to fewer searches and purchases at lower positions. I focus on this position-specific search cost mechanism because it is the standard one in prior work \citep[see][]{Ursu2024}.\footnote{More recent work proposes that position-specific beliefs about hidden attributes could also explain position effects \citep[][]{Kaye2024,Fong2024}. This alternative mechanism introduces more flexibility into the WM model, potentially allowing it to fit the constant conditional purchase probability. If that is the case, the limitation highlighted here could serve as an identification argument for parameters governing this additional flexibility.}

The difference in how the SD and WM models generate position effects is not only conceptual but also matters empirically \citep[][]{Greminger2021}: abstracting from product discovery can introduce biases in the inferred search costs and price elasticities because it assumes that consumers observe the entire list page, an assumption unlikely to hold unless the list displays only a few alternatives. I extend these results by showing how explaining position effects through product discovery affects the model-implied conditional purchase probabilities and, hence, the model's ability to fit position effects in purchases.

\subsection{Empirical Specification and Estimation Approach}

I use the same utility specification for the WM model as for the SD model. However, I adjust the distribution of the shocks to better fit specifications used in prior work \citep[e.g.,][]{Ursu2018, Ursu2024}. Specifically, I assume $\eta \sim N(0, \sigma_\varepsilon)$ and $\varepsilon_j \sim N(0, \sigma_\varepsilon)$, and estimate $\sigma_\varepsilon$. Estimating $\sigma_\varepsilon$ is possible because position on the list acts as a search cost shifter \citep[see][]{Yavorsky2021}. The results are similar (or even less favorable to the WM model) when assuming a standard uniform distribution for $\eta$ instead. Moreover, using the exact same specification as \cite{Compiani2023} also yields similar results.\footnote{The codes are available as part of the replication package of \cite{Compiani2023}.}

I introduce position-specific search costs by assuming that the search value depends on position $h$ through the following functional form for $h=1,\dots$: $\xi(h) =\xi + \rho \log(h)$. As in the SD model, I estimate the parameters $\xi$ and $\rho$, while the assumed shape allows the model to fit the nonlinear decrease in clicks and bookings across positions. Estimating the model this way is equivalent to imposing a functional form on how search costs depend on position and estimating cost parameters as in, e.g., \cite{Ursu2018}.

To estimate the WM model, I use the fact that it is a SD model where discovery costs are zero and consumers discover all alternatives. In this case, the likelihood can be calculated by adjusting the likelihood contribution in \eqref{eq:individual-likelihood-contribution} to
\begin{align}
    \mathcal{L}^i(\theta) = \log\int  \prod_{k \in S} \mathbb{P}(Z_{k}^{s} \geq \tilde{w}_{j} \cap U_{k}\leq \tilde{w}_{j})  \times \prod_{k \in J \backslash S} \mathbb{P}(Z_{k}^{s} \leq \tilde{w}_{j} )\text{d}H(\eta,\nu_{j}, \varepsilon_{j})\ , \label{eq:individual-likelihood-contribution-weitzman}
\end{align}
which can then be computed using the same simulation approach as for the SD model. Matching the specification for the SD model, I also condition the likelihood contribution for the WM model on searching at least once.

This estimation approach is easier to implement and computationally less demanding than the approaches proposed by \cite{Jiang2021} and \cite{Chung2023}. It effectively integrates at most over two dimensions $(\nu_{j}, \varepsilon_{j})$, instead of over as many dimensions as the number of searches each consumer makes. These approaches also cannot be easily used when the search order is not observed. Though integrating over all possible permutations of search sequences is possible when these are not observed \citep[][]{Compiani2023}, this becomes prohibitively costly when consumers make more than just a few searches. In contrast, my approach constructs the likelihood independent of the search sequence, making it straightforward to implement even when consumers make many searches.

\subsection{Estimation Results and Model Fit  \label{subsec:Weitzman-estimation-results}}

Table \ref{tab:results_pars_wm} reports the full parameter estimates for the WM model. The parameter estimates all have the expected sign, as for the SD model.

Table \ref{tab:fit} reports additional model fit statistics for both models. The results show that both models fit the attributes of the searched and booked hotels relatively well. However, in line with Figure \ref{fig:fit}, the table also reveals that the WM model predicts bookings occur on average two positions lower than in the data, whereas the SD model predicts the average position of bookings close to the one in the data. The WM model has a lower AIC and BIC, indicating a better overall model fit. However, these fit measures are based on the conditional likelihood, which does not capture how the models fit when consumers who visit the platform do not search any products. Hence, these fit measures should be interpreted with caution. Moreover, position effects in purchases and their heterogeneity are more relevant for the trade-offs between ranking objectives, which is why I focus these measures.

\begin{table}[h] \centering \footnotesize
\caption{Parameter Estimates: Weitzman Model} \label{tab:results_pars_wm}

\begin{threeparttable}
\begin{tabular}{lllc} 
\midrule

 && \multicolumn{1}{c}{Estimate} & \multicolumn{1}{c}{Std. error} \\
\midrule
\textbf{Preference parameters in $u_j^l = x_j^l{}'\beta + \nu_j^l$} \\
\hspace{1em}$\beta$: Price (in \$100)&&\hphantom{00}-0.312***&(0.024)\\
\hspace{1em}$\beta$: Star rating&&\hphantom{00}\hphantom{-}0.164***&(0.025)\\
\hspace{1em}$\beta$: Review score&&\hphantom{00}\hphantom{-}0.132***&(0.035)\\
\hspace{1em}$\beta$: No reviews&&\hphantom{00}\hphantom{-}0.255&(0.187)\\
\hspace{1em}$\beta$: Chain&&\hphantom{00}\hphantom{-}0.019&(0.034)\\
\hspace{1em}$\beta$: On promotion&&\hphantom{00}\hphantom{-}0.121***&(0.033)\\
\hspace{1em}$\beta$: Outside option&&\hphantom{00}\hphantom{-}6.416***&(0.150)\\
\midrule
\textbf{Preference parameters in $u_j^d = x_j^d{}'\kappa + \delta_j + \varepsilon_j$} \\
\hspace{1em}$\kappa$: Location score&&\hphantom{00}\hphantom{-}0.053***&(0.011)\\
\midrule
\textbf{Search value parameters in $z_j^s = u_j^l + x_j^l{}'\gamma +\tilde{\gamma}_j + \xi + \rho \log(h)$} \\
\hspace{1em}$\gamma$: Price (in \$100)&&\hphantom{00}\hphantom{-}0.108***&(0.023)\\
\hspace{1em}$\gamma$: Star rating&&\hphantom{00}\hphantom{-}0.054&(0.025)\\
\hspace{1em}$\gamma$: Review score&&\hphantom{00}-0.105***&(0.035)\\
\hspace{1em}$\gamma$: No reviews&&\hphantom{00}-0.194&(0.185)\\
\hspace{1em}$\gamma$: Chain&&\hphantom{00}-0.057&(0.033)\\
\hspace{1em}$\gamma$: On promotion&&\hphantom{00}-0.033&(0.033)\\
\hspace{1em}$\xi$&&\hphantom{00}\hphantom{-}2.201***&(0.126)\\
\hspace{1em}$\rho$&&\hphantom{00}-0.136***&(0.003)\\
\midrule
\textbf{Variance and search cost parameters} \\
\hspace{1em}$ \sigma_\varepsilon $&&\hphantom{00}\hphantom{-}0.722***&(0.014)\\
\hspace{1em}$c_s$&&\hphantom{00}\hphantom{-}0.032& \\
\midrule
\hspace{1em}Log likelihood&&-47,858.384& \\
\hspace{1em}N consumers&&\hphantom{-}11,467& \\
\midrule 
\end{tabular}
\begin{tablenotes}
\item \footnotesize{\emph{Notes:} Parameter estimates obtained using 3,000 simulation draws per region of the partitioned probability space. Product fixed effects $\delta_j$ and $\tilde{\gamma}_j$ are included for a limited set of products in the model but not shown in the table. Asymptotic standard errors are shown in parentheses. Search costs are computed using 10M simulation draws. Statistical significance is indicated by * p < 0.1, ** p < 0.05, *** p < 0.01.  }
\end{tablenotes} 
\end{threeparttable}
\end{table}

\begin{table}[htb] \centering \footnotesize
\caption{Model Fit Measures} \label{tab:fit}

\begin{threeparttable}
\begin{tabular}{llccccccc} 
\midrule

 && \multicolumn{3}{c}{Bookings}& \multicolumn{3}{c}{Clicks}\\
\cmidrule(rl){3-5} \cmidrule(rl){6-8} 
&& \multicolumn{1}{c}{Data}& \multicolumn{1}{c}{SD Model}& \multicolumn{1}{c}{WM Model}& \multicolumn{1}{c}{Data}& \multicolumn{1}{c}{SD Model}& \multicolumn{1}{c}{WM Model}\\
\midrule
N (per consumer)&&\hphantom{00}0.055&\hphantom{00}0.057&\hphantom{00}0.065&\hphantom{00}1.134&\hphantom{00}1.172&\hphantom{00}1.156\\
Mean position&&\hphantom{0}13.468&\hphantom{0}13.766&\hphantom{0}15.810&\hphantom{0}13.305&\hphantom{0}13.619&\hphantom{0}13.406\\
Price&&145.166&\hphantom{00}148.186&\hphantom{00}149.495&162.208&163.062&162.832\\
Star rating&&\hphantom{00}3.495&\hphantom{00}3.491&\hphantom{00}3.477&\hphantom{00}3.508&\hphantom{00}3.502&\hphantom{00}3.509\\
Review score&&\hphantom{00}3.965&\hphantom{00}3.960&\hphantom{00}3.966&\hphantom{00}3.917&\hphantom{00}3.904&\hphantom{00}3.921\\
No reviews&&\hphantom{00}0.009&\hphantom{00}0.009&\hphantom{00}0.010&\hphantom{00}0.018&\hphantom{00}0.017&\hphantom{00}0.017\\
Location score&&\hphantom{00}3.788&\hphantom{00}3.600&\hphantom{00}3.521&\hphantom{00}3.611&\hphantom{00}3.402&\hphantom{00}3.378\\
Chain&&\hphantom{00}0.667&\hphantom{00}0.654&\hphantom{00}0.671&\hphantom{00}0.643&\hphantom{00}0.625&\hphantom{00}0.647\\
On promotion&&\hphantom{00}0.413&\hphantom{00}0.431&\hphantom{00}0.397&\hphantom{00}0.359&\hphantom{00}0.376&\hphantom{00}0.358\\
\midrule
AIC&& &97638.713&95750.767& & & \\
BIC&& &97763.616&95875.670& & & \\
\midrule 
\end{tabular}
\begin{tablenotes}
\item \footnotesize{\emph{Notes:} This table compares moments in the estimation sample with those predicted by the SD and the WM model and reports the AIC and BIC fit statistics. The reported moments are the average for the number of clicks and bookings per consumer; the average position at which consumers clicked or booked a hotel; and the averages for the different attributes of hotels that consumers searched or chose. The simulated search paths were generated with 10,000 draws for each consumer, conditional on searching at least one alternative. }
\end{tablenotes} 
\end{threeparttable}
\end{table}

\clearpage

\subsection{Conditional Purchase Probabilities\label{subsec:Weitzman-discussion}}

\begin{figure}[htb]
    \FIGURE
    {\subfloat{\hspace{-2.em}\includegraphics[width=0.33\textwidth,trim=0 1em 0 1em,clip]{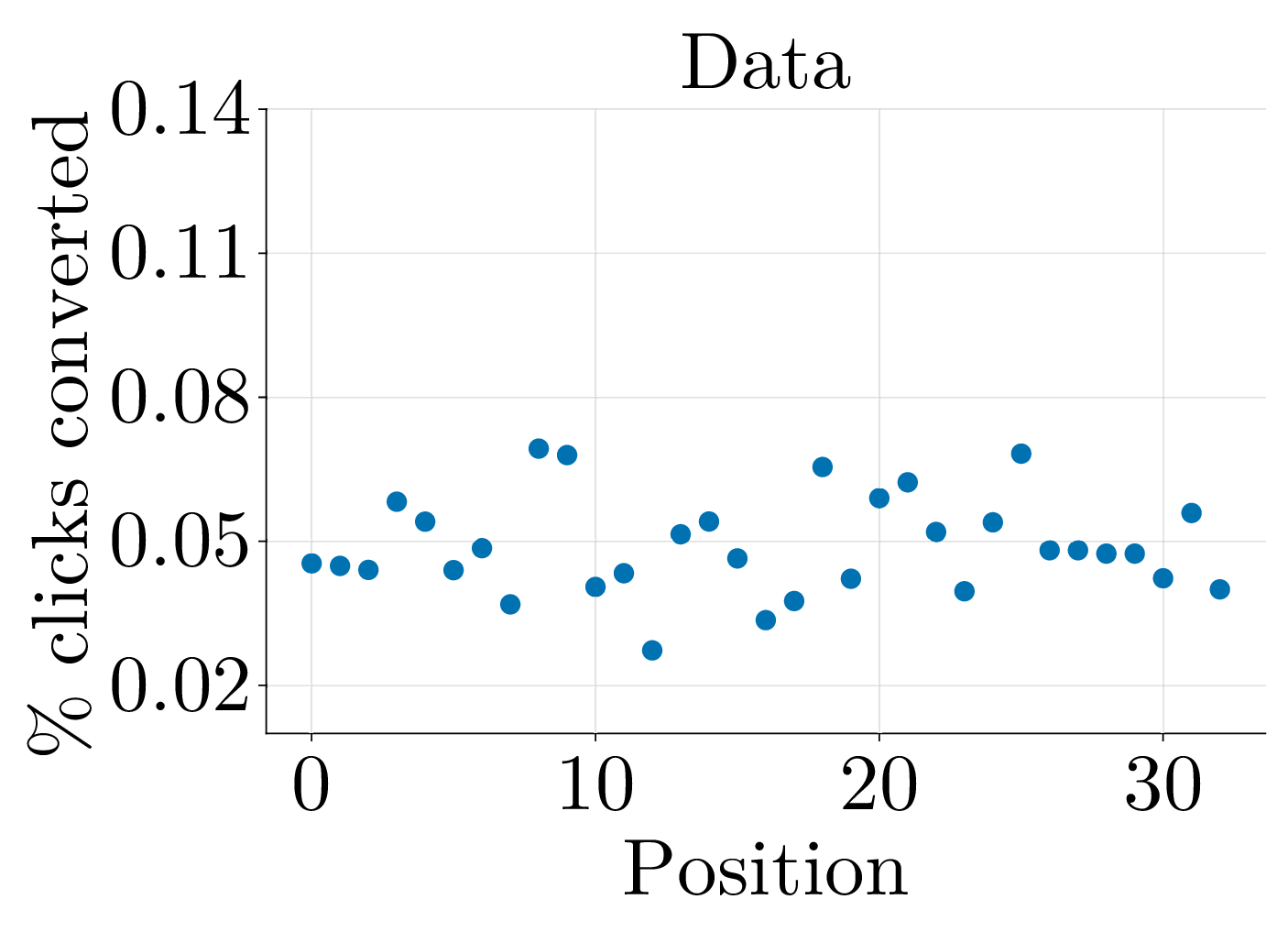}}\hspace{0.5em}
    \subfloat{\includegraphics[width=0.33\textwidth,trim=0 1em 0 1em,clip]{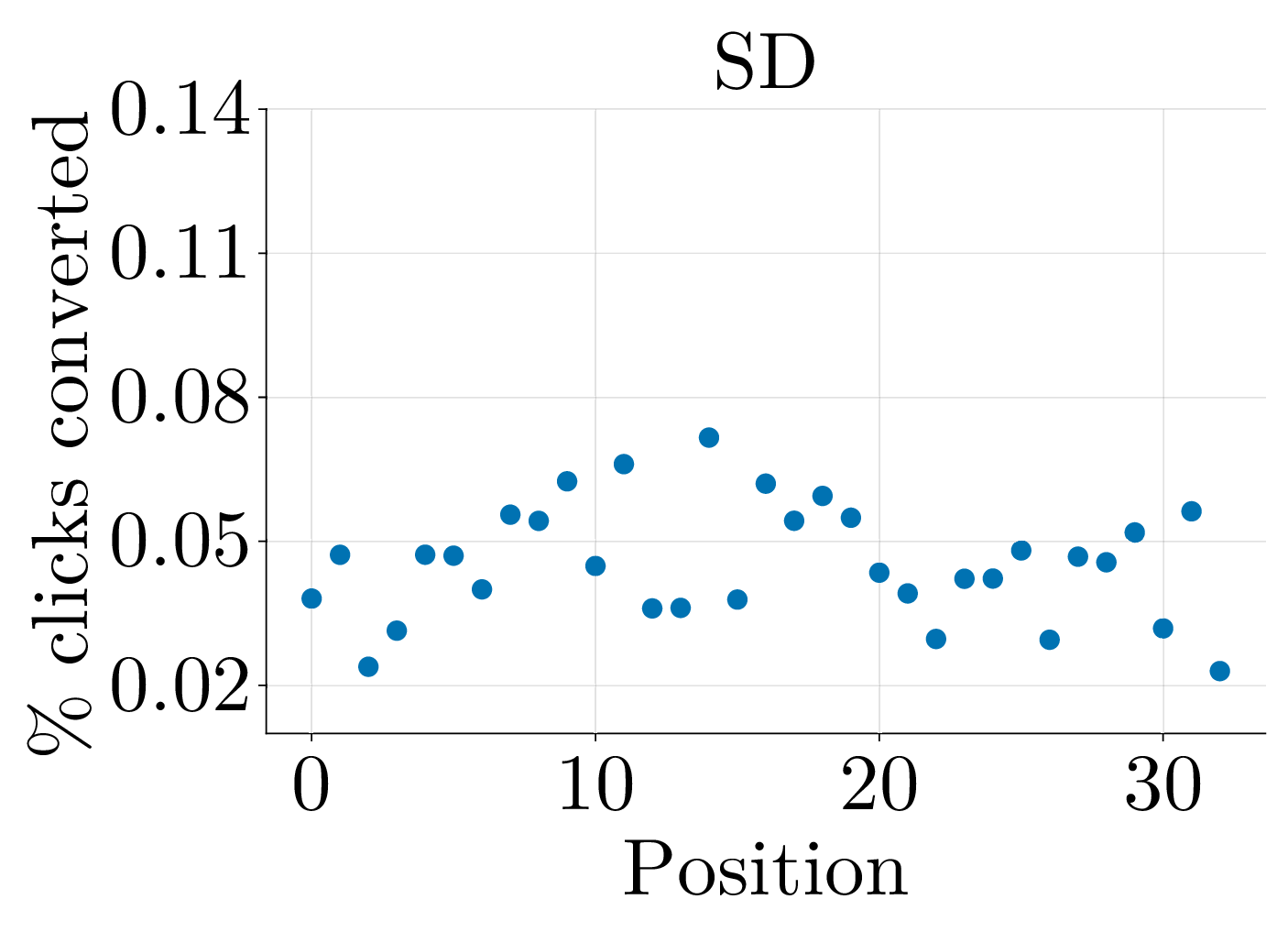}}\hspace{0.5em}
    \subfloat{\includegraphics[width=0.33\textwidth,trim=0 1em 0 1em,clip]{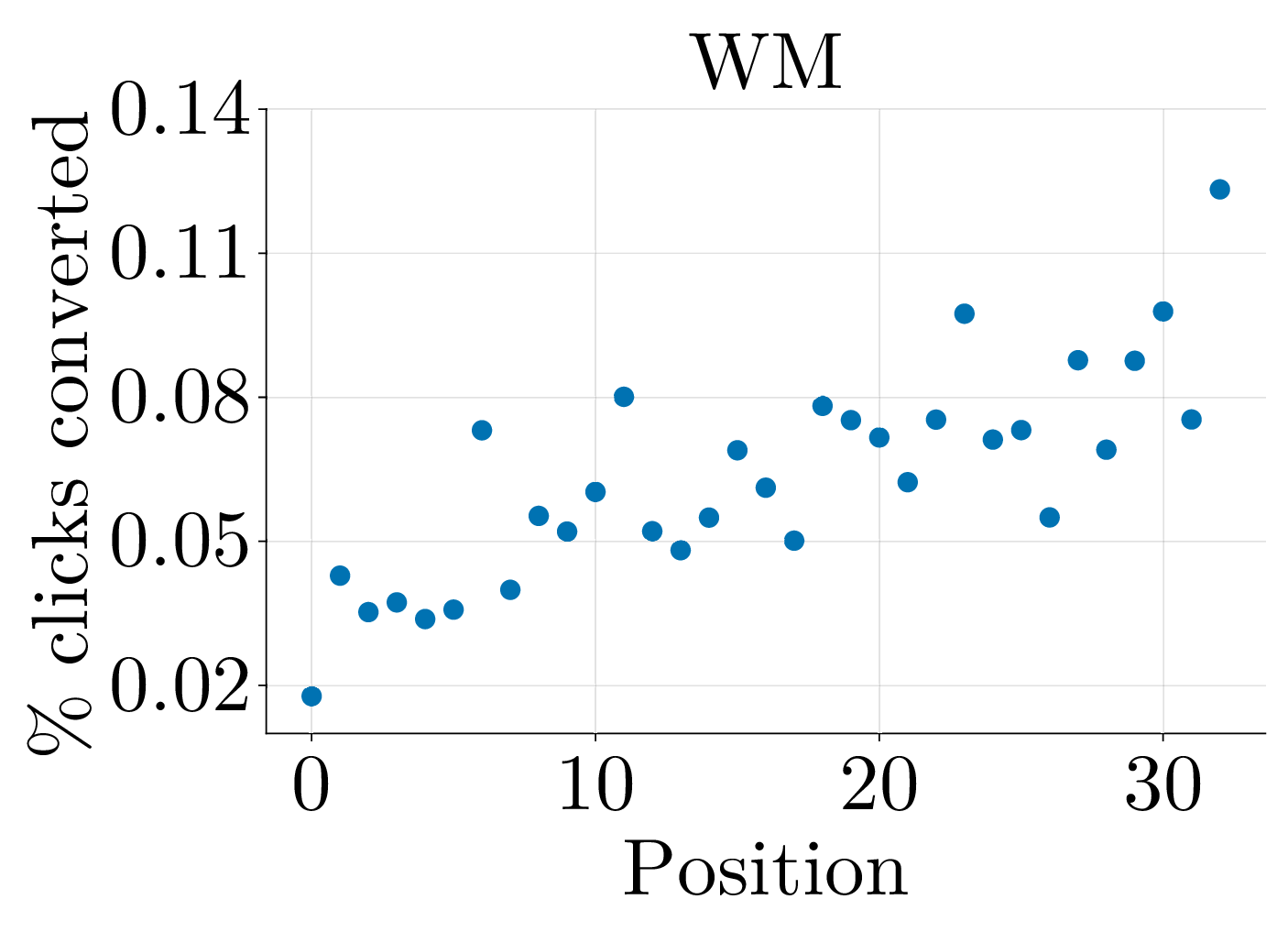}}}
    {Position-Specific Conversion Rate for Clicks \label{fig:position-conversion-rate}}
    {This figure shows the percent of clicks with a booking across the different positions in either the estimation sample or data simulated from the respective models. }
\end{figure}

As highlighted in Figure \ref{fig:fit} in Section \ref{subsec:Fit-measures}, the WM model does not capture well the positions from which consumers book hotels. Figure \ref{fig:position-conversion-rate} reveals the reason for this result. The panels show the percentage of clicks that result in a booking for the estimation sample and data simulated from the respective models. The first panel shows that this percentage is stable across positions in the data, as \cite{Ursu2018} highlighted. The middle panel shows that the SD model predicts a similarly stable conditional purchase probability across positions. In contrast, the right panel shows that the estimated WM model implies a conditional purchase probability that is three to four times larger for lower positions compared to the top ones. As a result, the WM model overpredicts bookings in lower positions despite matching clicks at those positions, creating the contrast highlighted in Figure \ref{fig:fit}.

To determine where the difference in the model-implied conditional purchase probabilities comes from, I derive and compare how an alternative's conditional purchase probability depends on its position in the two models.

To get a convenient expression for the conditional purchase probability in the WM model, I use that it is a special case of the SD model \citep[see][]{Greminger2021}. Hence, the implications in Proposition \ref{prop:implications} also apply. Let $\bar{w}_{-j}= \max_{k \in J\setminus j} \tilde{w}_k$ denote the maximum effective value across all alternatives except for some alternative $j$ in position $h$. The search and discovery implications require $z_j^s(h) \geq \bar{w}_{-j}(h)$ for $j$ to be searched, and the choice implication requires that $u_j \geq \bar{w}_{-j}(h)$ for $j$ to be chosen. The conditional purchase probability for an alternative $j$ in position $h$ therefore is given by
\begin{align}
\mathbb{P}_{\text{WM}}(\text{choose } j | \text{search } j ) & = \mathbb{P}(U_j \geq \bar{W}_{-j}| Z_j^s(h) \geq \bar{W}_{-j} ) \\
& = \mathbb{P}(U_j^l + U_j^d \geq \bar{W}_{-j} | U_j^l + x_j^l{}'\gamma + \tilde{\gamma}_j + \xi(h) \geq W_{-j}) \ , \label{eq:cond-choice-prob-wm}
\end{align}
where $u_j^l = x_j^l{}'\beta + \nu_j$ and the probability is taken only over the idiosyncratic shocks $\nu_j$ and $\varepsilon_j$, not over alternative-specific attributes that are not random across consumers.

Assuming continuous distributions for the idiosyncratic shocks and taking the derivative of \eqref{eq:cond-choice-prob-wm} with respect to $\xi(h)$ reveals that an alternative's conditional purchase probability (weakly) decreases in its search value and thus increases in its search cost.\footnote{Note that \eqref{eq:cond-choice-prob-wm} has the form $P(X+Y \geq c | X + \xi \geq c) = \frac{\int_{c - \xi}^{\infty} (1 - F_Y(c - t))f_X(t) \text{d}t }{\int_{c - \xi}^{\infty} f_X(t)\text{d}t}$, where $X$ and $Y$ are independent random variables and $F_Y$ is the CDF of $Y$ and $f_X$ the PDF of $x$. The derivative of the numerator and denominator with respect to $\xi$ are given by $1-F_Y(\xi) f_X(c - \xi)$ and $f_X(c - \xi)$, respectively. The derivative of the entire expression with respect to $\xi$ then is given by $\frac{f_X(c-\xi)}{(\int_{c - \xi}^{\infty} (1 - F_Y(c - t))f_X(t) \text{d}t)^2} \int_{c-\xi}^{\infty} (F_Y(c-t) - F_Y(\xi))f_X(t)\text{d}t \leq 0$.} This is because with a higher search cost, the alternative is only searched if it offers a sufficiently large list utility $u_j^l$, which also increases the probability that it is chosen after being searched.

As search costs increase with position $h$, this effect suggests that the conditional purchase probability increases in $h$. However, as $h$ increases, the distribution of the maximum effective value $\bar{w}_{-j}$ among alternatives other than $j$ also improves in a first-order-stochastic-dominance sense. Depending on the distributional assumptions, this effect can be either positive or negative. Nonetheless, this second effect will rarely offset the first effect because the change in the distribution of $\bar{W}_{-j}(h)$ tends to be small relative to the direct change in $\xi(h)$, unless there are very few alternatives or position effects in searches are small. As a result, the WM model typically predicts a conditional purchase probability that increases across positions, as the one in Figure \ref{fig:position-conversion-rate}.

The same does not hold for the SD model. In this model, an alternative's position determines its clicks and purchases by determining the likelihood of it being discovered. This probability depends only on the search values and utilities of the alternatives discovered before it rather than on the alternative's own search value and utility. Formally, the conditional purchase probability in the SD model is given by
\begin{align}
\mathbb{P}_{SD}(\text{choose } j | \text{search } j ) & = \mathbb{P}(U_j^l + U_j^d \geq \bar{V}_{-j} | U_j^l + x_j^l{}'\gamma + \tilde{\gamma}_j + \xi \geq \bar{V}_{-j}) \ , \label{eq:cond-choice-prob-sd}
\end{align}
where $\bar{v}_{-j} = \max_{k \in J \setminus j} \min\{\tilde{w}_k, z^d(h_k)\}$. The expression follows from the generalized eventual purchase theorem of \cite{Greminger2021}.

Unlike in \eqref{eq:cond-choice-prob-wm}, position $h$ does not enter the search value directly through $\xi$ in this expression. Instead, an increase in $h$ only improves the distribution of $\bar{V}_{-j}$ in a first-order-stochastic-dominance sense, which tends to have a small effect on the conditional purchase probability and can go in either direction depending on the distributional assumptions and parameters such as $\xi$. Hence, the SD model can generate a conditional purchase probability that is constant across positions and thus consistent with the Expedia data (see Figure \ref{fig:position-conversion-rate}).

\clearpage

\section{Consumers' Beliefs and Discovery Values\label{sec:Discovery-Values}}

This appendix provides details on how consumers' beliefs about the ranking affect the discovery values, how the discovery values are identified by the data, and how they then can be used to recover discovery costs and the belief parameter. Finally, the appendix shows that any functional form that implies that $\mu(h)$ weakly increases in $h$ contradicts the data.

\subsection{Position-Specific Discovery Values} \label{subsec:Position-Specific-Discovery}

\cite{Greminger2021} shows that the position-specific discovery value $z^d(h)$ is implicitly defined by
\begin{equation}
  c_d=\int_{z^d(h)}^{\infty}\left[1-\tilde{G}_h(t)\right]\text{d}t \ , \label{eq:discovery-value-appendix0}
\end{equation}
where $\tilde{G}_h$ is the CDF of the effective value $W_j(h) = X_j^l(h)'\beta + \nu_j + \min\{X_j^l(h)'\gamma + \tilde{\gamma}_j(h) + \xi, X_j^d(h)'\kappa + \delta_j(h), \varepsilon_j\} $. The distribution of this term given the position $h$ is determined by consumers' beliefs over the products' attributes and the shocks, reflected in the respective random variables depending on $h$.

In Online Appendix EC.2, \cite{Greminger2021} further shows that this is equivalent to $z^d(h) = \mu(h) + \Xi(h)$, where $\Xi(h)$ is implicitly defined by
\begin{equation}
  c_d=\int_{\Xi(h)}^{\infty}\left[1-G_h(t)\right]\text{d}t \ , \label{eq:discovery-value-appendix}
\end{equation}
and $G_h$ now is the CDF of the demeaned effective value $\tilde{W}_j(h) =  W_j(h) - \mu(h)$.

The variance and the shape of the distribution of these demeaned effective values does not change with positions. This is because consumers know the distributions of the two shocks and the overall distribution of attributes. Moreover, consumers only expect the mean of $W_j^l(h)$ to depend on the position on the list, whereas the shape of the distribution does not depend on the position. Hence, the distribution of the demeaned effective value does not change across the positions such that $\Xi(h)$ is constant. As a result, the discovery value $z^d(h)$ is given by $z^d(h) = \mu(h) + \Xi$, which is the expression given in the main text.

\subsection{Stopping Probabilities Uniquely Determine $z^d(h)$}

Given the search parameters, the probabilities of stopping discovery across the different positions determine the discovery value through the stopping rule. To see this, consider the case with only two alternatives, $A$ and $B$. In this case, the respective probability is given by
\begin{equation}
  \mathbb{P}(\text{Stop discovery on position }1) = \frac{1}{2} \sum_{j \in \{A, B\}}\mathbb{P}(\max\{\tilde{W}_j, U_0 \} \geq z^d(2))\ ,\label{eq:stopping-discovery-value-2simple}
\end{equation}
which is an expectation over the random ranking. Apart from $z^d(2)$, the search parameters (e.g., $\beta$) in \eqref{eq:stopping-discovery-value-2simple} are already determined. Hence, the stopping probability identified from the data fully determines $z^d(2)$ through \eqref{eq:stopping-discovery-value-2simple}. With more than two alternatives, the logic remains the same and the respective probability of stopping discovery on a position determines the associated discovery value.

\subsection{Discovery Values Uniquely Determine $c_d$ and $\rho$} \label{subsec:Identification-of-cd-rho}

The overall level of the discovery values $z^d(h)$ uniquely determines $c_d$ when consumers know the overall attribute distribution. Intuitively, the discovery values capture the net benefits of discovering more alternatives. These net benefits are determined by consumers' beliefs about the alternatives they will discover and the discovery costs. Hence, the discovery values could be large because consumers expect to discover good alternatives or because the cost of discovering alternatives is small. Whereas it is generally not possible to distinguish the two from the data, the assumption that consumers know the overall distribution they are sampling from fixes consumers' beliefs to the attribute distribution in the data.\footnote{Fixing beliefs by assuming rational expectations is the common approach in empirical search models. For example, virtually all studies that estimate a Weitzman model assume that consumers know the distribution of the shock that will be revealed when searching an alternative.} Given these beliefs, the discovery value fully determines  the discovery costs $c_d$. The procedure described in Online Appendix \ref{subsec:Post-estimation-Recovery-of} shows how the discovery costs can be recovered from the discovery values given the assumptions on consumers' beliefs.

Differences in $z^d(h)$ across positions $h$ then uniquely determine the belief parameter $\rho$. For example, the functional form \eqref{eq:functional_form_beliefs} that I impose implies
\begin{equation}
  \Delta z^d(h) = z^d(h) - z^d(h+1) = \mu(h) - \mu(h+1)  = \rho \frac{\log(h)}{\log(h+1)} \ , \label{eq:diff_zd}
\end{equation}
which pins down $\rho$ given the difference in the discovery values across the positions. Note, however, that this functional form is not necessary for the identification of $\rho$ because $z^d(h)$ is identified for each position $h > 1$ by the data. Hence, in principle, more flexible functional forms for the beliefs about the ranking could be used.

Combined, this shows that $c_d$ and $\rho$  are uniquely determined by the discovery values given the other parameters. Hence, whenever the discovery values are identified, $c_d$ and $\rho$ are also identified. Moreover, given some estimate for $\tilde{z}^d$ and the other parameters, $c_d$ can be recovered from $\tilde{z}^d$ without having to estimate it directly.

\subsection{Recovering Search and Discovery Costs \label{subsec:Post-estimation-Recovery-of}}

$c_d$ and $\rho$ are not estimated directly. Instead, they are recovered from the estimated discovery value for the first position $\tilde{z}^d = z^d(1)$ and the other parameters. Similarly, search costs $c_s$ are not estimated directly and instead recovered from the estimated search values. The following shows the procedure to recover these costs after estimating the other parameters of the model.

\paragraph{Recovering Discovery Costs.} To recover $c_d$ given estimates for $\tilde{z}^d = z^d(1)$ and the other parameters, I apply the following procedure:
\begin{enumerate}
  \item Obtain an estimate $\hat{\mu}_{\tilde{w}}$ for $\mu_{\tilde{w}} = \mathbb{E}[\tilde{W}_j]$. To obtain this estimate, I use the following Monte Carlo integration procedure:
  \begin{enumerate}
    \item Obtain estimate $\hat{\mu}_{x\kappa}$ by taking the average of $x_j^d{}'\hat{\kappa}$ across all products in the data.
    \item Take $N_{c_d}$ draws ($x_q^l$, $x_q^d$, $\hat{\tilde{\gamma}}_q$, $\nu_q$, $\varepsilon_q$) from the respective empirical and assumed distributions.
    \item For each draw, compute $\tilde{w}_q = x_q^l \hat{\beta} + \nu_q + x_q^l\hat{\gamma} + \hat{\tilde{\gamma}}_q+\min\{ \hat{\xi}, x_q^d \hat{\kappa} - \hat{\mu}_{x\kappa} + \varepsilon_q\}$.
    \item Compute $\hat{\mu}_{\tilde{w}}$ as the average across  $\tilde{w}_q$.
  \end{enumerate}

  \item Substitute \eqref{eq:functional_form_beliefs} in $\frac{1}{|J|}\sum_{h=1}^{|J|} \mu(h) = \hat{\mu}_{\tilde{w}}$ to obtain an estimate $\hat{\mu}_1$ given the estimate $\hat{\rho}$. In my application, I set $|J| = 34$, which is the maximum number of positions in the data.
  \item Obtain $\hat{\Xi}$ from $\hat{\mu}_1$ and $\hat{\rho}$ through $\hat{\tilde{z}}^d = \hat{\mu}_1 + \Xi$ implied by the definition of the discovery value and the functional form in \eqref{eq:functional_form_beliefs}.
  \item Compute the estimate $\hat{c}_d$ from $\hat{\Xi}$ by applying the definition of $\Xi$ given in \eqref{eq:discovery-value}, which is equivalent to the following:
  \begin{equation}
    c_d=\mathbb{E}\left[\max\{0,\tilde{W}_j - \hat{\mu}_{\tilde{w}} - \hat{\Xi} \}\right]\ . \label{eq:discovery-value-appendix1}
  \end{equation}
  To compute the right-hand side, I use the following standard Monte Carlo integration procedure:
  \begin{enumerate}
     \item Take $N_{c_d}$ draws ($x_q^l$, $x_q^d$, $\hat{\tilde{\gamma}}_q$, $\nu_q$, $\varepsilon_q$) from the empirical  distribution of attributes and the assumed distributions of the two taste shocks.
      \item For each draw, compute $\tilde{w}_q = x_q^l \hat{\beta} + \nu_q + x_q^l\hat{\gamma} + \hat{\tilde{\gamma}}_q+\min\{ \hat{\xi}, x_q^d \hat{\kappa} - \hat{\mu}_{x\kappa} + \varepsilon_q\}$.
      \item Compute $\hat{c}_d$ as the average of $\max\{0, \tilde{w}_q - \hat{\mu}_{\tilde{w}} - \hat{\Xi} \}$ across draws.
  \end{enumerate}
\end{enumerate}

Note that by directly sampling from the attribute distribution in the last step, I avoid having to impose parametric assumptions on the distribution of product attributes. Moreover, by computing $c_d$ from $\tilde{z}^d$ instead of estimating it directly, this procedure avoids having to do the costly computation of $\Xi$ given $c_d$ during estimation. Finally, note that consumers are assumed to believe that $U_{j}^d| x_j^l = x_j^l{}'\gamma + \tilde{\gamma}_j + \tilde{\varepsilon}_{j}$, which leads to the adjustment of the mean of the empirical distribution of $x_j^d{}' \kappa$ by $\hat{\mu}_{x\kappa}$ in the first and last step.

\paragraph{Recovering Search Costs.} Search costs $c_s$ are recovered from the search values $z_j^s$ implied by the estimates. The definition of the search values is given in equation \eqref{eq:search-value}, which integrates over the distribution of the shock $\tilde{\varepsilon}_j$ that has zero mean conditional on the list attributes in $x_j^l$. To compute the search costs, note that \eqref{eq:search-value} is equivalent to
\begin{equation} \label{eq:search-value-appendix}
  c_s=\mathbb{E}[\max\{0,\tilde{\varepsilon}_j - \xi \}]\ .
\end{equation}
The assumptions on consumers' beliefs imply that $\tilde{\varepsilon}_j = X_j^d{}' \kappa + \Delta_j + \varepsilon_j - \mu_{x\kappa}$. $\mu_{x\kappa}$ is the mean of $X_j^d{}' \kappa + \Delta_j + \varepsilon_j$, which is the sum of the random variables capturing consumers beliefs over $x_j^d$, $\delta_j$, and $\epsilon_j$. It is subtracted so that $\tilde{\varepsilon}_j$ has zero mean conditional on $x_j^l$.

To obtain the search cost estimate $\hat{c}_s$, I use the following Monte Carlo integration procedure to compute the right-hand side:
\begin{enumerate}
  \item Obtain estimate $\hat{\mu}_{x\kappa}$ by taking the average of $x_j^d{}'\hat{\kappa}$ across all products in the data.
  \item Take $N_{c_s}$ draws ($x_q^d$, $\varepsilon_q$) from the respective empirical and assumed distributions.
  \item For each draw, compute $u_q^d = x_q^d{}'\hat{\kappa} + \varepsilon_q - \hat{\mu}_{x\kappa}$.
  \item Compute the search cost estimate $\hat{c}_j^s$ as the average across $\max\{0, u_q^d - x_j^l{}'\hat{\gamma} - \xi\}$.
\end{enumerate}

\subsection{Weakly Increasing $\mu(h)$ \label{subsec:proof-nonincreasing-mu-h}}

Any functional form that implies that $\mu(h)$ weakly increases in $h$ contradicts the data. Hence, even if the model were estimated without the restriction that $\mu(h)$ weakly decreases in $h$, it would necessarily produce estimates that imply that the assumption holds. To show this, I first establish the following proposition:

\begin{proposition} \label{prop:non-increasing-mu-h}
  If $\mu(h)$ weakly increases in $h$ and the ranking is randomized, then there are no position effects on positions $h > 1$ for consumers who take the outside option.
\end{proposition}
\begin{proof}
The optimal policy is based on the three reservation values independent of how $\mu(h)$ depends on $h$, and $z^d(h)$ will continue to in- or decrease in $h$ depending on how $\mu(h)$ in- or decreases in $h$ \citep[see][]{Greminger2021}. The result then follows from the stopping rule. This rule implies that consumers who take the outside option continue discovering as long as $U_0 > z^d(h)$. As a result, if $\mu(h)$ and, hence, $z^d(h)$ weakly increase in $h$, consumers who take the outside option always discover either only the first---which is discovered for free---or all available alternatives. This already implies the result because position effects can only arise in the SD model through consumers stopping discovery before reaching the last position, unless the ranking is not randomized.
\end{proof}

The proposition makes a prediction for the case when $\mu(h)$ is weakly increasing in $h$. This prediction contradicts the data because there are position effects for consumers who take the outside option, including for positions $h > 1$. Specifically, re-running the analysis from Section \ref{subsec:Descriptive-evidence-of} while excluding the first position continues to produce significant position effects. Hence, a weakly increasing $\mu(h)$ contradicts the data. Moreover, note that when consumers know that they are facing a randomized ranking, $\mu(h)$ will be constant. Hence, the proposition also implies that consumers do not adjust their beliefs immediately to the randomized ranking; otherwise, the position effects would disappear for consumers who take the outside option.

\clearpage

\section{Software Implementation \label{subsec:software}}

I provide the estimation approach as Julia and Python packages. The packages are available on GitHub and can be installed using the usual package managers. The Julia package is the main implementation and includes all the functionalities, while the Python package provides a convenient interface to this package for Python users. Both packages are open-source and freely available. The links to the repositories are as follows:

\begin{itemize}[leftmargin=30pt, rightmargin=2cm, label=--]
    \item Link to GitHub repository for Julia package: \emph{\url{https://github.com/rgreminger/StructuralSearchModels.jl}}.
    \item Link to GitHub repository for Python wrapper: \emph{\url{https://github.com/rgreminger/structuralsearchmodels}}.
\end{itemize}

In its current form, the Julia package implements the estimation procedure developed in this paper, as well as methods to simulate data from the models. It is designed to be user-friendly, and comes with documentation and examples. Moreover, it is optimized for performance and efficiently estimates the two models using parallelization and automatic differentiation.

The Julia package builds on various open-source software and packages. In particular, the main dependencies are the following:
\begin{itemize}[leftmargin=30pt, rightmargin=2cm, label=--]
    \item \emph{Julia} by \cite{Bezanson2017}: high-level programming language used for the implementation.
    \item \emph{Distributions.jl} by \cite{Besancon2021}: implements distribution types that make it easy to work with different distributions.
    \item \emph{Optimization.jl} by \cite{Dixit2023}: flexible framework that allows to define optimization problems and solve them using different solvers and options.
    \item \emph{Optim.jl} by \cite{Mogensen2018}: provides numerical optimization routine used by default.
    \item \emph{ForwardDiff.jl} by \cite{Revels2016}: provides automatic differentiation to compute gradients and Hessians for gradient-based optimization.
\end{itemize}

The fit plots in this paper are not provided as part of the package to keep the dependencies limited. However, the respective codes are available as part of the replication package. These codes rely on the \emph{Makie.jl} plotting package by \cite{Danisch2021}.

\clearpage

\clearpage

\section{Data\label{subsec:Data-preparation}}

The original dataset from Kaggle.com contains 9,917,530 observations on a hotel-session level. Following \cite{Ursu2018}, I exclude sessions with at least one observation satisfying any of the following criteria: 
\begin{enumerate}
\item The implied tax paid per night either exceeds 30\% of the listed hotel price (in \$), or is less than \$1. 
\item The listed hotel price is below \$10 or above \$1,000. 
\item There are less than 50 consumers looking for hotels at the same destination throughout the sample period. 
\item The consumer observed a hotel in positions 5, 11, 17, or 23, i.e., the consumer did not have opaque offers (\citet{Ursu2018} provides a detailed description of this feature in the data). 
\end{enumerate}
The final dataset contains 4,503,128 observations, which is 85 fewer observations than \citet{Ursu2018} reports. The difference stems from three sessions (IDs 79921, 94604, 373518). For these sessions, the criteria above yields a different result due to differences in the numerical precision. Specifically, I evaluate the criteria with double precision calculations, whereas \citet{Ursu2018} uses the Stata default, single precision. 

Completing information on the dataset, Table \ref{tab:variable_descr} provides a detailed description of each variable. 

\begin{table}[!h] \centering \footnotesize
\caption{Variable Descriptions} \label{tab:variable_descr}
\begin{threeparttable}
\begin{tabular}{lll} 
\midrule
\bfseries{Hotel-level} \\
\hspace{1em} Price (in \$)&& Gross price in USD \\
\hspace{1em} Star rating&& Number of hotel stars \\
\hspace{1em} Review score&& User review score, mean over sample period \\
\hspace{1em} No reviews && Dummy whether hotel has zero reviews (not missing) \\
\hspace{1em} Chain&& Dummy whether hotel is part of a chain\\
\hspace{1em} Location score&& Expedia's score for desirability of hotel's location\\
\hspace{1em} On promotion&& Dummy whether hotel on promotion \\
\midrule
\bfseries{Session-level} \\ 
\hspace{1em} Number of items&& How many hotels in list for consumer, capped at first page\\
\hspace{1em} Number of clicks&& Number of clicks by consumer\\
\hspace{1em} Made booking&& Dummy whether consumer made a booking\\
\hspace{1em} Trip length (in days)&& Length of stay consumer entered \\
\hspace{1em} Booking window (in days)&&Number of days in future that trip starts\\
\hspace{1em} Number of adults && Number of adults on trip \\
\hspace{1em} Number of children && Number of children on trip\\ 
\hspace{1em} Number of rooms && Number of rooms in hotel\\
\midrule 
\end{tabular}
\end{threeparttable}
\end{table}

\end{document}